\documentclass[a4paper,10pt,xcolor=dvipsnames]{article}

\usepackage{cite}
\usepackage[sort&compress,numbers]{natbib}

\usepackage{geometry}
\usepackage{lscape}
\usepackage{array}
\usepackage{tikz}
\usepackage{tikz-qtree}

\usepackage{colortbl} 

\usepackage{graphicx}
\usepackage{amsmath}
\usepackage{bbm}
\usepackage{amsfonts}
\usepackage{amssymb}
\usepackage{amsmath}
\usepackage{latexsym}
\usepackage{xcolor}
\usepackage{url}
\usepackage{ntheorem}
\usepackage{braket}
\usepackage[normalem]{ulem}
\usepackage{enumerate}
\usepackage[colorlinks=true]{hyperref}
\usepackage{comment}
\usepackage{subcaption}
\usepackage{amsmath}

\newcommand{\fpsi}{\blue{f^*\psi}}

\newcommand{\ngkappa}{\red{\tau}}
\newcommand{\hkappa}{\blue{\hat\ngkappa}}
\newcommand{\ckappa}{\blue{\check\ngkappa}}
  
  \newcommand{\hN}{\blue{\hat N}}
  \newcommand{\chN}{\blue{\check N}}
  \newcommand{\interval}{\blue{I}}

\newcommand{\nomc}{\red{\sigma}}

\newcommand{\ptcnh}[1]{\mnote{{\bf ptc:}#1}}
\newcommand{\ptcrnh}[1]{\mnote{{\bf ptc:}#1}}

\DeclareFontFamily{U}{mathx}{}
\DeclareFontShape{U}{mathx}{m}{n}{<-> mathx10}{}
\DeclareSymbolFont{mathx}{U}{mathx}{m}{n}
\DeclareMathAccent{\widehat}{0}{mathx}{"70}
\DeclareMathAccent{\widecheck}{0}{mathx}{"71}

\newcommand{\sv}{\red{\tilde v }}

\newcommand{\bv}{\blue{ v }}

\newcommand{\hbeta}{\red{\hat\beta}}
\newcommand{\cbeta}{\red{\check\beta}}
\newcommand{\boost}{\blue{\hat C}}
\newcommand{\hboost}{\boost}

\newcommand{\shiftH}{\blue{\hat c}}
\newcommand{\newcshiftH}{\blue{\check \omega}}
\newcommand{\newhshiftH}{\blue{\hat \omega}}
\newcommand{\hshiftH}{\shiftH}

\newcommand{\mcHtwo}{\red{\mcH^2}}

\newcommand{\cboost}{\blue{\check C}}
\newcommand{\cshiftH}{\blue{\check c}}

\newcommand{\cpsi}{\blue{\check \psi}}
\newcommand{\hR}{\blue{\widehat{\mathbf{R}}}}
\newcommand{\cR}{\blue{\widecheck{\mathbf{R}}}}

\newcommand{\eqrefl}[1]{\eqref{#1}}

\newcommand{\chM}{\blue{\widecheck{M}}}
\newcommand{\chm}{\blue{\check m}}
\newcommand{\chg}{\blue{\check g}}
\newcommand{\chf}{\blue{\check f}}

\newcommand{\chpsi}{\blue{\check \psi}}

\newcommand{\mono}{\blue{\mathbf{M}}}

\newcommand{\cmonoz}{\widecheck{\mono}_0}
\newcommand{\chmonoz}{\widecheck{\mono}_0}
\newcommand{\chmono}{\widecheck{\mono}}
\newcommand{\cmono}{\chmono}

\newcommand{\hmono}{\widehat{\mono}}
\newcommand{\hmonoz}{\widehat{\mono}_0}
\newcommand{\hatM}{\blue{\widehat M}}
\newcommand{\hm}{\blue{\hat m}}
\newcommand{\hatg}{\blue{\hat g}}
\newcommand{\hf}{\blue{\hat f}}

\newcommand{\tfg}{\red{{\fourg}}}
\newcommand{\fourg}{\red{\mathbf g }}

\newcommand{\fourb}{\red{\mathbf{b}}}

\newcommand{\mcordBK}{\teal{\frak{m}}}

\newcommand{\teal}[1]{{\color{teal}#1}}

\newcommand{\Hzero}{\teal{H}}
\newcommand{\HZERO}{\Hzero}

\newcommand{\omofvarphi}{\red{\varphi(\bar \varphi)}}

\newcommand{\zerot}{\red{0}}
\newcommand{\twor}{\red{2}}
\newcommand{\onephi}{\red{1}}

\newcommand{\minHz}{\teal{\underline{\Hzero}}}

\newcommand{\nohvarphi}{\blue{\varphi}}

\newcommand{\hypmet}{\blue{\mathring b}}

\newcommand{\mcord}{\blue{M}}

\newcommand{\cNb}{\red{\mycal K_b}}
\newcommand{\Ricb}{\red{\mathring{R}}}

\newcommand{\HMext}{\red{ {M_\ext}}}

\newcommand\ben{\begin{enumerate}}
\newcommand\een{\end{enumerate}}
\newcommand\bit{\begin{itemize}}
\newcommand\eit{\end{itemize}}

\newcommand{\qedskip}{\qed\medskip}

\newcommand{\blue}[1]{{\color{blue}#1}}
\newcommand{\red}[1]{{\color{red}#1}}

\newcounter{mnotecount}[section]

\renewcommand{\themnotecount}{\thesection.\arabic{mnotecount}}

\newcommand{\mnote}[1]
{\protect{\stepcounter{mnotecount}}$^{\mbox{\footnotesize
$
\bullet$\themnotecount}}$ \marginpar{
\raggedright\tiny\em
$\!\!\!\!\!\!\,\bullet$\themnotecount: #1} }

\newcommand*{\rr}{{r}}

\newtheorem{theorem}{\sc  Theorem\rm}[section]

\newtheorem{conjecture}[theorem]{\sc  Conjecture\rm}
\newtheorem{corollary}[theorem]{\sc  Corollary\rm}

\newtheorem{lemma}[theorem]{\sc Lemma\rm}
\newtheorem{Lemma}[theorem]{\sc Lemma\rm}

\newtheorem{proposition}[theorem]{\sc Proposition\rm}
\newtheorem{Proposition}[theorem]{\sc Proposition\rm}

\theorembodyfont{}
\newtheorem{remark}[theorem]{\sc Remark\rm}
\newtheorem{Remark}[theorem]{\sc Remark\rm}

\newcommand{\jlcax}[1]{}
\newcommand{\eean}{\nonumber\end{eqnarray}}

\newcommand{\mcP}{{\mycal P}}

\newcommand{\kk}[1]{}

\newcommand{\mcH}{{\mycal H}}

\newcommand{\beq}{\begin{equation}}

\newcommand{\hpsi}{\hat \psi}

\newcommand{\FS}       
                  {F}

\newcommand{\HS} 
       {H_{\mbox{\scriptsize volume}}}

{\ptc{this should be removed in the oberwolfach version}}%

\newcommand{\mc}{\red{m_{\mathrm c}}}

\newcommand{\zD}{\mathring{D}}%
\newcommand{\mcE}{{\mycal E}}%

\newcommand{\ourU}{\mathbb U}%
\newcommand{\eeal}[1]{\label{#1}\end{eqnarray}}
\newcommand{\id}{\textrm{id}}
\newcommand{\C}{{\mathbb C}}
\newcommand{\bed}{\begin{deqarr}}
\newcommand{\eed}{\end{deqarr}}
\newcommand{\bedl}[1]{\begin{deqarr}\label{#1}}
\newcommand{\eedl}[2]{\arrlabel{#1}\label{#2}\end{deqarr}}

\newcommand{\bel}[1]{\begin{equation}\label{#1}}
\newcommand{\bea}{\begin{eqnarray}}
\newcommand{\bean}{\begin{eqnarray}\nonumber}
\newcommand{\beal}[1]{\begin{eqnarray}\label{#1}}
\newcommand{\eea}{\end{eqnarray}}

\newcommand{\qed}{\hfill $\Box$}
\newcommand{\proof}{\noindent {\sc Proof:\ }}
\newcommand{\be}{\begin{equation}}
\newcommand{\eeq}{\end{equation}}
\newcommand{\ee}{\end{equation}}
\newcommand{\beqa}{\begin{eqnarray}}
\newcommand{\eeqa}{\end{eqnarray}}
\newcommand{\beqan}{\begin{eqnarray*}}
\newcommand{\eeqan}{\end{eqnarray*}}
\newcommand{\ba}{\begin{array}}
\newcommand{\ea}{\end{array}}

\newcommand{\Id}{\mbox{\rm Id}} 

\newcommand{\hyp}{\mycal S}

\newcommand{\warn}[1]
{\protect{\stepcounter{mnotecount}}$^{\mbox{\footnotesize
$
\bullet$\themnotecount}}$ \marginpar{
\raggedright\tiny\em
$\!\!\!\!\!\!\,\bullet$\themnotecount: {\bf Warning:} #1} }

\newcommand{\R}{\mathbb R}

\newcommand{\N}{\mathbb N}
\newcommand{\Z}{\mathbb Z}

\newcommand{\eq}[1]{(\ref{#1})}

\newcommand{\ext}{\mathrm{ext}}

\newcommand{\ptc}[1]{\mnote{{\bf ptc:}#1}}

\newcommand{\ptcheck}[1]{\mnote{{\bf ptc :}checked on  #1}}

\newcommand{\beqar}{\begin{deqarr}}
\newcommand{\eeqar}{\end{deqarr}}

\newcommand{\beaa}{\begin{eqnarray*}}
\newcommand{\eeaa}{\end{eqnarray*}}

\newcommand{\tr}{\mbox{tr}}

\newcommand{\hg}{{\hat g}}

\newcommand{\chcoshx}{\cosh (\sqrt{\chm} \pi )}
\newcommand{\hcosx}{\cos (\sqrt{\left| \hm\right| }\pi)}
\newcommand{\hsinx}{\sin (\sqrt{\left| \hm\right| }\pi)}
\newcommand{\chcosh}{\widecheck\cosh{}\,}

\newcommand{\chcos}{\widecheck\cos{}\,}
\newcommand{\chsin}{\widecheck\sin{}\,}
\newcommand{\hcos}{\widehat\cos{}}
\newcommand{\hsin}{\widehat\sin{}}

\newcommand{\sigmar}{\textcolor{red}{\sigma}}

\DeclareFontFamily{OT1}{rsfs}{}
\DeclareFontShape{OT1}{rsfs}{CGNPm}{n}{ <-7> rsfs5 <7-10> rsfs7 <10-> rsfs10}{}
\DeclareMathAlphabet{\mycal}{OT1}{rsfs}{CGNPm}{n}

{\catcode `\@=11 \global\let\AddToReset=\@addtoreset}
\AddToReset{equation}{section}

{\catcode `\@=11 \global\let\AddToReset=\@addtoreset}
\AddToReset{figure}{section}

{\catcode `\@=11 \global\let\AddToReset=\@addtoreset}
\AddToReset{table}{section}

\renewcommand{\ptcheck}[1]{}
\renewcommand{\red}[1]{#1}
\renewcommand{\blue}[1]{#1}
\renewcommand{\teal}[1]{#1}
\renewcommand{\ptcnh}[1]{}
\renewcommand{\ptcrnh}[1]{}

\begin{document}
\title{Gluing-at-infinity of two-dimensional asymptotically locally hyperbolic manifolds\protect\thanks{Preprint UWThPh-2024-1}}
\author{Piotr T. Chru\'{s}ciel\thanks{University of Vienna, Faculty of Physics, and Beijing Institute for Mathematical Sciences and Applications} \thanks{
{\sc Email} \protect\url{piotr.chrusciel@univie.ac.at}, {\sc URL} \protect\url{homepage.univie.ac.at/piotr.chrusciel}}
\\
{Raphaela Wutte}\thanks{Department of Physics and Beyond: Center for Fundamental Concepts in Science, Arizona State University, and Physique Math\'{e}matique des Interactions Fondamentales, Universit\'{e} Libre de Bruxelles}
\thanks{
{\sc Email} \protect\url{rwutte@hep.itp.tuwien.ac.at} {} \protect\url{}}
}
\maketitle

\begin{abstract}
We review notions of mass of asymptotically locally Anti-de Sitter three-dimensional spacetimes, and apply them to some known solutions.
For two-dimensional general relativistic initial data sets the mass is \emph{not} invariant under asymptotic symmetries, but a unique mass parameter can be obtained either by minimisation, or by a monodromy construction, or both.
We give an elementary proof of positivity, and of a Penrose-type inequality, in a natural gauge.
We apply the ``Maskit gluing construction'' to time-symmetric asymptotically locally hyperbolic vacuum initial data sets and derive mass/entropy formulae for the resulting manifolds.
Finally, we show that all mass aspect functions can be realised by constant scalar curvature metrics on complete manifolds which are smooth except for at most one conical singularity.
\end{abstract}

\tableofcontents

\section{Introduction}

One of the challenges of mathematical relativity is the construction of general relativistic initial data sets with interesting physical properties. In~\cite{ILS} the ``Maskit gluing'' method was introduced for such a purpose, allowing one to construct new initial data sets with a negative cosmological constant out of old ones, via a connected-sum construction at the conformal boundary at infinity. A refinement of this method has been presented in~\cite{ChDelayExotic}.

A natural question that arises in the Maskit gluing context is that of the total energy of the glued initial data sets. This has been analysed in~\cite{CDW,ChDelayGluing} in space-dimensions $n\ge 3$. The object of this work is to extend the analysis there to two-dimensional time-symmetric general relativistic initial data sets. Indeed, we derive formulae for the total mass of a Maskit-glued two-dimensional asymptotically locally hyperbolic (ALH) manifold in terms of the total masses of the summands.

Recall (see Section~\ref{s1XI23.1} below) that  many (in fact most, in a specific sense) complete vacuum time-symmetric initial data sets with a finite number of locally hyperbolic ends describe spacetimes with Killing horizons.
The mass formulae we derive have then an entropy interpretation. Indeed, one of our mass formulae
(Equation \eqrefl{3XII23.2} below) can be rephrased using thermodynamical language: the entropy $S$ of a black hole obtained by Maskit-gluing of two static black holes with entropies $\check S$ and $\hat S$ is determined by the equation
\begin{equation}\label{19XII23.1}
    \cosh(2S )
    =   2\,  \newcshiftH\,
  \newhshiftH \cosh   (2 \check S) \cosh   (2 \hat S)- \cosh  \big (2 (\check S - \hat S)\big)
  \,,
\end{equation} %
where $\newcshiftH$ and   $\newhshiftH $ are parameters, introduced
in the gluing procedure, satisfying $\newcshiftH>1$ and $\newhshiftH>1$ and which are arbitrary otherwise.

At first sight the problem in vacuum and time-symmetry appears to be much simpler in space-dimension two, when compared to the remaining dimensions, as the space-metric is then locally isometric to hyperbolic space. However,  space-dimension two turns out to be more interesting from the mass perspective, as asymptotic symmetries act in a complicated way on energy-momentum-type integrals. So while one avoids the technical intricacies related to gluing, which had to be addressed in~\cite{CDW,ChDelayGluing}, one faces the challenge of extracting a mass-type invariant from the mass-aspect function of the glued manifold.

 Before we proceed to gluing, it is relevant to review what is known about mass of asymptotically locally hyperbolic three-dimensional spacetimes, or two-dimensional general relativistic initial data sets.
 For this, we start in Section~\ref{s19XII23.1} with the standard examples of  asymptotically locally hyperbolic manifolds with constant scalar curvature (ALH CSC), and discuss there the simplest notions of mass. In Section~\ref{s19XII23.2} we review the definition  of the Hamiltonian mass, analyse its behaviour under asymptotic symmetries, and show that in favorable situations an invariant can be extracted from the mass-aspect function by minimising its integral.
 In Section~\ref{s16I22.3} we establish both positivity of the mass and a Penrose-type inequality for specific classes of initial data sets.
 In Section~\ref{s1XI23.1} we review known results on vacuum, geometrically finite, time-symmetric, complete initial data sets, and point out that the mass-aspect function can always be made constant for such data.
 In Section~\ref{secgluingconical} we show how to glue together, near the conformal boundary at infinity, two ALH
 CSC metrics.
 Our main result, namely the formulae for the mass of the glued manifold, can be found in Section~\ref
 {ss10X23.1a}.
It should be stressed that the gluing construction here only requires the metric to be locally hyperbolic near the conformal boundary at infinity, no
further global assumptions are needed.
In Section~\ref{s27X23.1} we
show
that all mass-aspect functions
   can be obtained from time-symmetric, asymptotically locally hyperbolic, vacuum general relativistic initial data sets
on a geometrically finite topological manifold with a metric which is  either smooth  or has one conical singularities.  
In Appendix~\ref{boostappendix}
we analyse the behaviour of the mass-aspect function under boosts.

 \bigskip

 {\noindent \sc Acknowledgements:} PTC is grateful to Mark Troyanov and Greg Galloway for useful discussions, and Marius Oancea for bibliographic advice. He thanks the BIMSA for hospitality and support during part of work on this paper. RW is grateful to Glenn Barnich, Blagoje Oblak, Jakob Salzer, Georg Stettinger and Friedrich Schöller for useful conversations. She is supported by the
 Heising-Simons Foundation under the “Observational Signatures of Quantum Gravity”
 collaboration grant 2021-2818 and the U.S. Department of Energy, Office of High Energy Physics,
 under Award No. DE-SC0019470. She also acknowledges support of the
 Fonds de la Recherche Scientifique F.R.S.-FNRS (Belgium) through the PDR/OL C62/5 project
 “Black hole horizons: away from conformality” (2022-2025) and
of the Erwin Schr\"odinger Institute for Mathematics and  Physics, Vienna, during part of work on this paper.

\section{Asymptotically hyperbolic mass in
two dimensions}
\label{s19XII23.1}

In vacuum and in time symmetry, after normalising the negative cosmological constant $\Lambda$ to $-1$,  the $n$-dimensional general relativistic constraint equations become
\begin{equation}
 R= -n(n-1)
 \,.
\end{equation}
In space-dimension $n=2$, we are thus left to consider two-dimensional Riemannian manifolds with metrics of  constant scalar curvature equal to
$-2$. In the mathematics literature such surfaces are usually referred to as
\emph{hyperbolic}.

See, e.g.,  \cite{Barrow3d,GarciaDiazBook} for a wealth of three-dimensional spacetimes with matter fields.

Here some comments are in order.
Whatever the dimension, the following terminology is often used:

\begin{enumerate}
 \item  A metric $g$  is said to be \emph{hyperbolic} if all its sectional curvatures are equal to $-1$.

      In dimension $n=2$, in time-symmetry, and in vacuum, for a general relativistic initial data set this is the same as constant scalar curvature equal to $-n(n-1)$, but is not in higher dimensions.

 \item A Riemannian manifold $(M,g)$ is said to be \emph{hyperbolic} if $g$ is.
     \item
\emph{Hyperbolic space} is the unique, up to isometry, complete simply connected hyperbolic manifold.

     Note that the metric on  a quotient of hyperbolic space is hyperbolic.
 \item A metric is said to have an \emph{asymptotically locally hyperbolic end} (ALH) if it contains a region in which the metric approaches a hyperbolic metric as one recedes to infinity.  The rates of approaches will be made precise wherever relevant. In our gluing results here, we will actually assume that the scalar curvature of $g$ is constant in the gluing region.
      \item An end  is said to be \emph{asymptotically hyperbolic} (AH) if the metric approaches the metric on hyperbolic space as one recedes to infinity. In the conformally compactifiable case, this corresponds to the requirement that the conformal boundary be diffeomorphic to a sphere, and that the metric on the conformal boundary be conformal to the canonical round metric on the sphere.
\end{enumerate}

While these distinctions are clear in higher dimensions, the distinction between AH and ALH   in dimension two requires clarification.
Indeed, the conformal boundary of a two-dimensional, conformally compactifiable, hyperbolic metric is necessarily diffeomorphic to $S^1$, and the conformal metric there is necessarily conformal to the canonical round metric on $S^1$. To avoid the need to differentiate between the notions  for two-dimensional metrics, we will only use the terminology ALH in two dimensions.

\subsection{The coordinate mass}
\label{s16I22.11}

The $(n+1)$-dimensional Birmingham-Kottler (BK)
metrics with \emph{spherical conformal infinity}, $n\ge 3$ are usually written in the form
\begin{equation}
    \label{16I22.21}
       \tfg  = - \big(
        r^2 +1 -
        \frac{2\mcord}{r^{n-2}}
        \big)  dt^2 +
      \frac{ dr^2 }{r^2 +1 -
        \frac{2\mcord}{r^{n-2}}}+ r^2 d\Omega^2
        \,,
    \end{equation}
    where $d\Omega^2$ is the unit round metric on $S^{n-1}$. The case $\mcord=0$ corresponds to hyperbolic space.
    In spacetime dimension four, the parameter $\mcord$ is usually thought of as the total mass of the metric. Keeping this same notation for $n=2$, we   write
\begin{equation}
    \label{16I22.22}
      \tfg  = - \big(
        r^2 +1 - 2\mcord
        \big)  dt^2 +
      \frac{dr^2}{ r^2 +1 - 2\mcord  }  + r^2 d\varphi^2
      \,,
       \quad
       \mbox{where}\  e^{i\varphi} \in S^1
       \,,
    \end{equation}
    and refer to the parameter $\mcord$ as the \emph{coordinate mass}.

On the other hand, the $(n+1)$-dimensional Birmingham-Kottler (BK)
metrics, $n\ge 3$, with \emph{Ricci-flat conformal infinity} are usually written in the form
\begin{equation}
    \label{16I22.21a}
       \tfg  = - \big(
        r^2  -
        \frac{2\mcordBK}{r^{n-2}}
        \big)  dt^2 +
      \frac{dr^2}{r^2   -
        \frac{2\mcordBK}{r^{n-2}}}  + r^2 d\Omega^2
        \,,
    \end{equation}
where $d\Omega^2$ is now a Ricci-flat metric on an  $(n-1)$-dimensional manifold. Since $S^1$ is certainly Ricci-flat, when $n=2$ one is tempted to write
\begin{equation}
    \label{16I22.22a}
       \tfg  = - \big(
        r^2  - 2\mcordBK
        \big)  dt^2 +
      \frac{dr^2}{ r^2  - 2\mcordBK  }  + r^2 d\varphi^2
      \,,
       \quad
       \mbox{where}\  e^{i\varphi} \in S^1
       \,.
    \end{equation}

So, when $n=2$, we need  to decide whether a circle $S^1$ at infinity qualifies as a special case of a spherical conformal infinity, in which case the notation \eqref{16I22.22} applies,
or a special case of a Ricci-flat conformal infinity,  in which case the notation \eqref{16I22.22a} applies.

We resolve the dilemma by opting for yet another notation  for the definition of coordinate mass
$\mc$, namely we will write
\begin{equation}
    \label{27VII23.91}
       \tfg  = - \big(
        r^2  -\mc
        \big)  dt^2 +
      \frac{dr^2}{ r^2  - \mc }  + r^2 d\varphi^2
      \,,
       \quad
       \mbox{where}\  e^{i\varphi} \in S^1
       \,,
    \end{equation}
because this has been used in several papers building on~\cite{BTZ,Banados:1992gq}.
In this notation hyperbolic space has
\emph{coordinate
 mass equal to $-1$},
\begin{equation}
    \label{16I22.22ba}
       \tfg  = - \big(
        r^2 +1
        \big)  dt^2 +
      \frac{ dr^2}{ r^2  +1 } + r^2 d\varphi^2
      \,,
       \quad
       \mbox{where}\  e^{i\varphi} \in S^1
       \,.
    \end{equation}

As  first pointed out by Ba\~nados, Teitelboim and Zanelli~\cite{BTZ}, for
$
 \mc >0
$
the metric \eqref{27VII23.91}
 can be extended by continuity across an event horizon, which provides the
rationale for the choice of normalisation
\eqref{27VII23.91}.

Riemannian metrics with leading-order behaviour given by
the space-part of \eqref{27VII23.91} will be  said to have coordinate mass $\mc$.

We have the trivial relations
\begin{equation}\label{27VII23.92}
2  \mcordBK =  \mc = 2 \mcord -1
  \,.
\end{equation}
While the transition from $\mc $ to $\mcordBK$ is innocuous, the one from $\mc$ to $\mcord$ is not, because it determines the background which is used to determine the zero-point of the mass. Indeed, a metric which asymptotes to a background metric $\fourb$ defined as \eqref{27VII23.91}
with $\mc=0$ will be naturally expressed in terms of the $\fourb$-orthonormal coframe
\begin{equation}\label{27VII23.93}
 \theta^{\zerot}= r dt
\,, \quad
 \theta^{\onephi}= r d\varphi
\,, \quad
 \theta^{\twor} = \frac{dr}{r}
 \,,
\end{equation}
while using \eqref{16I22.22ba} as the asymptotic background leads naturally to
\begin{equation}\label{27VII23.94}
 \theta^{\zerot}=\sqrt{r^2+1} \, dt
\,, \quad
 \theta^{\onephi}= r \, d\varphi
\,, \quad
 \theta^{\twor} = \frac{dr}{\sqrt{r^2+1}}
 \,.
\end{equation}

\subsection{Mass as a deficit angle}
 \label{s16I22.2}
 \newcommand{\phic}{\blue{\varphi_{\mathrm{c}}}}
The definition of mass as a deficit angle has been advocated in~\cite[Remark~3.1]{ChHerzlich} (see also \cite{Wong2d,ashtekar:varadarajan}).
This proceeds as follows:

Consider the constant-time slices
of the $(2+1)$-dimensional Birmingham-Kottler metrics,
\be\label{twoK}
 g=
 \frac{dr^2}{r^2 -\mc} +r^2 d\varphi^2
  \,,
 \quad
  \varphi\in[0,2\pi]\ \mbox{mod\ } 2\pi \,,
\ee
with
$\mc\in\R$.
When
$$
 \mc<0
$$
 the  coordinate transformation
$$
 \bar r = \lambda r
 \,,
 \qquad
  \bar \varphi = \varphi/\lambda
  \,,
$$
with
$$
  \lambda =
 \frac{1}{\sqrt{|\mc|}}
\,,
$$
brings the metric \eq{twoK} to one of the standard forms of the
hyperbolic-space metric,
\be\label{twoK2} g=\frac{d\bar r^2}{\bar r^2+1} +\bar r^2
d\bar \varphi^2\,, \quad \bar \varphi\in[0,2\pi/\lambda]\ \mbox{mod\ }
2\pi/\lambda \,,\ee
where now the new
angular variable $\bar \varphi$ does \emph{not} range over $[0,2\pi]$; that range will coincide with the
standard one
if and only if   $\mc = -1$.

So, in this sense the deficit angle, say $\phic$, of the metric \eqref{27VII23.91} with $\mc<0$ equals
\begin{equation}\label{16I22.13}
 \phic:=   2\pi \lambda^{-1} - 2\pi
  =   2\pi \big((-\mc)^{1/2} -1
   \big)
  \,.
\end{equation}

It turns out that when $\mc>0$, there is another natural  procedure to define the deficit angle, as follows:
For positive $\mc$, the curve  $r^2=\mc$ is a closed, outermost minimalising geodesic for the metric  \eqref{27VII23.91},
corresponding to the bifurcation ``surface'' of a Killing horizon in the associated static spacetime.
(Here ``outermost'' refers to the connected component of conformal infinity under consideration.)
A natural coordinate, say $u$, is the distance to this geodesic, which is given by the formula
\begin{equation}\label{27X23.1}
  r=  \sqrt\mc\cosh(u)
  \,,
\end{equation}
so that the space-part of the metric \eqref{27VII23.91} becomes
\begin{equation}
     \label{27VII23.9110}
      g =
       \frac{dr^2}{ r^2  - \mc }  + r^2 d\varphi^2
       =
       d u^2 + \mc \cosh^2(u) \, d\varphi^2
       \,,
        \quad
        \mbox{where}\  e^{i\varphi} \in S^1
        \,.
     \end{equation}
One can absorb the prefactor $\sqrt{\mc}$ into a new  angle $\bar \varphi:= \sqrt{\mc} \varphi$, which brings the metric into a
canonical form:
\begin{equation}
     \label{27VII23.9111}
      g =
       d u^2 +  \cosh^2(u) \, d\bar\varphi^2
       \,,
        \quad
        \mbox{where}\  e^{i\bar \varphi/ \sqrt{\mc}} \in S^1
        \,.
     \end{equation}
     The coordinate $\bar\varphi$ is therefore $2\pi \sqrt{\mc}$-periodic, leading to a deficit angle
\begin{equation}\label{27X23.4}
   \phic
:=   2\pi \big((\mc)^{1/2} -1
   \big)
  \,.
\end{equation}
Comparing with \eqref{16I22.13}, we obtain a formula for $\phic$ which is independent of the sign of $\mc$:
\begin{equation}\label{27X23.5}
   \phic
  =   2\pi \big((|\mc|)^{1/2} -1
   \big)
  \,.
\end{equation}

We emphasise that this definition has an invariant geometric character when $\mc>0$, since then \emph{the length of the smooth curve lying a distance $u$ from the outermost minimising closed geodesic $\{u=0\}$  equals}
$$
 2 \pi \sqrt{\mc}\cosh(u) =  (2\pi +\phic)  \cosh(u)
 \,.
$$

In particular, the length $\ell$ of this closed geodesic in terms of the mass is
\begin{equation}\label{10XII23.4}
  \ell =
  2 \pi \sqrt{\mc}
   \,.
\end{equation}
We note that the area of cross-sections of the event horizon, so in our context the length of the outermost closed geodesic, is related to the entropy $S$ of the black hole as $S=\ell/4$ (see e.g.~\cite[Equation~(5.5)]{Barnich:2012xq}, where we have set Newton's constant to 1), leading to
\begin{equation}\label{10XII23.3}
  S = \frac{\pi \sqrt{\mc}}{2}
   \,.
\end{equation}

\subsection{Hamiltonian mass}
 \label{s28VII23.1}

 We specialise the definition of hyperbolic mass of~\cite{ChHerzlich} to space-dimension $n=2$.
 While the definition given there was based on an ad-hoc procedure, the resulting functional coincides with the Hamiltonian mass of~\cite{CJL} in many situations of interest, which justifies the name.

For this we consider metrics which asymptote to the  metric
   $b$  given by the space-part of  \eqref{27VII23.91} with
 $\mc=0$,
\begin{equation}
     \label{27VII23.91s}
      b=
       \frac{dr^2}{ r^2  }  + r^2 d\varphi^2
       \,,
        \quad
        \mbox{where}\  e^{i\varphi} \in S^1
        \,.
     \end{equation}
      In the coordinates of
     \eqref{27VII23.91s}  we set
     \begin{equation}\label{3VII21.2}
     \HMext:= [R,\infty)\times S^1
      \,,
     \end{equation}
     for some large $R\in \R^+$, and we consider there the  $b$-orthonormal coframe \eqref{27VII23.93}. Letting $f_i$ be the dual frame to $\theta^i$, we set
     \be \label{m2} g_{ij}:=g(f_i,f_j)
      \,,
      \quad b_{ij}:=b(f_i,f_j)
      = \left\{
          \begin{array}{ll}
            1, & \hbox{$i=j$;} \\
            0, & \hbox{otherwise,}
          \end{array}
        \right.
     \quad
      e_{ij}:=g_{ij}-b_{ij}
       \,.
     \ee
  We will assume that
     \begin{subequations}
      \label{Hm3}
     \begin{gather}\label{Hm3a}
      \displaystyle
      \int_{\HMext} \big( \sum_{i,j} |g_{ij}-b_{ij}|^2 + \sum_{i,
      j,k} |f_k(g_{ij})|^2 \big)r
      \;d\mu_g<\infty\,,
     \\
      \displaystyle
      \int_{\HMext} |R_g-R_b|\;r\;d\mu_g<\infty\,,\label{Hm3b}
    \end{gather}
  \end{subequations}
     \begin{equation}
     \label{m0} \exists \ C > 0 \ \textrm{ such that }\
     C^{-1}b(X,X)\le g(X,X)\le Cb(X,X)\,,
     \end{equation}
     where
     $d \mu_{g}$ denotes the measure associated with
     the metric $g$.

The background metric $b$ is equipped with a one-dimensional set of solutions of the \emph{static Killing Initial Data (KID)}
     equations, denoted by  $
     \cNb$,  spanned by the function $V=r$
      (cf., e.g., \cite[Appendix~B.1.3]{CDW}):
     \begin{eqnarray}
      &  \label{eq:1}
       \red{\mathring \Delta}  V -2V =0\,,
      &
     \\
      &
       \label{eq:2} \zD_i\zD_j V = V( \Ricb_{ij} +2
     b_{ij})
      \,,
      &
     \end{eqnarray}
     where   $\Ricb_{ij}$ denotes the Ricci tensor of the
     metric $b$, $\zD$ the Levi-Civita connection of $b$, the operator
     $\red{\mathring \Delta}:=b^{k\ell}\zD_k\zD_\ell $ is the Laplacian of $b$, and we use the symbol $D$ for the Levi-Civita connection of $g$. Nontrivial
     triples $(M,b,V)$, where $V$ solves \eq{eq:1}-\eq{eq:2}, are called
     \emph{static Killing Initial Data (KIDs)}.

     The \emph{Hamiltonian mass}
       takes the form~\cite{CJL} (compare ~\cite{ChHerzlich})
      \be \label{mi12I}
     H(V,b):=\lim_{R\to\infty}
      \frac{1 }{2\pi}
      \int_{r=R} \ourU^i(V) \red{dS}_i \ee
     where $V\in
     \cNb$, with
     \begin{eqnarray} \label{eq:3.312I} & {}\ourU^i (V):=  2\sqrt{\det
     g}\;\left(Vg^{i[k} g^{j]l} \zD_j g_{kl}
     +D^{[i}V 
     g^{j]k} (g_{jk}-b_{jk})\right)
     \,,
     \end{eqnarray}
     and
     where $dS_i$ are the hypersurface  forms $\partial_i\rfloor dx^1\wedge \cdots \wedge dx^n$.
     The arbitrary normalisation factor $1/(2\pi)$ has been introduced for further
convenience.%
 \footnote{The normalisation of $\Hzero$ in spacetime-dimension four is dictated by physical considerations, which have no clear-cut counterpart in other dimensions.}
     We will write
     $$\Hzero:=H(r,b)
      \,.
      $$
Under the conditions spelled out above the functional
$ \HZERO$ is finite and  can be rewritten as~\cite[Remark~2.4]{ChHerzlich}
          \begin{eqnarray}
       \displaystyle \HZERO
          =
          \lim_{R\to\infty} \frac{R^3}{2\pi}
            \displaystyle \int_{\{r=R\}} \left(- \frac {
                   \partial e_{\onephi \onephi}}{\partial r} +\frac {e_{\twor \twor}}{r}
            \right)
            d \varphi
            \,.
            \label{massequation1}
         \end{eqnarray}

     A class of ALH metrics of interest are those for which
     \begin{equation}\label{29VII21.3}
       e_{ij} = r^{-2} \mu_{ij} + o (r^{-2})
     \,,
      \quad
       \partial_r e_{ij} = \partial_r \big(
        r^{-2} \mu_{ij}
        \big)  + o (r^{-3})
     \,,
     \end{equation}
     where the $\mu_{ij}$'s depend only upon the coordinate $\varphi$ on $S^1$. One can further specialise the coordinates so that $\mu_{\twor \twor}\equiv 0$, but this choice might be unnecessarily restrictive for some calculations.  The tensor $\mu_{ij}$ will be referred to as the \emph{mass aspect tensor}.

     For metrics in which \eqref{29VII21.3} holds, Equation~\eqref{massequation1} reads
      \begin{eqnarray}
      \displaystyle \HZERO
      = \frac{1}{2\pi} \int_{ \partial M}
     \Big(
        \mu_{\twor \twor}
     +  2  \mu_{\onephi \onephi}
        \Big) d\varphi
        =: \frac{1}{2\pi}  \int_{ \partial M}
        \mu \,d\varphi\,,
        \label{29VII21.4}
     \end{eqnarray}
     where $\mu$ is the \emph{mass-aspect function}.

Had we chosen instead  the background   $b$ to be  the space-part of the   metric \eqref{16I22.22ba},
\begin{equation}
     \label{16I22.22bas}
       \hypmet=
       \frac{ dr^2}{ r^2  +1 } + r^2 d\varphi^2
       \,,
        \quad
        \mbox{where}\  e^{i\varphi} \in S^1
        \,,
     \end{equation}
 the space of static KIDs would be three-dimensional, spanned by
 $$
  \{\sqrt{r^2+1}, r\cos\varphi, r\sin\varphi\}
   \,.
 $$
This leads to two further global charges associated with the static potentials $r\cos\varphi$ and  $r\sin\varphi$,  which will not be considered further in this work.

\subsection{Asymptotic symmetries, coordinate dependence}
 \label{ss19VII23.31}

In space-dimension $n\ge 3$, the hyperbolic mass functional defines an energy-momentum vector, whose length is an absolute geometric invariant.
As pointed out in ~\cite{Brown:1986nw}, this is not the case in $n=2$,
which finds its roots in the different structure of the set of asymptotic symmetries.
To see this, in dimension $n\ge2$, consider for definiteness metrics which asymptote to the hyperbolic metric as
\begin{equation}\label{29VII23.65}
  e_{ij} = O(r^{-2})
  \,,
  \quad
  f_k(
  e_{ij} )= O(r^{-3})
  \,,
\end{equation}
which by the way is compatible with \eqref{29VII21.3} when $n=2$.
 It has been shown in~\cite{ChNagyATMP} that coordinate transformations which preserve the above are of the form
 \begin{equation}\label{27VII23.6}
   (r, x^a)\mapsto
    \big(\bar r = \psi (x^a) r + O(r^{-1})
   \,,
   \
   \bar x^a = \psi^a(x^b)  + O(r^{-2})
   \big)
   \,,
 \end{equation}
 where $\psi$ is a conformal transformation of the conformal metric at infinity, with derivatives behaving in the obvious way. In dimensions $n\ge 3$, the set of such conformal transformations forms a
finite-dimensional Lie group, while when $n=2$, the maps $\psi$ are arbitrary diffeomorphisms of   $S^1$.

 It turns out that under such transformations the mass-aspect function transforms in a non-trivial way. By way of example,  consider a metric which, to leading order, takes the form
 \begin{equation}
      g = b +  r^{-2}  \mu_{i j}  \theta^i \theta^j
       + O(r^{-3}) \,,
        \label{29VII23.81}
   \end{equation}
   with $b$ given by \eqref{27VII23.91s},
   where the size of the error terms is understood as the size of the metric functions in the $b$-orthonormal coframe \eqref{27VII23.93},
   \begin{equation}
      \{{\theta}^2 = \frac{d {r}}{{r}}\,,  {\theta}^1 = {r} d {\varphi}\}
      \,,
       \label{11XI23.2}
   \end{equation}
and where the $\mu_{ij}$'s depend only on the coordinate $\varphi$ on $S^1$.

Let $f:S^1\mapsto S^1$ be any diffeomorphism of the circle, and consider the following coordinate transformation:
  \begin{align}
     \label{change2}
      \varphi = f(\hat{\varphi}) - \frac{f''(\hat{\varphi})}{2 {\hat r}^2 } \,, \qquad  r = \frac{\hat{r}} {f^\prime(\hat \varphi)}
      \,.
  \end{align}
In the $\hat{b}$-orthonormal frame
  \begin{equation}
     \{\hat{\theta}^2 = \frac{d\hat{r}}{\hat{r}}\,,  \hat{\theta}^1 = \hat{r} d \hat{\varphi}\}\,,
  \end{equation}
one finds
  \begin{align}
   g - \hat b
     &=
       \left(
    \frac{f''(\hat{\varphi})^2+\mu_{\twor \twor}(f(\hat{\varphi}))
    f'(\hat{\varphi})^4}{\hat{r}^2
    f'(\hat{\varphi})^2}
   + O(\hat{r}^{-3})
   \right) (\hat{\theta}^2)^2
     + 2 \left(\frac{\mu_{12}(f(\hat{\varphi}))
     f'(\hat{\varphi})^2}{\hat{r}^2} + O(\hat{r}^{-3})  \right)\hat{\theta}^1\hat{\theta}^2 \nonumber \\
     &+ \left(\frac{f''(\hat{\varphi})^2+\mu_{\onephi \onephi}(f(\hat{\varphi})) f'(\hat{\varphi})^4-f^{(3)}(\hat{\varphi}) f'(\hat{\varphi})}{\hat{r}^2  f'(\hat{\varphi})^2}  +O(\hat{r}^{-3})\right) (\hat{\theta}^1)^2\,.
      \label{5IX23.3}
  \end{align}
  The new mass-aspect function \eqref{29VII21.4}  is therefore equal to
  \begin{align}
     \hat \mu &= \hat  \mu_{\twor \twor} + 2\hat \mu_{\onephi \onephi} = \frac{3 f''(\hat{\varphi})^2+ (2 \blue{\mu}_{\onephi \onephi}(f(\hat{\varphi})) + \blue{\mu}_{\twor \twor}(f(\hat{\varphi}))) f'(\hat{\varphi})^4-2
     f^{(3)}(\hat{\varphi})f'(\hat{\varphi})}{f'(\hat{\varphi})^2}
      \nonumber
       \\
     &=  \blue{\mu}(f(\hat{\varphi}))f'(\hat{\varphi})^2 - 2\red{S(f)(\hat \varphi)}
     \,,
     \label{finalmucc}
  \end{align}
  where $S(f)$ denotes the Schwarzian derivative:
  \begin{equation}
  \red{S(f)(\hat \varphi)} =
    \frac{f^{(3)}(\hat \varphi)}{f'(\hat \varphi)} - \frac{3}{2} \left( \frac{f''(\hat \varphi)}{f'(\hat \varphi)}\right)^2\,.
    \label{10IX23.31}
  \end{equation}
The new Hamiltonian mass $\hat \HZERO$ equals
      \begin{eqnarray}
      \displaystyle
      \hat \HZERO
      = \frac{1}{2\pi} \int_{S^1}
       \hat  \mu \,d\hat \varphi
      = \frac{1}{2\pi}  \int_{S^1}
      \big(
        \blue{\mu}(f(\hat{\varphi})) f'(\hat{\varphi})^2 - 2\red{S(f)(\hat \varphi)}
       \big)
        \,d\hat \varphi
        \,,
        \label{29VII21.4asdf}
     \end{eqnarray}
      with a numerical value without any obvious relation to that of $\HZERO$.

We note that the choice $O(r^{-3})$ of the error terms in \eqref{29VII23.81} leads to a coherent setup, but so would as well the choice $o(r^{-2})$.

\section{Properties of $\Hzero$}
 \label{s19XII23.2}

\subsection{Minimising $\Hzero$}
 \label{ss30VII23.1}

It appears natural to enquire about the properties of $\Hzero$. The results presented in this section are based on~\cite{Oblak:2016eij,Barnich:2014zoa}.

Let us start by rewriting  \eqref{29VII21.4asdf} as
\begin{align}
      \Hzero[\mu;f] &
      =  \frac{1}{2\pi} \int_{S^1}
      \big(
        \blue{\mu}(f ) (f')^2 - 2 \red{S(f)}
       \big)
        \,d  \varphi
         \nonumber
        \\
        & = \frac{1}{2\pi}  \int_{0}^{2 \pi}
        \left(
 \blue{\mu(f)}  (f')^2- 2 \left(\frac{f''}{f'}\right)^\prime
        + \left(\frac{f''}{f'}\right)^2
        \right) d \nohvarphi
 \nonumber
\\
         &= \frac{1}{2\pi}
        \int_{0}^{2 \pi}
        \left(
 \blue{\mu(f)}(f^\prime)^2
                 +  \left(\frac{f''}{f'}\right)^2 \right)
        d \nohvarphi\,.
\label{1VIII23.2}
\end{align}
      Thus, the Hamiltonian energy $\Hzero$ with a mass-aspect function $\mu$ equals $ \Hzero[\mu;\id]$, where $\id:S^1\to S^1$ is the identity map.

It should be clear from \eqref{1VIII23.2}
that $\Hzero$ is unbounded from above.

One is then tempted to minimise $\Hzero$ to obtain an invariant whenever $\Hzero$ is bounded from below.
It turns out that whether or not minimisation succeeds depends upon the function $\mu$.

For this it is useful to recall a result of Schwartz~\cite{Schwartz}, which concerns $\Hzero[\mu;f]$ with $\mu \equiv -1$:
for all diffeomorphisms $f$ of $S^1$ it holds that
\begin{equation}\label{21IX23}
       \Hzero[-1;f]
     \equiv  \frac{1}{2\pi}  \int_{0}^{2 \pi}
        \left(
             -(f^\prime)^2 +  \left(\frac{f''}{f'}\right)^2 \right)
        d \nohvarphi
 \ge -1\,,
\end{equation}
with equality attained precisely    on  diffeomorphisms  arising from $S^1$-preserving M\"obius transformations
\begin{equation}\label{21IX23.1}
  e^{i f(\varphi) } = \frac{a e^{i\varphi} + \bar b}{b e^{i\varphi} +\bar a}
\,,
\end{equation}
where $a,b\in \C$ satisfy  $|a|^2 - |b|^2 =
1$.
Note that $\mu=-1$ corresponds to the Hamiltonian mass of hyperbolic space.
So if we set
 \begin{equation}
  \label{29VII23.71}
   \minHz[\mu]:= \inf _{f}  \Hzero[\mu;f]
        \,,
\end{equation}
where $\inf$ is taken over all differentiable functions satisfying  $f'>0$ together with
$$
 \int_{0}^{2\pi} f^\prime d\nohvarphi =
  2 \pi
 \,,
$$
then we obtain
\begin{equation}\label{21IX23.2}
  \minHz[\blue{-1}] \equiv \Hzero[\blue{-1}; \id]
  = -1
 \,,
\end{equation}
and the minimum is attained on the identity map $f=\id$, as well as on any of the remaining diffeomorphisms defined by \eqref{21IX23.1}, e.g.\ rotations.

It turns out that $\mu=\blue{-1}$ is a threshold in the following sense: If  $\mu <-1$, there exists diffeomorphisms of $S^1$ for which $\Hzero$ can take any values. This can be seen using the following family of functions~\cite[Section~7.3.3]{Oblak:2016eij}:
\begin{equation}\label{21IX23.1a}
  e^{i f_\gamma(\varphi) }:= \frac{ \cosh(\gamma/2) e^{i\varphi} + \sinh(\gamma/2)}{ \sinh(\gamma/2)e^{i\varphi} + \cosh(\gamma/2) }
\,.
\end{equation}
Indeed, set $\mc:= \max \mu < -1$. Clearly
\begin{equation}
 \int_{S^1}
      \Big(
 \blue{\mu(f)}(f^\prime)^2+  \left(\frac{f''}{f'}\right)^2
       \Big)
        \,d\nohvarphi
 \le
  \int_{S^1}
      \Big(
        \mc(f^\prime)^2 +  \left(\frac{f''}{f'}\right)^2
       \Big)
        \,d\nohvarphi
        \,.
 \label{21IX23.3}
\end{equation}
A calculation gives
$$
  \int_{S^1}
     \Big(
        \mc(f^\prime_\gamma)^2 +  \left(\frac{f''_\gamma}{f'_\gamma}\right)^2
       \Big)
        \,d\nohvarphi
 =
2\pi\big( ( \mc+1) \cosh \gamma -1
\big)
 \to_{\gamma\to\infty} -\infty \,,
$$
and it follows that the left-hand side of the inequality  \eqref{21IX23.3} is (also) unbounded from below.

Now, for a general mass-aspect function $\mu$ we can write
\begin{align}
      \Hzero[\mu;f]
         &= \frac{1}{2\pi}
        \int_{0}^{2 \pi}
        \left(
               \big(
 \blue{\mu(f)}+1-1\big)
                (f^\prime)^2
                 +  \left(\frac{f''}{f'}\right)^2 \right)
        d \nohvarphi
        \nonumber
        \\
         &= \frac{1}{2\pi}
        \int_{0}^{2 \pi}
          \big(
 \blue{\mu(f)}+1 \big)
                (f^\prime)^2
        d \nohvarphi
        +  \Hzero[\blue{-1};f]
        \nonumber
        \\
         &= \frac{1}{2\pi}
        \int_{0}^{2 \pi}
          \big(
 \blue{\mu(f)}+1 \big)
                (f^\prime-1+1)^2
        d \nohvarphi
        +  \Hzero[\blue{-1};f]
        \nonumber
        \\
         &= \frac{1}{2\pi}
        \int_{0}^{2 \pi}
        \Big(
          \big(
 \blue{\mu(f)}+1 \big)
                (f^\prime-1)^2
                 +
                 \underbrace{
                 2
                  \big( \blue{\mu(f)} +1 \big)(f'
                  }_{\text{change  variables}\to 2 H[\mu;\id]+ 2}
                  -1)
                  + \blue{\mu(f)}+1
                  \Big)
        d \nohvarphi
        +  \Hzero[\blue{-1};f]
        \nonumber
        \\
         &= 2   \Hzero[\mu ;\id]
       +
    \underbrace{  \Hzero[\blue{-1};f]}_{ \ge- 1}
       +    \frac{1}{2\pi}   \int_{0}^{2 \pi}
        \Big(
          \big(
 \blue{\mu(f)}+1 \big)
                (f^\prime-1)^2
                 -\blue{\mu}(f)
                  \Big)
        d \nohvarphi
        + 1
        \,.
\label{1VIII23.2nw}
\end{align}
If
\begin{equation}\label{23IX23.2}
\min \mu \ge -1\,,
\end{equation}
we obtain a lower bound
\begin{equation}\label{23IX23.1}
  \Hzero[\mu;f] \ge 2  \Hzero[\mu ;\id]
      -   \max \mu
        \,.
\end{equation}
It then follows that, under \eqref{23IX23.2},  the infimum $
   \minHz[\mu]:= \inf _{f}  \Hzero[\mu;f]$ exists
   and provides a geometric invariant. An example where the infimum exists but is not attained can be found in~\cite[Section~4.4]{Balog}.

A case of interest is one where $\mu$ is a constant, $\mu \equiv \mc$ with $\mc\ge -1 $. Then
$$\min\mu = \max \mu = \Hzero[\mu;\id]
\,,
$$
and \eqrefl{1VIII23.2nw} becomes
\begin{align}
      \Hzero[\mu;f]
         &=
          \mc
       +
    \underbrace{
      \Hzero[\blue{-1};f]
       }_{ \ge- 1}
      +     \frac{1}{2\pi}  \int_{0}^{2 \pi}
          \big(                \blue{\mc+1
}
\big)
              (f^\prime-1)^2
        d \nohvarphi
        + 1
        \,.
\label{1VIII23.2xs}
\end{align}
The explicit integral is minimised by $f'\equiv 1$, which also minimises  $  \Hzero[\blue{-1};f]$. Thus, for constants satisfying \eqref{23IX23.2},
the minimum of the left-hand side is attained on the identity map, or in fact on any rotation of the circle, and
we have $ \minHz[\mc] = \mc$.

\medskip

Summarising, it holds that:

\begin{proposition}[\cite{Oblak:2016eij,Barnich:2014zoa}]
  \label{P23IX23.1}
\begin{enumerate}
\item  $\Hzero[\mu;f]$ is always unbounded from above.
  \item Suppose that $\min \mu \ge -1$. Then $\minHz[\mu] > -\infty$ exists and provides a geometric invariant. If moreover $\mu=\mc$ is a constant, then $ \minHz[\mc] = \mc$ and is attained on the identity map.
  \item Suppose that $\max\mu < -1$, then $\Hzero[\mu;f]$ is unbounded from below.
\end{enumerate}
\end{proposition}


\subsection{Critical points of $\Hzero$}

More generally, one can enquire about stationary points of $\Hzero$. For this,
introduce a new function $h$ by the equation
\begin{equation}\label{1VIII23.1}
  f'=e^h
  \,,
\end{equation}
with $h$ satisfying the constraint
\begin{equation}\label{1VIII23.3}
  \int_{0}^{2\pi} e^h d \nohvarphi =  2\pi
  \,.
\end{equation}

The gauged-Hamiltonian $\Hzero[\mu;f]$ can be rewritten as a functional of $h$, which we denote by the same symbol
\begin{align}
       \Hzero[\mu;h]
         &: = \frac{1}{2\pi}
        \int_{0}^{2 \pi}
        \left(
                \blue{\mu}  e^{2h}
        + (h^\prime)^2  \right)
        d \nohvarphi\,.
         \label{10IX23.55}
\end{align}
Critical points of $\Hzero[\mu;h]$ on the constraint set \eqref{1VIII23.3} are solutions of the equation
\begin{equation}\label{1VIII23.6}
  h^{\prime\prime} =  \mu e^{2h} +  \lambda e^h
  \,,
\end{equation}
where  $\lambda \in \R$ arises from Lagrange's theorem on constrained stationary points.

\subsubsection{Constant mass aspect}
 \label{ss5X23.1}

Clearly, when $\mu$ is constant then $h=0$ is always a solution of \eqref{1VIII23.6}, with  $\lambda = - \mu$. So, for constant mass-aspect functions $\mu$, the identity map is always a critical point of $\Hzero$.

\begin{figure}
  \centering
  \includegraphics[width=.4\textwidth]{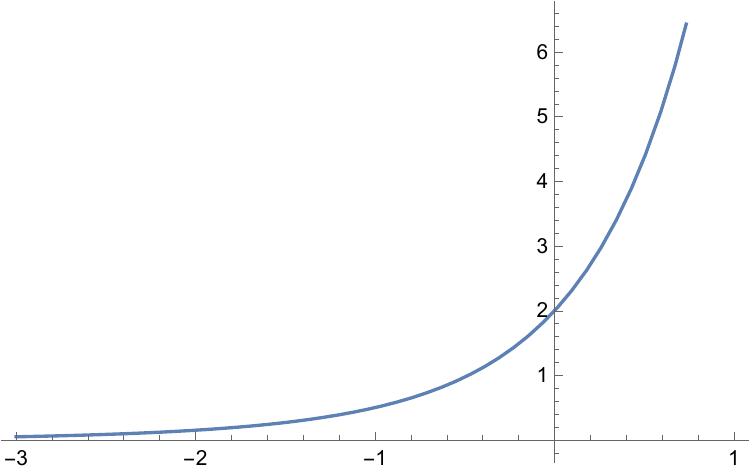} \quad
  \includegraphics[width=.4\textwidth]{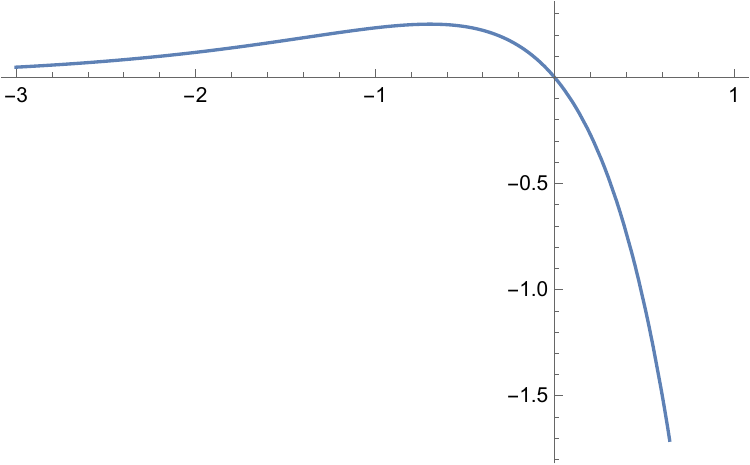}

  \includegraphics[width=.3\textwidth]{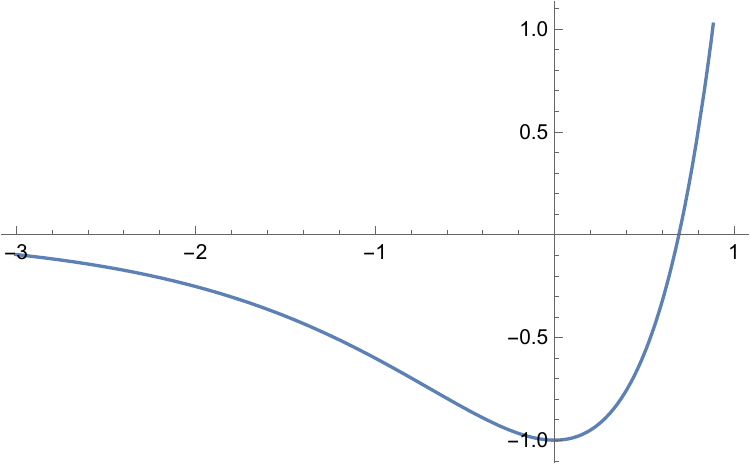} \quad
  \includegraphics[width=.3\textwidth]{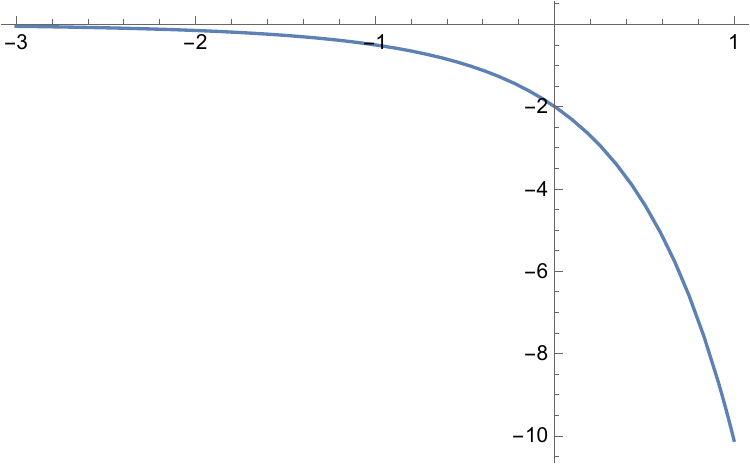}
  \caption{Representative plots of the potential $V$ with
  a) $\lambda\le 0$, $\mc \le 0$, $\lambda^2 + \mc^2 \ne 0$ (upper left plot);
  b) $\lambda<0$, $\mc>  0$,  (upper right plot);
  c) $\lambda>0$, $\mc < 0$,  (lower left plot);
  d) $\lambda>0$, $\mc >0$,  (lower right plot).
   }\label{F5X23.1}
\end{figure}

The question then arises whether there are other solutions
for constant $\mu:=\mc$. In this case, we are looking for  periodic solutions of an equation describing the motion of an object of mass one,  on a trajectory $\varphi \to x(\red{x})\equiv h(\red{x})$, moving in a Morse-type potential
\begin{equation}\label{5X23.1}
  V(x) = -\lambda e^x - \frac{\mc}{2} e^{2x}
   \,.
\end{equation}
Typical plots of $V$, depending upon the sign of $\lambda$ and $\mc$, are shown in Figure~\ref{F5X23.1}.
It should be clear, by inspection of the plots, that a) constant solutions
\begin{equation}
 e^h \equiv -\frac{\lambda}{\mc}
  \label{5X23.6}
\end{equation}
 exist for all values of $\mc\in \R$ (then $\lambda$ must have a sign opposite to $\mc$, and in fact must  be equal to $-\mc$ when the normalisation condition is imposed), with $f'\equiv 1$ providing the only critical points of $\Hzero[\mc;f]$ when $\mc>0$, and b) nontrivial periodic solutions only exist when $\lambda>0$ and $\mc<0$, and then $V(h)<0$  on the orbit, thus
\begin{equation}\label{5X23.2}
  -\lambda e^h + \frac{|\mc|}{2} e^{2h} < 0
  \quad
   \Longleftrightarrow
  \quad
  e^{h} < \frac{2  \lambda }{|\mc|}
    \,.
\end{equation}
\ptcnh{maximum principle argument commented out}

Note that if $h$ is a solution  of \eqref{1VIII23.6} with a constant $\mu=\mc$, and if $\mc$ is replaced by $e^\alpha \mc$ for some $\alpha\in \R$, then the function $h -\alpha/2$ is again a solution after changing $\lambda$ to $\lambda e^{\frac{1}{2} \alpha}$.
However, the normalisation condition \eqref{1VIII23.1} will not be satisfied anymore, so this property of \eqref{1VIII23.6} is irrelevant for our purposes.

For $\mu$ equal to a negative constant $\mu=-|\mc|<0$, in addition to the constant solutions we find two families of solutions parameterised by  $n\in \Z^*$,
\begin{equation}\label{26IX23.1}
 e^{h_{\pm,n}(\red{x})}:=  \frac{  n^2}{\lambda \pm \sqrt{\lambda ^2- n^2 |\mc|}
 \cos
   \left( n\varphi\right)}
   \,,
\end{equation}
with
\begin{equation}\label{27IX23.7}
    \lambda  >  |n|   \sqrt{|m_c|}
    \,,
\end{equation}
and with the proviso that  \eqref{1VIII23.3} holds.

It is useful to note that \eqref{26IX23.1} provides all non-constant solutions
of \eqref{1VIII23.6}
which are $2\pi$-periodic
and which have a critical point at $\varphi =0$.
Indeed, to find all such solutions it suffices to show that the above family, where $n$ is now allowed to be an arbitrary real number satisfying \eqref{27IX23.7}, provides all solutions for which
$h'(0)=0$
 and, in view of \eqref{5X23.2},
 for which  $e^{h(0)} \equiv f'(0)$ takes an arbitrary value in
 $(0, 2\lambda / |\mc|)$.
 Now, we have
\begin{equation}
  \label{eq251023}
 e^{h_{\pm,n} (0)}
 =\frac{1
 }{|\mc|}
  \big(
   \lambda \mp   \sqrt{\lambda^2-n^2|\mc|}
 \big)
 \,.
\end{equation}
%
As $n$ varies in the allowed range at any fixed $\lambda>0$, the value of $f'$ at $0$
ranges over
$$
 f' (0) \in \left ( 0,\frac{\lambda}{|\mc|}\right)
  \cup
  \left (\frac{\lambda}{|\mc|},   \frac{2\lambda}{|\mc|}\right)
  \,.
 $$
Keeping in mind the constant solutions  \eqref{5X23.6},
we conclude that \eqref{26IX23.1} provides all non-trivial periodic solutions.

The solutions $h_{-,n}$ differ from $h_{+,n}$ by a rotation of the circle by $\pi$
for odd $n$ and $\frac{\pi}{2}$ for even $n$,
and can be considered to be identical to $h_{+,n}$ in this sense. Likewise the solutions $h_{+,-n}$ can be considered identical to the solutions $h_{+,n}$. Hence, up to a rotation of the circle, we are left with the functions
\begin{equation}\label{5X23.8}
 e^{h_{n}(\red{x})}:=  e^{h_{+,n}(\red{x})} \equiv  \frac{  n^2}{\lambda + \sqrt{\lambda ^2- n^2 |\mc|} \cos
   \left( n\varphi\right)}
   \,,
    \qquad
     n\in \N^*
     \,.
\end{equation}

Next, we pass to  the normalisation condition \eqref{1VIII23.3}.
For this, we note that from the residue theorem we have, for $a>1$,
\begin{equation}\label{27IX23.2}
  \int_0^{2\pi} \frac{d\varphi}{a+\cos(\red{x})} =
   \frac{2\pi}{\sqrt{a^2-1}}
   \,,
\end{equation}
which gives
\begin{equation}\label{27IX23.2}
  \int_0^{2\pi}  e^{h_{n}(\red{x})} d\varphi  =
   \frac{2n \pi}{\sqrt{|\mc|}}
   \,.
\end{equation}
Hence, \eqref{5X23.8} satisfies the normalisation condition \eqref{1VIII23.3} provided that
\begin{equation}
  \label{periodicmccondition}
\mc = -n^2\,.
\end{equation}

\label{Hofspecialpart}
We now compute the Hamiltonian mass  \eqref{10IX23.55} of the solutions \eq{5X23.8}. For this, we use \cite[GW(332)(38) and LI(64)(14)]{Gradshteyn:1943cpj}: for $a^2>b^2$:
\begin{equation}
  \int_{x = 0}^{2 \pi} \frac{d x}{(a + b \cos(x))}
    = \frac{2 \pi }{(a^2 - b^2)^{1/2}}\,,
    \label{Gradshteyn1}
\end{equation}
\begin{equation}
  \int_{x = 0}^{2 \pi} \frac{d x}{(a + b \cos(x))^2}
    = \frac{2 \pi a}{(a^2 - b^2)^{3/2}}\,.
    \label{Gradshteyn2}
\end{equation}
Then
\begin{align}
\int_{\varphi=0}^{2 \pi} \frac{d \varphi}{(a+ b\cos(n\varphi))^2}
&= \frac{1}{n} \int_{x=0}^{2 \pi n} \frac{dx}{(a + b\cos(x))^2}
= \int_{x=0}^{2 \pi} \frac{dx}{(a + b \cos(x))^2}\,,
\end{align}
and the first contribution to \eqref{10IX23.55}
reads
\begin{equation}
- \int_0^{2\pi} |\mc| e^{2 h_{n}}  d \varphi
= - \int_0^{2 \pi} |\mc|
\left(
\frac{  n^2}{\lambda + \sqrt{\lambda ^2- n^2 |\mc|} \cos
\left( n\varphi\right)}
\right)^2 d \varphi
= - \frac{2 \pi \lambda  n}{\left| \mc\right| ^{1/2} }\,.
\end{equation}
Using
\begin{eqnarray}
 \int_{0}^{2\pi} \frac{b^2 \sin^2(x)}{\big(a+ b \cos(x)\big)^2}
  d x
  &= &
 \int_{0}^{2\pi} \frac{b^2-\big(a+ b \cos(x)\big)^2 + a^2 + 2ab \cos (x)}{\big(a+ b \cos(x)\big)^2}d x
  \nonumber
 \\
& = & - 2 \pi +  \int_{0}^{2\pi} \frac{b^2  + a^2  -2 a^2 + 2a\big(a+b\cos (x)\big)}{\big(a+ b \cos(x)\big)^2}d x \nonumber \\
&= & 2 \pi \left(\frac{a}{\sqrt{a^2-b^2}}-1 \right)\,,
\end{eqnarray}
together with \eqref{Gradshteyn1} and \eqref{Gradshteyn2},
we find that
\begin{align}
  \int_{0}^{2 \pi} (h_{n}'(\red{x}))^2
  &= n^2 \int_{0}^{2 \pi}
  \frac{\sin ^2(n \varphi ) \left(\lambda ^2-n^2 \left| \mc\right| \right)}{\left(\cos (n
   \varphi ) \sqrt{\lambda ^2-n^2 \left| \mc\right| }+\lambda \right)^2} d\varphi \nonumber \\
   &= n^2 \int_{0}^{2 \pi}
   \frac{\sin ^2(\varphi ) \left(\lambda ^2-n^2 \left| \mc\right| \right)}{\left(\cos (
    \varphi ) \sqrt{\lambda ^2-n^2 \left| \mc\right| }+\lambda \right)^2} d\varphi
    =  -2 \pi n^2 + \frac{2 \pi n\lambda}{|\mc|^{1/2}}\,.
\end{align}
Thus,  the Hamiltonian mass \eqref{10IX23.55}  of the solutions \eq{5X23.8} is $\lambda$-independent, equal to
\begin{align}
  \Hzero[\mu = n^2;h_{n}]
  &= \frac{1}{2\pi} \int_0^{2\pi} \left(- |\mc| e^{2 h_{n}}  + (h_{n}'(\red{x}))^2\right) d \varphi =
 - n^2 = \mc \nonumber \\
 &=  \Hzero[\mu = n^2; \id]\,,
\end{align}
where the last equality in the first line holds due to the normalisation condition \eqref{periodicmccondition}. Hence, the Hamiltonian mass of \eqref{5X23.8} equals the Hamiltonian mass evaluated on the identity map.

\medskip

Summarising, we have shown:

\begin{proposition}
  \label{p27IX23.1}  Let $\mc\in\R$. The functional
  $$
  f\mapsto \Hzero[\mc;f]
  $$
   has no critical points other than rotations of the circle, unless
\begin{equation}\label{27IX23.3}
  \mc = -   n ^2
  \,,
  \quad
   n \in \N
   \,.
\end{equation}
For these last values of $\mc$, all critical maps  are obtained by integrating the equation
$f'= e^{h_{n}}$, with the functions $ e^{h_{n}}$, $n\in \N^*$, given by \eqref{5X23.8}; two distinct such solutions differ by a rotation of the circle. All metrics with the  mass-aspect functions so-obtained have the same Hamiltonian energy $\Hzero = -n^2$.
\end{proposition}

\subsection{A spacetime approach}
\label{asymptoticsymmmass}

So far, we considered the Riemannian version of the problem, which can be thought of as an analysis of two-dimensional Cauchy surfaces in a three-dimensional spacetime. The aim of this section is to revisit the problem from a space-time perspective.
For this, we turn our attention to asymptotically AdS$_3$ metrics of the form
\begin{equation}
  \label{Feff1}
{\fourg } = \frac{d\rr^2}{\rr^2} +
  \underbrace{ g_{A B}(\rr, x^C) dx^A dx^B}_{=:g}
\end{equation}
in a Fefferman-Graham-type coordinate system
in which
\begin{equation}
  \label{Feff2}
  g(\rr, x^C) = \rr^2 \eta + O(1)\,,
\end{equation}
where $\eta$ is the flat metric on the cylinder, $\eta = - dt^2 +d \varphi^2$,
and $(x^A)=(t,\varphi)$, where $\varphi$ is assumed to be $2\pi$-periodic.
The reader might wish to note that here, as opposed to the remainder of this paper, both metrics $\fourg$ and $g_{AB}dx^Adx^B$ are Lorentzian.

Assuming that the metric $r^{-2} \fourg$ extends smoothly to the conformal boundary $\{1/r=0\}$,
 solving the vacuum Einstein equations with negative cosmological constant order by order,  one is led to
  metrics of the form  \cite{Banados:1998gg, Barnich:2010eb}
   \ptcrnh{what is the global structure? mention solutions like that with localised matter sources by Troyanov's construction? when are two such solutions isometric? (locally always; and answer given by ).  Banados says: The point here is that
the coordinate transformations which change the values of L and ¯L are not
generated by constraints and therefore they are not gauge symmetries. ??? }
\begin{equation}
  \label{metrics}
  {\fourg }  = \frac{d \rr^2}{\rr^2}  - \left( \rr d x^+ - \frac{\mathcal{L}_-(x^-)}{\rr} d x^- \right)
  \left( \rr d x^- - \frac{\mathcal{L}_+(x^+)}{\rr} d x^+ \right)
  \,,
\end{equation}
where $x^\pm = t \pm \varphi$,
with arbitrary $2\pi$-periodic functions $\mathcal{L}_\pm$.

Here $r\ne 0$ (otherwise \eqref{metrics} is not defined) and
\begin{equation}\label{24IX23.1}
  \mathcal{L}_+
   \mathcal{L}_- \ne r^4
\end{equation}
(otherwise the tensor field \eqref{metrics} has vanishing determinant).

 The global structure of the  metrics \eqref{metrics} is well-understood
  when the functions $\mathcal{L}_\pm$ are constant.
  The character of the singularities
  cannot be probed with the curvature tensor, as these metrics are all locally isometric to Anti-de Sitter space.
  Our gluing results below show that all time-symmetric such metrics can be obtained by evolving time-symmetric data on complete hyperbolic manifolds with no more than two conical singularities.

Writing the metric in the $\fourb $-orthonormal frame
\begin{equation}
   \{{\theta}^2 = \frac{d \rr}{r}\,,  {\theta}^1 = \rr d \varphi\,,  {\theta}^0 = \rr d t\,\}\,,
\end{equation}
we obtain
\begin{align}
    {\fourg }  &= \theta^2 \theta^2
    +\left(-1 + \left(\mathcal{L}_-(x^-)+ \mathcal{L}_+(x^+) \right)\rr^{-2} - \mathcal{L}_+(x^+)\mathcal{L}_-(x^-) \rr^{-4}\right) \theta^0 \theta^0
     \nonumber
 \\
    &~~~+\Big(
     1 + \left(\mathcal{L}_-(x^-)+\mathcal{L}_+(x^+)\right) \rr^{-2} + \mathcal{L}_+(x^+)\mathcal{L}_-(x^-)\rr^{-4}
      \Big) \theta^1 \theta^1
     \nonumber
 \\
    &~~~+ 2 \left( \mathcal{L}_+(x^+)-\mathcal{L}_-(x^-) \right) \rr^{-2}\theta^0 \theta^1
     \,. \label{bansolutionsONB}
\end{align}
A remark on the periodicity of the functions $\mathcal{L}_\pm$ is in order, had it not been assumed. Supposing that $\varphi$ is $2\pi$-periodic, inspection of the $r^{-2}$-terms in the  last equation at $t=0$ shows that both functions $\mathcal{L}_-(-\varphi)+ \mathcal{L}_+(\varphi)$ and $\mathcal{L}_-(-\varphi)- \mathcal{L}_+(\varphi)$ are $2\pi$-periodic. It then easily follows that both   $\mathcal{L}_+$  and   $\mathcal{L}_-$ must be $2\pi$-periodic.

The extrinsic curvature of a constant time  surface $t = t_c$ vanishes if
\begin{equation}
  \label{timesymmBanados}
 \mathcal{L}_+(x^+ = t_c+ \varphi) = \mathcal{L}_-(x^- =t_c- \varphi)
  \ \mbox{and}
  \
 \mathcal{L}'_+(x^+ = t_c+ \varphi) = \mathcal{L}'_-(x^- =t_c- \varphi)
 \,.
\end{equation}

The usual Hamiltonian mass of
three-dimensional gravity \cite[eq. (2.30)]{Barnich:2010eb}
(compare \cite{Brown:1986nw, deHaro:2000xn},
cf.\ also \cite[eq. (9)]{Barnich:2012aw})
is given by
%
\begin{equation}
  \label{Hamiltonianmassspacetime}
  H = 
  \frac{1}{8\pi G}\int_{S^1}
     \big(\mathcal{L}_-(t_c-\varphi)+\mathcal{L}_+(t_c+\varphi)
      \big)
       d\varphi\,.
  \end{equation}
Inspection of  \eqref{bansolutionsONB} shows that the mass-aspect function $\mu$ of \eqref{29VII21.4} reads
\begin{equation}
  \mu = \mu_{22} + 2 \mu_{11} = 2 (\mathcal{L}_-(t_c-\varphi)+\mathcal{L}_+(t_c+\varphi))
   \,,
  \end{equation}
hence $H$ coincides with the Hamiltonian mass of Section~\ref{s28VII23.1} when $G$ is suitably chosen.
In the time-symmetric case \eqref{timesymmBanados} this can be rewritten as
\begin{equation}
  \mu =   4 \mathcal{L}_+(t_c+\varphi)
  \equiv  4 \mathcal{L}_-(t_c-\varphi)
   \,.
  \end{equation}

\subsubsection{Asymptotic symmetries revisited}
 \label{ss10IX23.1}

Consider the coordinate transformations
\begin{align}
  \label{xprr}
    x^+ &= \bv_+({\bar x}^+)
    + \frac{\bv^\prime_+({\bar x}^+) \bv^{\prime \prime}_-({\bar x}^-)}{2 {\bar\rr}^2 \bv^{\prime }_-({\bar x}^-)}
    + \frac{\bv_+''({\bar x}^+) \left(4
    \mathcal{L}_-(\bv_-({\bar x}^-))
    \bv_-'({\bar x}^-)^4+\bv_-''({\bar x}^-)^2
    \right)}{8 {\bar\rr}^4 \bv_-'({\bar x}^-)^2}
    + O\left(\frac{1}{{\bar\rr}^6} \right)
    \,, \nonumber \\
    x^- &= \bv_-({\bar x}^-)
    +\frac{\bv_-'({\bar x}^-) \bv_+''({\bar x}^+)}{2 {\bar\rr}^2
    \bv_+'({\bar x}^+)}
    + \frac{\bv_-''({\bar x}^-)
    \left(4 \mathcal{L}_+(\bv_+({\bar x}^+))
    \bv_+'({\bar x}^+)^4+
    \bv_+''({\bar x}^+)^2\right)}{8 {\bar\rr}^4
    \bv_+'({\bar x}^+)^2}
    + O\left(\frac{1}{{\bar\rr}^6} \right)
    \,, \nonumber
\\
    \rr &=  \frac{{\bar\rr}}{\sqrt{\bv_+'({\bar x}^+)\bv_-'({\bar x}^-) }} -\frac{\bv_-''({\bar x}^-) \bv_+''({\bar x}^+)}{4
    {\bar\rr}  \sqrt{\bv_-'({\bar x}^-)
    \bv_+'({\bar x}^+)}^3}
    + O\left(\frac{1}{{\bar\rr}^3} \right)
\,,
\end{align}
where $\bv_\pm$ are, say smooth, functions satisfying
$$
 \bv_-({\bar x}^-+ 2\pi) =
 \bv_-({\bar x}^-)+ 2\pi
  \,,
   \qquad
 \bv_+({\bar x}^++ 2\pi) =
 \bv_+({\bar x}^+)+ 2\pi
  \,.
$$
Under this transformation, 
the metric retains the  form \eqref{metrics} up 
 to the order induced by \eqref{xprr}, with $\mathcal{L}_\pm$ replaced by new functions $\bar{\mathcal{L}}_\pm$ given by

\ptcnh{should one expect an alternative formula with a finite number of powers of r?}
\begin{subequations}
    \label{barLs}
    \begin{align}
        \bar{\mathcal{L}}_+(\bar{x}^+) &= \mathcal{L}_+(\bv_+({\bar x}^+)) \bv_+'^2({\bar x}^+) - \frac{1}{2}
        \red{S(\bv_+) (\bar x^+)}
         \,,\\
        \bar{\mathcal{L}}_-(\bar{x}^-) &= \mathcal{L}_-(\bv_-({\bar x}^-)) \bv_-'^2({\bar x}^-) - \frac{1}{2}\red{S(\bv_-) (\bar x^-)}\,;
    \end{align}
    \end{subequations}
%
recall that $S(\cdot)$ denotes  the Schwarzian derivative
\eqref{10IX23.31}.
Changing coordinates as $\bar{x}^\pm = \bar{t} \pm \bar{\varphi}$, this can be compared with the results of Section~\ref{ss19VII23.31} by
 restricting to time-symmetric initial data as in \eqref{timesymmBanados} and setting $t = \bar{t}= 0$.
Computing $\bar{\mu}$ in the frame
\begin{equation}
    \{\bar{\theta}^2 = \frac{d \bar{\rr}}{\bar{\rr}}\,,  \bar{\theta}^1 = \bar{\rr} d \bar \varphi\,
    \}.
 \end{equation}
 yields
\begin{equation}
    \label{massaspectspacetime}
    \bar{\mu} = 
    2 (\bar{\mathcal{L}}_+(\bar{\varphi})+\bar{\mathcal{L}}_-(-\bar{\varphi}))
     \,.
\end{equation}
This allows us to compare the transformed mass-aspect function \eqref{massaspectspacetime} with the original one.
At $\bar t =0=t$
 we have  $f(\bar \varphi) = \bv_+(\bar\varphi) =- \bv_-(- \bar\varphi)$, yielding
\begin{subequations}
    \label{xprrn}
\begin{align}
    \varphi  &= \bv_+({\bar \varphi}) + O \left(\frac{1}{{\bar\rr}^2} \right)
    = - \bv_-(-{\bar \varphi})+ O\left(\frac{1}{{\bar\rr}^2} \right)
    \,,\\
    \rr &=  \frac{{\bar\rr}}{\bv_+'({\bar \varphi}) } + O \left(\frac{1}{{\bar\rr}} \right)\,.
\end{align}
\end{subequations}
Thus
\begin{align}
    \bar{\mu} (\bar \varphi) &=  2 (\mathcal{L}_+ + \mathcal{L}_- ) (f'(\bar\varphi))^2
    - 2
     \red{S(f) (\bar\varphi)} =   2 (\mathcal{L}_+ + \mathcal{L}_- ) (f'(\bar\varphi))^2 - 2\red{S(f) (\bar\varphi)} \nonumber \\
    &=   \mu\big(f(\bar \varphi)\big) (f'(\bar\varphi))^2 - 2\red{S(f) (\bar\varphi)}
     \,,
\end{align}
in accordance with \eqref{finalmucc}.

We note that the transformation \eqref{xprr} is closely related to the following vector fields, known as  \emph{asymptotic Killing vectors},  which preserve the   form \eqref{metrics} of the metric
 to all orders in $r$,
 \emph{including the vanishing of terms falling-off faster in $r$}
(cf., e.g. \cite{Barnich:2010eb}):
 \ptcheck{18XII;  with mathematica file Banados-metrics-Killing.nb}
\begin{align}
\xi =& - \frac{1}{2} \rr (\sv_-^\prime(x^-) + \sv_+^\prime(x^+)) \partial_\rr
+ \left(\sv_+(x^+) +
\frac{\rr^2  \sv_-''(x^-) + \mathcal{L}_-(x^-) \sv_+''(x^+)}{2 \rr^4 - 2 \mathcal{L}_-(x^-) \mathcal{L}_+(x^+)}
\right) \partial_{+}\nonumber \\
&+ \left(\sv_-(x^-) + \frac{\rr^2  \sv_+''(x^+) + \mathcal{L}_+(x^+) \sv_-''(x^-)}{2 \rr^4 - 2 \mathcal{L}_+(x^+) \mathcal{L}_-(x^-)}\right) \partial_{-}\,,
\label{Killingallorders}
\end{align}
with arbitrary $2\pi$-periodic functions $\sv_\pm$. Indeed, the Lie derivative of the metric \eqref{metrics} in the direction of the
vector fields  \eqref{Killingallorders} is the tensor field
 \ptcheck{18XII;  with mathematica file Banados-metrics-Killing.nb}
 \begin{align}\label{10IX23}
   \mathrm{Lie}_{\xi} g &= -\frac{
    \mathcal{L}_-(x^-) \delta_\xi \mathcal{L}(x^+)
   +
   \mathcal{L}_+(x^+) \delta_\xi \mathcal{L}_-(x^-)}{r^2} dx^- dx^+ +\delta_\xi \mathcal{L}_-(x^-) (dx^-)^2 \nonumber \\
   &~~~+ \delta_\xi \mathcal{L}(x^+) (dx^+)^2\,,
 \end{align}
where
\begin{subequations}
\begin{align}
  \delta_\xi \mathcal{L}_+(x^+)&= 2 \mathcal{L}_+(x^+) \sv_+'(x^+) + \mathcal{L}_+^\prime(x^+) \sv_+(x^+) - \frac{\sv_+'''(x^+)}{2}\,,\\
  \delta_\xi \mathcal{L}_-(x^-)&= 2 \mathcal{L}_-(x^-) \sv_-'(x^-) + \mathcal{L}_-^\prime(x^-) \sv_-(x^-) - \frac{\sv_-'''(x^-)}{2}.
\end{align}
\end{subequations}
 \ptcheck{18XII;  with mathematica file finite-coordinate-transformations.nb}
   We note that \eqref{xprr} can be obtained from the flow of the asymptotic Killing vectors \eqref{Killingallorders}.

\section{A Penrose-type  inequality}
 \label{s16I22.3}

 \ptcnh{can one settle Oblak's question here, which is causal character with the zero and pm one modes? do positivity without horizons?}
Consider a two-dimensional general relativistic initial data set $(\hyp,g,K)$ on a maximal surface (i.e., $\tr K =0$). Assume that the positive energy condition is satisfied:
\begin{equation}\label{24XII23.11}
  R = 2 \Lambda + |K|^2 -
   \underbrace{ (\tr K)^2}_{=0}
   + \underbrace{2 \rho}_{\ge 0}
   \,.
\end{equation}
We normalise the metric so that $\Lambda =-1$, thus
\begin{equation}\label{24XII23.12}
  R = -2  + |K|^2
   + 2\rho
   \,.
\end{equation}
We assume that the initial data set contains a minimal surface, i.e.\ a closed geodesic $\gamma$ with length $\ell$.
Suppose that $\gamma$  separates $M$ into two regions, one called ``outwards'', and that in
the outwards region we can introduce global coordinates  $(r,\varphi)$, $r\in [r_0,\infty)$, $e^{i\varphi}\in S^1$,
so that $\gamma =\{r=r_0>0\}$, with the metric taking the form
\begin{equation}
g= f^{-1}dr^2 + r^2 d\varphi^2
\,,
      \label{18VIII23.1}
    \end{equation}
with $f\ge0 $ vanishing precisely on $\gamma$.
The Ricci scalar $R(g)$ of $g$ reads
\begin{equation}\label{18VIII23.2}
  R(g)= -\frac{1}{r    } \frac{\partial f  }{\partial r}
  + \frac 1 {r^2 f} \frac{\partial^2 f}{\partial \varphi^2 }
   - \frac{3}{2f^2 r^2} \big( \frac{\partial f}{\partial \varphi}\big)^2
   = -\frac{1}{r    } \frac{\partial f  }{\partial r}
 -\frac{1
   }{2
   r^2 f^2}
   \big( \frac{\partial f}{\partial \varphi}\big)^2
   + \partial_\varphi \big(\frac{1}{r^2 f}\frac{\partial f}{\partial \varphi} \big)
  \,.
\end{equation}
On these level sets of $r$ on which  $f$ has no zeros  we find
 \begin{eqnarray}
   \frac{d}{dr} \int_{S^1}
   ( f -r^2)
     \, d\varphi
       &=&
    - \int_{S^1}
    \Big( R + 2
    +
   \frac{1}{2 r^2 f^2} \big( \frac{\partial f}{\partial \varphi}\big)^2
   \Big)
   r \, d\varphi
   \nonumber
 \\
       &=&
    - \int_{S^1}
    \Big( 2\rho + |K|^2
    +
   \frac{1}{2 r^2f^2} \big( \frac{\partial f}{\partial \varphi}\big)^2
   \Big)
   r \, d\varphi
   \,.
    \label{25XII23.1}
 \end{eqnarray}
Suppose that for large $r$ we have the expansion
\begin{equation}\label{18VII23.40}
f(r,\varphi)= r^{2} - \mu(\varphi)+ o(1)
 \,.
\end{equation}
Integrating \eqref{25XII23.1} one obtains
 \begin{align}
- \int_{r_0}^{\infty}
 \Big(
   &
   \frac{d}{dr} \int_{S^1}
   ( f -r^2)
     \, d\varphi
     \Big) dr
=
-\lim_{r\to\infty}  \int_{S^1}
   ( f -r^2)
     \, d\varphi  +
     \int_{S^1}
     \big(\underbrace{f|_{r_0}}_{0} -r_0^2
     \big)
     \, d\varphi
 =
  \int_{S^1} \mu
     \, d\varphi - 2\pi r^2_0
   \nonumber
\\
       &=
       2 \pi (\Hzero -r_0^2 )
       =
    \int_{r_0}^{\infty} \int_{S^1}
    \Big( 2\rho + |K|^2
    +
    \frac{1}{2 r^2f^2} \big( \frac{\partial f}{\partial \varphi}\big)^2
   \Big)
   r \, dr\,d\varphi
   \,.
    \label{25XII23.2}
 \end{align}
Hence
 \begin{eqnarray}
 \fbox{$
 \displaystyle
 \Hzero
  =
   r_0^2  +
   \frac{1}{2\pi}
    \int_{r_0}^{\infty} \int_{S^1}
    \Big( 2\rho + |K|^2
    +
    \frac{1}{2 r^2f^2} \big( \frac{\partial f}{\partial \varphi}\big)^2
   \Big)
   r \, dr\,d\varphi
   \,,
   $}
    \label{25XII23.3}
 \end{eqnarray}
and we conclude that
 \begin{eqnarray}
 \Hzero \ge
   r_0^2  =
    \big(\frac{\ell}{2\pi}\big)^2
   \,,
    \label{25XII23.3p}
 \end{eqnarray}
 with equality if and only if
 \begin{eqnarray}
  \mbox{$f=f(r)$ and} \
    \rho=0 = K
   \,.
    \label{25XII23.4}
 \end{eqnarray}
In this last case, \eqref{24XII23.12} and \eqref{18VIII23.2} give
 \begin{equation}\label{25XII23.5}
   f (r)= r^2 - r_0^2
   \,.
 \end{equation}
So far we assumed existence of an outermost minimal surface. Suppose instead that we have a global polar-type coordinate system \eqref{18VIII23.1} with $r\in [0,\infty)$ and with $f(0)=1$. One obtains instead
 \begin{eqnarray}
 \fbox{$
 \displaystyle
 \Hzero
  =
   -1 +
   \frac{1}{2\pi}
    \int_{r_0}^{\infty} \int_{S^1}
    \Big( 2\rho + |K|^2
    +
    \frac{1}{2 r^2f^2} \big( \frac{\partial f}{\partial \varphi}\big)^2
   \Big)
   r \, dr\,d\varphi
   \,,
   $}
    \label{25XII23.3a}
 \end{eqnarray}
and we conclude that
 \begin{eqnarray}
 \Hzero \ge -1
   \,,
    \label{25XII23.3ap}
 \end{eqnarray}
 with equality if and only if  $(M,g)$ is the hyperbolic space:
 \begin{equation}\label{25XII23.5a}
   f (r)= r^2 +1
   \,.
 \end{equation}
It should be kept in mind that the interest of the inequalities \eqrefl{25XII23.3} and  \eqrefl{25XII23.3a} is somewhat unclear, as for any solution $\Hzero$ can be made arbitrarily large by an asymptotic symmetry. Indeed, \eqrefl{25XII23.3} and \eqref{25XII23.3a} are only statements about the properties of $\Hzero$ in the gauge \eqref{18VIII23.1}. We note that the function $r$ is a solution of the equation
 \ptcrnh{but: this shows that we are in the minimisation regime, perhaps this could be exploited; also I think that the mass aspect must be constant in these coordinates, needs a comment and rewording}
\begin{equation}\label{31XII23.1}
  D_i \big(
   \frac{D^i r}{r|Dr|}
    \big) =0
    \,,
\end{equation}
with suitable conditions at the outermost closed geodesic  or at the center of the coordinates $(r,\varphi)$.
 \ptcnh{convince Mazzieri or somebody such to clean this up}

It should be emphasised  that the measures in \eqref{25XII23.3} and in  \eqref{25XII23.3a} are \emph{not} the geometric measure of the metric $g$; integration of $R$ against the latter leads  to a renormalised-volume identity, see~\cite{CormickDahl}, compare~\cite{GSW2d}.

\section{Vacuum time-symmetric geometrically finite initial data sets}
 \label{s1XI23.1}

 A two-dimensional manifold is called \emph{geometrically finite} if it has finite Euler characteristic.

 The classification of two-dimensional non-compact geometrically finite  CSC models is well understood, a pedagogical presentation can be found in~\cite{Borthwick}, compare~\cite{Aminneborg:1997pz}.
 We review those results which are relevant for the understanding of the mass, following the terminology of~\cite{Borthwick}.

\emph{Elementary} hyperbolic manifolds are the hyperbolic space (namely, $\R^2$ with the metric \eqref{twoK2}, with $(r,\varphi)$ understood as the usual polar coordinates), and the  (complete) manifold $\R\times S^1$ with the metric
\begin{equation}\label{26XII23.1}
 \frac{dr^2}{r^2}+ r^2 d\varphi ^2\,,
 \quad
  e^{i\lambda \varphi}\in S^1\,,
  \quad
  \lambda \in (0,\infty)
  \,.
\end{equation}
We will refer to this surface as the  \emph{hyperbolic trumpet} (compare~\cite{Hannam:2009ib}).
(Rescaling $\varphi$ and $r$, without loss of generality, one can assume $\lambda = 1$.)

Given $r_0\in \R$, the region $r\le r_0$ with a metric   \eqref{26XII23.1} will be referred to as a \emph{hyperbolic cusp}, and the region $r\ge r_0$ will be called a \emph{hyperbolic end}. Further \emph{hyperbolic ends} are defined as
the manifolds $[r_0,\infty)\times S^1$ with metrics of the form
\begin{equation}\label{26XII23.2}
 \frac{dr^2}{r^2- \mc }+ r^2 d\varphi ^2\,,
 \quad
  e^{i\lambda \varphi}\in S^1\,,
  \quad
  \lambda \in (0,\infty)
  \,,
\end{equation}
where $\mc \in \R$, with $r_0 >0$ and $r \ge \mc$, where the second form
 of the metric in \eqref{27VII23.9110} should be used when $r_0= \sqrt{\mc}$. Note that a rescaling of $r$, $\mc$ and $\varphi$ leads again to $\lambda =1$.

A \emph{funnel} is defined as $[r_0,\infty)\times S^1$ with a metric \eqref{26XII23.2} with $\mc >0$, $r_0 = \sqrt{\mc}$, and $\lambda \ge 1$.
Note that a rescaling of $r$, $\mc$ and $\varphi$ leads to $\lambda =1$.
The boundary $\{r=\sqrt{\mc}\}$ is the shortest closed geodesic.  In the physics literature funnels are  known as \emph{non-rotating BTZ black holes};
or perhaps time-symmetric slices of non-rotating BTZ black holes.
The boundary $\{r= \sqrt{\mc}\}$ is   referred to as \emph{event horizon}, or  \emph{apparent horizon}, or \emph{outermost apparent horizon}.
 In the associated vacuum spacetime, the surface $r=\sqrt{\mc}$ becomes the bifurcation surface of a bifurcate Killing horizon.

A \emph{hyperbolic bridge} is defined as the doubling of a funnel across its minimal boundary; equivalently, this is the rightermost metric \eqref{27VII23.9110} on $\{u\in \R \,,\, e^{i\varphi}\in S^2\}$.

The fundamental result is (cf., e.g., \cite[Theorem~2.23]{Borthwick}):

\begin{theorem}
  \label{t26XII23.1}
Consider a complete non-compact two-dimensional hyperbolic manifold $(M,g)$ with finite Euler characteristic. Then either $(M,g)$ is hyperbolic space, or a hyperbolic trumpet, or it is the union of a compact set with a finite number of cusps and a finite number of funnels.
\end{theorem}

Since funnels admit a coordinate system with constant positive mass-aspect function, we conclude that:

\begin{corollary}
  \label{C26XII23.1}
  Under the hypotheses of Theorem~\ref{t26XII23.1}, the mass-aspect function can always be transformed to a constant, equal to the
  infimum $\minHz$ of the
  Hamiltonian mass $\Hzero$ of $(M,g)$,  with
  \begin{equation}\label{26XII23.12}
    \minHz
  \in \{-1\}\cup [0,\infty)
  \,,
  \end{equation}
  and we have:
  \begin{enumerate}
   \item If $\minHz=-1$, then $(M,g)$ is the hyperbolic space.
   \item If $\minHz=0$, then $(M,g)$ is isometric to $\R\times S^1$ with the metric \eqref{26XII23.1}.
   \item If $\minHz>0$, then each hyperbolic end is contained in, or coincides with a funnel.
   \end{enumerate}
\end{corollary}

We note that  all manifolds as in point 3.\ above, with an arbitrary number of hyperbolic ends, can be constructed by PDE methods, as follows:
Let $(M,\mathring g)$ be any compact Riemannian manifold and let $p_i\in M$ be any finite collection of points. Solving the two-dimensional Yamabe equation one can find a metric $g$ conformal to $\mathring g$ which has constant Gauss curvature $-1$ and such that original neighborhoods of the points $p_i$ become asymptotically hyperbolic ends~\cite{TroyanovHulin}.

We also note that one can allow for a finite number of conical singularities in the last construction~\cite{TroyanovConical,MazzeoConical}.

\section{``Maskit gluing'' of two asymptotically time-symmetric solutions}
 \label{secgluingconical}

In this section we show how to glue together,  near the conformal boundary at infinity,  two  ALH  metrics, i.e.\
asymptotically locally hyperbolic metrics, which have  constant scalar curvature near a point on the conformal boundary at infinity; compare~\cite{MazzeoPacardMaskit}.
The idea is to use asymptotic symmetries to render the metric manifestly hyperbolic, in a half-disc within the half-plane model, near the conformal boundary. The gluing can then be done using isometries of hyperbolic space as in~\cite{ChDelayHPETv1}.


Consider, thus, a point $p$ on the conformal boundary at infinity of an  ALH metric with negative constant scalar curvature equal to $-2$. Let $\mu$ be the mass-aspect function. As a first step, we claim that we can find a diffeomorphism of $S^1$ so that $\mu=-1$ near $p$. Indeed,   \eqref{finalmucc} shows that this can be done by solving the equation

\begin{equation}\label{7X23.1b}
  -1 = (f')^2 \mu \circ f    -
    \frac{2f^{(3)} }{f'}
    +3  \left( \frac{f''}{f'}\right)^2
  \,.
\end{equation}
This is a third order ODE which can always be solved locally near $p$ for a function satisfying $f'>0$ in a (possibly small)
neighborhood of $p$, and
then extended to a diffeomorphism of $S^1$ in an arbitrary way.

In this way we obtain a mass-aspect function, still denoted by $\mu$, equal to  minus one for
$\hat \varphi \in [-\varepsilon,\varepsilon]$ for some $\varepsilon >0$, so  that the metric becomes there
\begin{equation}
  g = \frac{d{\hat r}^2}{{\hat r}^2+1} +{\hat r}^2 d {\hat\varphi}^2\,.
\end{equation}

In order to make contact with the approach in \cite{CDW}, we go to the  Poincar\'e half-space model
\begin{equation}
  \label{Poincarehalf1}
g = \frac{dz^2 +d w^2}{z^2}
\end{equation}
using the map
\begin{align}
  &\hat r = \frac{\sqrt{\left(w^2+z^2-1\right)^2+4 w^2}}{2 z}\,, \quad \cos  (\hat \varphi-\pi) = \frac{w^2+z^2-1}{\sqrt{\left(w^2+z^2-1\right)^2+4
  w^2}}\,,
  \label{17VIII23.1}
  \end{align}
  with inverse
  \begin{align}
  &w = \frac{\hat r \sqrt{\sin
   ^2({\hat \varphi }) \left(\hat r
   \left(-2 \sqrt{\hat r^2+1} \cos
   ({\hat \varphi})+\hat r \cos
   ^2({\hat \varphi})+\hat r\right)+1\right)}}{-\hat r^2
   \cos ^2({\hat \varphi})+\hat r^2+1}\,, \quad
   z =\frac{1}{\sqrt{{\hat r}^2+1}+\hat r \cos ({\hat \varphi})}\,.
  \label{17VIII23.2}
\end{align}
In this way, $(\hat \varphi = 0, \hat r \to \infty)$, is mapped to $(z=0, w = 0)$.
Next, we consider a half-disc
\begin{equation}\label{20VI23.main}
  D^+(\epsilon):=\{z>0\,,\
  z^2 + w^2 <\epsilon^2\}
  \,
\end{equation}
centered around $(z=0, w=0)$, with
\begin{equation}
  \label{relationepsvareps}
  \epsilon:=\tan\left(
    \frac{{\varepsilon}}{2 }\right)
  \,.
\end{equation}
This choice guarantees that the $\epsilon$-half-disc is contained
within  the region $- \varepsilon \leq \hat \varphi \leq \varepsilon$, as follows from the equations below.
 We define a hyperbolic hyperplane as the boundary of the half-disc
\begin{equation}
  \label{halfdiscbefore}
  \mathfrak{h} = \{(z, w) | z^2+ w^2 = \epsilon^2 \}\,.
\end{equation}
By the transformation \eqref{17VIII23.1}, this hyperbolic hyperplane is mapped to
\begin{equation}
  \label{halfdiscafter}
   \mathfrak{h} = \Bigl\{({\hat r}, \hat \varphi)|
   \frac{{\hat r}^2+1- {\hat r}^2 \cos ^2(\hat \varphi)}{\left(\sqrt{{\hat r}^2+1}+{\hat r}
   \cos (\hat \varphi)\right)^2} =  \epsilon^2
  \Bigr\}\,.
\end{equation}
The intersection 
of the hyperbolic hyperplane
and of the conformal boundary $\hat r \to \infty$ is described by the equation
\begin{equation}
  \mathfrak{h}_\infty = \Bigl\{ \hat \varphi|
  \tan^2\left(\frac{{\hat \varphi}}{2}\right) = \epsilon^2 \Bigr\}\,,
\end{equation}
with the two solutions
\begin{equation}
  \tan\left(\frac{{\hat \varphi}}{2}\right)  = \pm \epsilon\,.
   \label{5IX23.1}
\end{equation}

We now apply a ``boost'' (an isometry of hyperbolic space) so that the boundary of the half-disc is mapped to the diagonal of the Poincar\'e-disc model, equation \eqref{Poincaredisc}. To analyse the behaviour under boosts, we give the explicit relation between the parametrisaton of the hyperboloid in
$$
 \{t^2 = 1+ x^2 + y^2 \}\subset \mathbb{R}^{1,2}
 $$
 and the Poincaré half-space model coordinatised as in \eqref{Poincarehalf1}
\begin{align}
   t = \frac{ \left(w^2+z^2+1\right)}{2
   z}\,, \quad x = \frac{ \left(w^2+z^2-1\right)}{2
   z}\,,\quad y = \frac{ w}{z}\,, \\
   w = \frac{y \left(x+ \sqrt{x^2+y^2+1}\right)}{y^2+1}\,, \quad
   z =
   \frac{x+\sqrt{x^2+y^2+1}}{y^2+1}\,.
\end{align}
From this,  and using that $z = 1/(t-x)$ and $w = y/(t-x)$,
 \ptcheck{23VI}
one readily finds that a boost in the $t-x$ plane
corresponds to the following map in terms of  $z$ and $w$:
\begin{align}
   \label{zwtransform}
(z,w)\mapsto (\bar z,\bar w)
\ \mbox{with}
\
\bar{z} &= \frac{z}{\ngkappa}\,, \  \bar{w} = \frac{w}{\ngkappa}\,,
\end{align}
with
\begin{equation}
   \label{defgammat}
\ngkappa
:= \gamma + \gamma \beta
= \sqrt{\frac{1+\beta}{1 - \beta}}
 \,.
\end{equation}
 Under the boost, the hyperplane \eqref{halfdiscbefore} becomes
 \begin{equation}
  \mathfrak{h}_b = \{(\bar z, \bar w) | {\bar z}^2+ {\bar w}^2 = \frac{\epsilon^2}{\ngkappa^2} \}\,,
\end{equation}
 which using the transformation \eqref{17VIII23.2}
 (now between $(\bar z, \bar w)$ and $(\bar r, \bar \varphi)$),
 reads
  \begin{equation} \label{boostednoninf}
    \mathfrak{h}_b = \Bigl\{({\bar r}, \bar \varphi)|
    \frac{{\bar r}^2+1- {\bar r}^2 \cos ^2(\bar \varphi)}{\left(\sqrt{{\bar r}^2+1}+{\bar r}
    \cos (\bar \varphi)\right)^2} =  \frac{\epsilon^2}{\ngkappa^2}
   \Bigr\}\,,
 \end{equation}
 and
  \begin{equation} \label{boostedinf}
  \mathfrak{h}_{\infty, b} = \Bigl\{ \bar \varphi|
  \tan^2\left(\frac{{\bar \varphi}}{2}\right) = \frac{\epsilon^2}{\ngkappa^2} \Bigr\}\,.
\end{equation}
To continue, we invoke the relation of the metric \eqref{Poincarehalf1} to the Poincaré disc model
\begin{equation}
  \label{Poincaredisc}
   g = \frac{4}{(1- y_1^2-y_2^2)^2} (dy_1^2+ dy_2^2)\,,
   \end{equation}
where
\begin{align}
  w = \frac{2 y_1}{y_1^2 + (y_2-1)^2}\,,
   z = \frac{1 - y_1^2 - y_2^2}{y_1^2 + (y_2-1)^2}
   \quad
  \Longleftrightarrow
  \quad
 y_1  = \frac{2
  w}{w^2+(z+1)^2}\,,  y_2 =\frac{w^2+z^2-1}{w^
  2+(z+1)^2}\,.
\end{align}
In the disc model, the boosted hyperplane takes the form
\begin{equation}
  \mathfrak{h}_b =  \Big\{(\bar{y}_1, \bar{y}_2) |
   1 +  \frac{ 4 \bar{y}_2 }{\bar{y}^2_1+(-1+\bar{y}_2)^2} = \frac{\epsilon^2}{\ngkappa^2}\Big\}\,.
 \end{equation}
It follows that the boost  parameter needed  to map the hyperbolic hyperplane into a diagonal
of the Poincaré disc is a solution of the equation
\begin{equation}
  \label{relationgammatepsilon}
  \ngkappa
  \equiv
  \gamma + \gamma \beta
  \equiv
   \sqrt{\frac{1+\beta}{1-\beta} }= \epsilon
  \quad
  \Longleftrightarrow
  \quad
  \beta = \frac{\epsilon ^2-1}{\epsilon ^2+1}
  \quad
  \Longrightarrow
  \quad
  \gamma = \frac{\epsilon^2+1}{2\epsilon}
   \,.
\end{equation}

\section{Controlling the mass}
 \label{ss10X23.1a}

We wish, now, to determine the type of the mass-aspect function after the gluing has been done. For this, some preliminary comments will be useful.

\ptcnh{27X23;  leads nowhere, toying.tex commented out}

\subsection{Monodromy}
 \label{ss27X23.1a}

Given a smooth  function $\mu:S^1\to \R$, we will denote by the same symbol its lift  to a $2\pi$-periodic function on $\R$.

Consider a pair $(\psi_1,\psi_2):\R\to \R$ of solutions
of the Hill equation, namely
\begin{equation}\label{10X23.1}
  \psi'' =\frac  \mu 4 \psi
  \,,
\end{equation}
normalised so that the Wronskian equals 1:
\begin{equation}\label{10X23.3}
    \psi_1' \psi_2 - \psi_2' \psi_1 =1
 \,.
\end{equation}

In view of the periodicity of $\mu$, if $\psi$ solves the Hill equation, then so does $\psi(2\pi +\cdot)$. It follows that there exists a matrix $\mono$, called
  the \emph{monodromy matrix}, such that
\begin{equation}\label{10X23.2}
  \left(
    \begin{array}{c}
      \psi_1
  (2\pi +\red{x}) \\
      \psi_2
  (2\pi +\red{x}) \\
    \end{array}
  \right)
  =
  \mono  \left(
    \begin{array}{c}
      \psi_1
 (\red{x}) \\
      \psi_2
 (\red{x})\\
    \end{array}
  \right)
  \,.
\end{equation}
Differentiating \eqref{10X23.2} leads to the matrix equation
\begin{equation}\label{10X23.}
  \left(
    \begin{array}{cc}
      \psi'_1
  (2\pi +\red{x})
  &    \psi_1
  (2\pi +\red{x}) \\
      \psi'_2
  (2\pi +\red{x})
  &   \psi_2
  (2\pi +\red{x}) \\
    \end{array}
  \right)
  =
  \mono  \left(
    \begin{array}{cc}
      \psi'_1
 (\red{x})
 &  \psi_1
 (\red{x}) \\
      \psi'_2
 (\red{x})
  &  \psi_2
 (\red{x})\\
    \end{array}
  \right)
  \,,
\end{equation}
 which shows that $\mono$ has determinant one.

Let $f $ be a positively oriented diffeomorphism of $S^1$,  let us denote by $\mu_f $ the transformed mass-aspect function:
$$
\mu_{f} =(f' )^2  \mu \circ f -2 S[f ]
\,.
$$
A standard calculation shows that the   functions 
\begin{equation}\label{15X23.50}
(\fpsi_{1},\fpsi_{2})
:=(f')^{-1/2}(\psi_1\circ f, \psi_2\circ f)
\end{equation}
are solutions of the Hill equation with $\mu$ replaced by $\mu_f$:
\begin{equation}\label{10X23.4a}
  \frac{d^2(\fpsi)}{dx^2} =\frac{\mu_{f}} 4  \fpsi
  \,.
\end{equation}
We note that \eqref{10X23.3} is preserved by \eqref{15X23.50}.
More significantly, we have:
\begin{eqnarray}
  \left(
    \begin{array}{c}
      \fpsi_{1}
  (2\pi +\red{x}) \\
      \fpsi_{2}
  (2\pi +\red{x}) \\
    \end{array}
  \right)
   &=&
  \frac{1}{\sqrt{f'(2\pi +\red{x})}} \left(
    \begin{array}{c}
      \psi_1 \big(f(2\pi +\red{x})\big)
 \\
      \psi_2 \big(f(2\pi +\red{x})\big)
    \end{array}
  \right)
  \nonumber
\\
   &=&
  \frac{1}{\sqrt{f'(\red{x})}} \left(
    \begin{array}{c}
      \psi_1 \big(2\pi +f(\red{x})\big)
 \\
      \psi_2 \big(2\pi +f(\red{x})\big)
    \end{array}
  \right)
  \nonumber
\\
   &=&
   \frac{1}{\sqrt{f'(\red{x})}}
  \mono
 \left(
    \begin{array}{c}
      \psi_1 \big( f(\red{x})\big)
 \\
      \psi_2 \big( f(\red{x})\big)
    \end{array}
  \right)
  \nonumber
\\
   &=&
  \mono \left(
    \begin{array}{c}
      \fpsi_{1}
  (\red{x}) \\
      \fpsi_{2}
  (\red{x}) \\
    \end{array}
  \right)
  \,.
  \nonumber
\end{eqnarray}
Thus,  the monodromy matrix is invariant under diffeomorphisms of the circle.

Suppose that a Wronskian-normalised pair $(\psi_1,\psi_2)$ is replaced by another such pair of solutions of the Hill equation $(\bar\psi_1,\bar\psi_2)$,  thus there exists a constant-coefficient matrix $A$ such that:
\begin{equation}\label{10X23.4b}
  \left(
    \begin{array}{c}
      \psi_1  \\
      \psi_2  \\
    \end{array}
  \right)
  =
  A \left(
    \begin{array}{c}
     \bar \psi_1  \\
     \bar \psi_2  \\
    \end{array}
  \right)
  \,.
\end{equation}
Denoting by $\bar \mono$ the new monodromy matrix, we find
\begin{equation}\label{15X23.8}
  \bar \mono = A^{-1} \mono A
  \,.
\end{equation}
Since
\begin{equation}\label{10X23.6}
  \left(
    \begin{array}{cc}
      \psi'_1
      & \psi_1 \\
      \psi'_2
       & \psi_2
      \  \\
    \end{array}
  \right)
  =
  A \left(
    \begin{array}{cc}
     \bar \psi'_1
     &\bar \psi_1  \\
     \bar \psi'_2
     &\bar \psi_2   \\
    \end{array}
  \right)
  \,,
\end{equation}
the requirement of preservation of \eqref{10X23.3} gives
\begin{equation}\label{15X23.7}
  \det A = 1
  \,.
\end{equation}

It turns out
(cf., e.g., \cite{Oblak:2016eij,Balog}) that the type of the mass-aspect function $\mu $ is determined by the trace of the monodromy matrix $\mono$,
after taking into account  the number of zeros of the associated solutions of the Hill equation.
In this context some terminology will be useful:  A two-by-two matrix $\mono$ with determinant equal to $1$ will be called \emph{hyperbolic} if it has two distinct real eigenvalues; equivalently
\begin{equation}\label{15XII23.1}
  |\tr\mono| > 2
  \,.
\end{equation}
The matrix $\mono$ will be called \emph{elliptic} if
$\mono = \pm \mathbb{I}$, where $\mathbb I$ is the identity matrix, or if $\mono$ has two distinct imaginary eigenvalues; equivalently
\begin{equation}\label{15XII23.2}
 \mbox{$\mono = \pm \mathbb{I}$ or $| \tr\mono| < 2$.}
\end{equation}
Finally, the matrix $\mono$ will be called \emph{parabolic} if $\mono$ has repeated real eigenvalues $1$ or $-1$; equivalently
\begin{equation}\label{20I24.1}
 \mbox{ $ | \tr\mono| = 2$
and $\mono \ne \pm \mathbb{I}$.
 }
\end{equation}

The terminology here and below has been tailored to the problem of mass of initial data sets. The reader is warned that it  is related to, but not identical with that of \cite{Balog}.

An ALH  manifold, or the mass-aspect function of such a manifold, will be called \emph{of hyperbolic/elliptic/parabolic type} whenever the associated monodromy matrix is.

\subsubsection{Elliptic type}
 \label{ss27X23.1b}

It is shown e.g.\ in \cite{Balog} that for  mass-aspect functions $\mu$ of elliptic type there exist diffeomorphisms of the circle which transform  $\mu$ to a negative constant $\mc$. After performing such a transformation, solving the Hill equation is straightforward: a
  Wronskian-normalised basis of solutions of Hill's equation is provided by the functions
 \begin{equation}\label{27X23.31ha}
   \red{ \psi_1(x}) = \frac { \sqrt{2}}{|\mc|^{1/4}} \sin\big(\frac{\red{\sqrt{|\mc|}x}}2\big)
   \,,
   \quad
   \red{ \psi_2(x}) =  \frac { \sqrt{2}}{|\mc|^{1/4}}  \cos\big(\frac{\red{\sqrt{|\mc|}x}}2\big)
   \,.
 \end{equation}
 We then say that $\mu$ is of
 $\mcE_{\mc}$-type.

 We have
 \begin{eqnarray*}
   \red{ \psi_1(x} + 2\pi )
    & = &  \frac { \sqrt{2}}{|\mc|^{1/4}}  \sin\big(\frac{\red{\sqrt{|\mc|}x}}2+  \sqrt{|\mc|}\pi\big)
    \nonumber
 \\
    & = &
    \frac { \sqrt{2}}{|\mc|^{1/4}} \Big( \sin\big(\frac{\red{\sqrt{|\mc|}x}}2\big) \cos\big( \sqrt{|\mc|}\pi\big)
    + \cos\big(\frac{\red{\sqrt{|\mc|}x}}2\big) \sin\big(\sqrt{|\mc|}\pi\big)
    \Big)
   \,,
\\
   \red{ \psi_2(x}+ 2\pi ) %
   & = & \frac { \sqrt{2}}{|\mc|^{1/4}}
    \left(
    \cos\big(\frac{\red{\sqrt{|\mc|}x}}2\big) \cos\big( \sqrt{|\mc|}\pi\big)
 -
 \sin \big(\frac{\red{\sqrt{|\mc|}x}}2\big) \sin\big(\sqrt{|\mc|}\pi\big)
 \right)
   \,,
 \end{eqnarray*}
 leading to the elliptic  monodromy matrix
 \begin{equation}\label{27X23.31e}
   \mono
   =
   \left(
\begin{array}{cc}
     \cos \left(\pi  \sqrt{\left| \mc\right|    }\right)
      & \sin \left(\pi  \sqrt{\left|
   \mc\right| }\right) \\
 -\sin
   \left(\pi  \sqrt{\left| \mc\right|
   }\right) & \cos \left(\pi  \sqrt{\left|
   \mc\right| }\right) \\
\end{array}
\right)
 \,.
 \end{equation}

\ptcrnh{Balog's E type commented out, there is no need to make a distrinction from our point of view? perhaps I am missing something, though}

 A natural way to read off the mass from the matrix $\mono$ in the current case is to calculate its trace:
 \begin{equation}\label{31X23.1}
   \tr\, \mono = 2
     \cos \left(\pi  \sqrt{\left| \mc\right|    }\right)
     \,.
 \end{equation}
Together with  the number  $N = \lceil \sqrt{|\mc|}\rceil$   of zeros of $\psi_1$ on $[0,2\pi)$, 
this trace
determines $\mc$ uniquely. 
 We conclude that $ \sqrt{\left| \mc \right| } $, viewed as a function of $\tr\, \mono$, is the solution of \eqref{31X23.1}
 lying in $(N-1, N]$. 

\subsubsection{Hyperbolic type}

Suppose that the mass-aspect function can be transformed to a positive constant  $\mc$. Then   a normalised basis of solutions of the Hill equation is provided by the functions
 \begin{equation}\label{27X23.31hb}
   \psi_1(x) = \frac1 {|\mc|^{1/4}} \exp\big(\frac{\sqrt{\mc}x}2\big)
   \,,
   \quad
   \psi_2(x) =   \frac1 {|\mc|^{1/4}} \exp\big(-\frac{\sqrt{\mc}x}2\big)
   \,,
 \end{equation}
 with hyperbolic monodromy matrix
 \begin{equation}\label{27X23.33h}
   \mono
   =
   \left(
     \begin{array}{cc}
       \exp\big(\sqrt{\mc}\pi\big)& 0
        \\
       0& \exp\big(-\sqrt{\mc}\pi\big)
     \end{array}
   \right)
     \,.
 \end{equation}
   In this case, the mass can be read off from the conjugation-invariant formula
 \begin{equation}\label{3XII23.1}
   \tr\, \mono = 2 \cosh(\sqrt{\mc}\pi)
   \qquad
   \Longleftrightarrow
   \qquad
   \mc = \left(\frac{\cosh^{-1}\left(\frac{\tr\, \mono}{2}\right)}{\pi}\right)^2
   \,.
 \end{equation}
Surprisingly enough, there exist mass-aspect functions $\mu$ with hyperbolic monodromy which cannot be transformed to a constant by an asymptotic symmetry. Indeed, it is shown in~\cite{Balog} that a  function $\mu$ of hyperbolic type either can be transformed to a constant, or there exists $1\le  n\in \N$ such that $\mu$ can be transformed to
%
\begin{equation}\label{16I24.1}
   \mu_n =
     \nomc + 2 \frac{ n^2 + \nomc }{F(x)}
    - 3 \frac{n^2}{F(x)^2}
\end{equation}
where 
\begin{align}
  F(x) &= \cos^2 \big(\frac{n x }{2} \big)
  + \big( \sin \big(\frac{n x }{2} \big)
  + \frac{\sqrt{\nomc }}{n} \cos \big(\frac{n x }{2} \big)
    \big)^2
    \equiv a + b \cos(n x + \alpha)
    \,,
\end{align}
with 
\begin{equation}
  a = 2 \cot ^2(\alpha )+1\,,\quad  b= 2 \cot (\alpha ) \csc (\alpha )\,, \quad \sqrt{\nomc} = - 2 n \cot (\alpha )
  >0
  \ \big(\mbox{thus}\ 
    \alpha \in (-\pi/2,0)
     \big)
  \,.
\end{equation}
The Hamiltonian mass $\Hzero$, as defined in \eqref{29VII21.4}, associated with \eqref{16I24.1},
equals
%
\begin{equation}
 \HZERO =
 \frac{3 \sigma }{2}-n^2\,.
\end{equation}
Here,  we have used that
\begin{align}
  \int_0^{2 \pi} \frac{dx}{F(x)^\rho} &= \int_0^{2 \pi}
  \frac{dx}{(a + b \cos(n x + \alpha))^\rho} =
  \frac{1}{n}\int_0^{2 \pi n}
  \frac{dy}{(a + b \cos(y + \alpha))^\rho}\nonumber \\
  &= \int_0^{2 \pi }
  \frac{dy}{(a + b \cos(y ))^\rho}\,,
\end{align}
with $\rho= \{1, 2\}$
  to compute the integrals.

It is shown in \cite[eq. (4.36)]{Balog},
using an explicit  one-parameter family of mass-aspect functions (``Virasoro densities'' in the terminology there),
that $\Hzero$ can be made arbitrarily negative by applying  an asymptotic symmetry to   \eqref{16I24.1}.

The following functions provide a Wronskian-normalised basis of solutions of the Hill equation
\begin{eqnarray}\label{16I24.2}
  \psi_{1,n} (x)
   & = &
   \frac{e^{\frac{\sqrt{\nomc }}{2} x}}{\sqrt{F(x)}}
    \sqrt{\frac{n}{2}} \left( \frac{\sqrt{\nomc }}{n^2} \cos \left(
      \frac{n x}{2}
    \right) + \frac{2}{n} \sin \left(
      \frac{n x}{2}
    \right)\right)
    \,,
\\
  \psi_{2,n} (x)
   & = &
   \frac{e^{- \frac{\sqrt{\nomc }}{2} x}}{\sqrt{F(x)}}
    \sqrt{\frac{2}{n}}  \cos \left(
      \frac{n x}{2}
    \right)
    \,,
        \label{16I24.3}
\end{eqnarray}
with monodromy matrix
 \begin{equation}\label{27X23.33hq}
   \mono
   = (-1)^n
   \left(
     \begin{array}{cc}
       \exp\big(\sqrt{\nomc }\pi\big)& 0
        \\
       0& \exp\big(-\sqrt{\nomc }\pi\big)
     \end{array}
   \right)
     \,.
 \end{equation}
 \begin{figure}
  \centering
  \includegraphics[scale = 0.6]{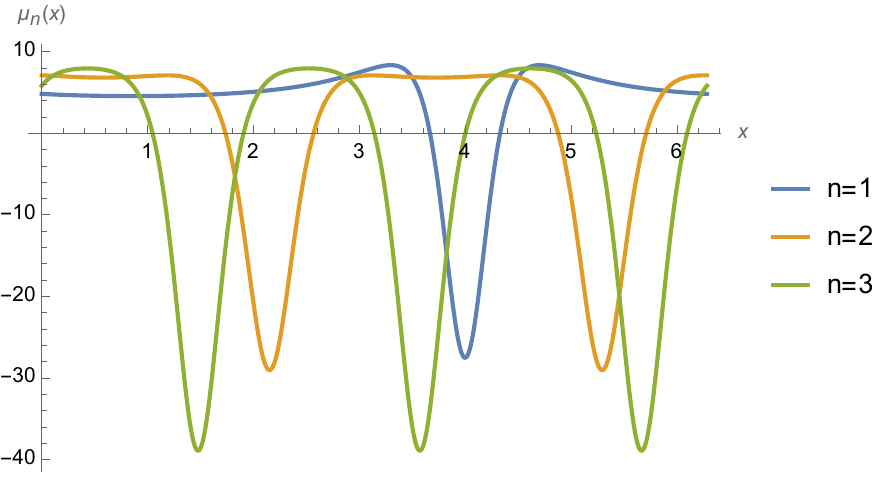}
  \caption{The functions  $\mu_n$ of \eqref{16I24.1} with $n=1,2,3$ and $\nomc =3$ provide  examples of   mass aspect  functions, of $\mcH_n$-type, which \emph{cannot} be transformed to constants.}\label{F17124.1}
 \end{figure}
Note that each of the functions \eqref{16I24.2}-\eqref{16I24.3} with $n\ge 1$ has exactly  $n$  zeros on $[0,2\pi)$. Examples of functions $\mu_n$ are plotted in Figure~\ref{F17124.1}.

We will say that $\mu$ is $\mcH_0$-type if it can be transformed to a
 positive
constant, and  of $\mcH_n$-type, $n\ge 1$, if it can be transformed to the function  \eqref{16I24.1}.

\subsubsection{Parabolic type}

Lastly, there exist mass-aspect functions of parabolic type.
The simplest example is $\mu\equiv 0$, with a basis of solutions of the Hill equation given by
\begin{equation}\label{10I24.1}
  \psi_1 (x) = \frac{x}{\sqrt{2\pi}}
  \,,
  \quad
  \psi_2 (x) = \sqrt{2\pi}
  \,,
\end{equation}
and  with parabolic monodromy matrix
\begin{equation}
  \mono =    \left(
  \begin{array}{cc}
    1& 1
     \\
    0& 1
  \end{array}
  \right)
  \,.
   \label{18I24.p2}
\end{equation}
(Our matrices are transpose to these of~\cite{Balog}, because the definition of monodromy in \cite{Balog} uses the transpose of our equation \eqref{10X23.2}.)

We will say that $\mu$ is of $\mcP_0$-type if it can be transformed to zero by an asymptotic symmetry.

According to \cite{Balog},
the remaining ``parabolic''  mass-aspect functions $\mu$    cannot be transformed to a constant, but can be transformed into
\begin{equation}
  \mu_{n,q} = \frac{2 n^2}{ J(x)} - \frac{3 n^2 (1+\frac{q}{ 2\pi})}{J(x)^2}\,,
   \label{18I24.1}
\end{equation}
where $q \in \{\pm 1\}$ and $1\le n \in \mathbb{N}$, with
\begin{equation}
  J(x) = 1 + \frac{q}{2\pi} \sin^2 ( \frac{n x}{2})\,.
\end{equation}

The functions
\begin{eqnarray}\label{17I24.2}
  \psi_{1,n, q} (x)
   & = &
   \frac{1}{\sqrt{J(x)}} \left(
\frac{q x }{2 \pi} \sin (\frac{n x}{2}) - \frac{2}{n}
\cos (\frac{n x}{2})
    \right)
    \,,
\\
  \psi_{2,n, q} (x)
   & = & \frac{1}{\sqrt{J(x)}}
   \sin (\frac{n x}{2})
    \,,
        \label{17I24.3}
\end{eqnarray}
provide a Wronskian-normalised basis of solutions of the Hill equation, with parabolic monodromy matrix
\begin{equation}
  \mono =  \eta \left(
  \begin{array}{cc}
    1& q
     \\
    0& 1
  \end{array}
  \right)\,, \quad \mathrm{with} \
  \eta = (-1)^n  \ \mbox{and} \
  q = \pm 1\,.
   \label{18I24.p1}
\end{equation}
A mass-aspect function $\mu$ will be said to be of $\mcP^q_{n}$-type, $n\ge 1$, if it can be transformed to \eqref{18I24.1} by an asymptotic symmetry.
(The type $\mcP_0$ can thus be thought of as $\mcP_0^{+1}$.)

\begin{figure}
  \centering
  \includegraphics[scale = 0.6]{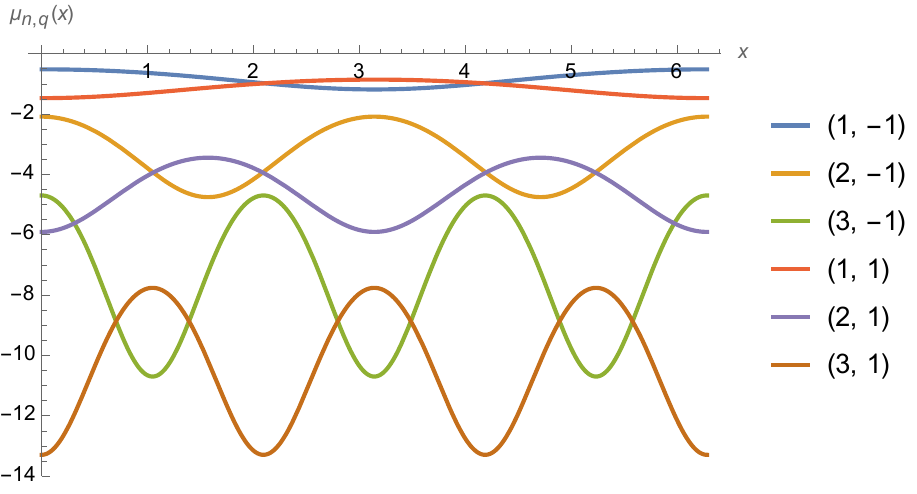}
  \caption{The functions
  $\mu_{n, q}$
  of \eqref{18I24.1}, with $(n, q)$ as specified by the legend, of type $\mcP^{q}_{n}$,  provide further examples of
  mass aspect  functions which cannot  be transformed to a constant.}\label{F17124.3}
 \end{figure}

Examples of  functions $\mu_{n,q}$ are plotted in Figure~\ref{F17124.3}.

The Hamiltonian mass associated with \eqref{18I24.1} is given by
\begin{equation}
  \Hzero = -\frac{n^2 (3 q+4 \pi )}{2 \sqrt{2 \pi  q+4 \pi ^2}} \,.
\end{equation}
%

Extrema of $\mu_{n,q}$ are located at $\sin(nx)=0$ and are equal to
\begin{equation}\label{22I24.1}
-\frac{n^2}{\frac{q}{2 \pi }+1}
\ \mbox{and} \
-\frac{n^2 (3 q+2 \pi )}{2 \pi }
\end{equation}
(with the ordering depending upon $q$).
It follows that $\mu_{n,q}<-1$ unless $n=1$ and
$q=\pm1$.
Hence $\minHz = -\infty$ by Proposition~\ref{P23IX23.1}, except possibly in the last case.
It is shown in \cite[eq. (4.33)]{Balog} that the equality $\minHz = -\infty$ remains true in the case $n= q=1$.
According to \cite[Section~4.4]{Balog} one has
 $\minHz=-1$ when $n= -q =1$ (keeping in mind that their
normalisation of the energy differs by a factor $4$), and the bound is not attained.

\subsection{Positivity}

Whether or not some version of the positive energy theorem holds   in $(2+1)$ dimensions remains to be seen. The proof of the following related result  could perhaps be less demanding:

\begin{conjecture}
  \label{C20I23.1}
  Mass aspect functions of $\mcH_n$- and $\mcP_n^q$-type   with $n\ge 1$,
  and of $\mcE_{\mc}$-type with $\mc<-1$,
   do \underline{not} arise on smooth, conformally compact, geometrically finite general relativistic initial data sets satisfying the energy condition $\rho \ge |J|$, where $\rho$ is the energy density and $J$ the momentum density.
\end{conjecture}

\subsection{Zeros of Hill functions}
 \label{ss12I24.1}

The classification in~\cite{Balog} of mass-aspect functions, presented above, is related to the number of zeros of Hill functions \emph{after a canonical form of the holonomy matrix  $\mono$ has been achieved}. The transformation of $\mono$ to canonical form  requires a change of basis of the Hill functions, and the question arises, what happens with the number of zeros of Hill functions under such changes.

It is relevant to start  with a warning. While the mass-aspect functions are periodic, the Hill functions are not.
This implies that, in spite of naive expectations,  the number of zeros of a Hill function
on an interval of length equal to $2\pi$, e.g.\
on $[0,2\pi]$, or on $[0,2\pi)$, or on $(0,2\pi]$,
might change under the transformation
\begin{equation}\label{15X23.5}
 \psi \mapsto \frac{\psi\circ f}{\sqrt{ f'}}
 \,.
\end{equation}
This is due to  an abuse of notation inherent in this equation.  Indeed, while the mass-aspect function $\mu$ is defined on $S^1$,  when solving the Hill equation we view $\mu$ as a $2\pi$-periodic function on $\R$.
The Hill function $\psi$ is defined on $\R$, so for
this equation to make sense one needs to make precise how
a diffeomorphism of  $S^1$ becomes a function on $\R$. We will
always assume that the function $f$ in \eqref{15X23.5} is smooth,
lifted to $\R$ so that its derivative is  $2\pi$-periodic, and that $f$ increases by $2\pi$ after a rotation by $2\pi$. A rotation of $S^1$ by an angle $\psi_0$ corresponds to a shift of the interval $[0,2\pi]$ in $\R$ by $\psi_0$, which \emph{can} change the number of zeros of solutions of the Hill equation.

The following standard result is useful in our further considerations:

\begin{Lemma}
\label{L14III24.1}
Between two zeros of one of  two linearly-independent Hill functions there is exactly one zero of the other one.
\qed
\end{Lemma}

\proof
Let $W\ne 0$ be the Wronskian of two independent solutions $\psi_1$ and $\psi_2$, thus so $W'=0$.
If $\psi_2(x_0)=0$,  we have
\begin{eqnarray}
W = (\psi_1 ' \psi_2 - \psi_2' \psi_1)|_{x_0}
 =  - (\psi_2' \psi_1)|_{x_0}
    \,,
\end{eqnarray}
which
shows that the sign of $\psi_1$ 
changes between any two successive zeros of $\psi_2$. The  symmetry $(\psi_1,\psi_2)\leftrightarrow (\psi_2,-\psi_1)$ shows that this sign changes only once.
\qedskip

In what follows we will
be interested in the number of zeros on \emph{half-closed} intervals $I \subset \R$, i.e. closed from the left, or from the right but not both,  of length $2\pi$.
A word of justification for this is in order. Given a $2\pi$-periodic function $\chi$ on $\R$ and a closed interval $\bar I\subset \R$ of length $2\pi$, the number of zeros of $\chi$ will only be independent of shifts of $\bar I$ when zeros occurring  at both end points are counted once.
Now,  solutions $\psi$ of the Hill equation have no reason to be $2\pi$-periodic, and the number of zeros of $\psi$ will in general depend upon $I$. However, if it happens that there exists a set of zeros of $\psi$ which are   $2\pi$-distant from their nearest neighbours within the set (which is clearly the case for, e.g.,  the second Hill function  from a Wronskian-normalised pair with parabolic
monodromy \eqref{18I24.p1}), then taking $I$ to be half-closed renders the counting independent of its shifts.

We continue with:

\begin{Proposition}
   \label{P11I24}
 Let $I\subset \R$ be a half-closed interval of length $2\pi$.  Then:

\begin{enumerate}
  \item
   \label{P11I24(b1)}
   If $\mu$ is of $\mcH_n$-type     with $n\ge 1$,  or of $\mcP_n^q$-type
   with $n\ge 1$, $q = \pm 1$,
   then
  all Hill functions have
  $n-1$, $n$ or $n+1$ zeros on $\interval $.

  \item
  \label{10VI24pa}
   If $\mu$ is of elliptic $\mcE_{\mc}$-type, then all Hill functions have either $\lfloor\sqrt{|\mc|}\rfloor$ or $\lfloor\sqrt{|\mc|}\rfloor+1 $ zeros on $\interval $; precisely $\lfloor\sqrt{|\mc|}\rfloor$ zeros  when $\lfloor\sqrt{|\mc|}\rfloor\in \N$.
   \label{P11I24(a)}

  \item
   \label{P11I24(c)}
   If  $\mu$ is of $\mcH_0$- or $\mcP_0$-type,
   then
  all Hill functions have either one or no zeros on $\interval $.
\end{enumerate}
\end{Proposition}

\proof
We start by noting that on any interval $[a,a+2\pi)$, where $a\in \R$,  it
suffices to consider mass-aspect functions $\mu $ in canonical form as proposed in~\cite{Balog}, with the Hill functions being linear combinations of the basis functions  presented there and listed above.
Indeed, let $f_{S^1}:[a,a+2\pi)\mapsto [a,a+2\pi)$ be a diffeomorphism of $[a,a+2\pi)$ which extends to a diffeomorphism of $S^1$ and which transforms the mass aspect to canonical form. Let us denote by  $f'$ the $2\pi$-periodic lift of $f'_{S^1}$ to $\R$, and let 
$$
 f(x)=a+\int_{a}^{x}f'(y)dy
\,.
$$
Then the function, say $\psi_f$, obtained from a Hill function $\psi$ using \eqrefl{15X23.5},  is a Hill function which has the same number of zeros on $[a,a+2\pi)$ as $\psi$.

To continue, consider  a Hill function $\psi$ with  $\mu$ of parabolic type $\mcP^q_n$ with $n\ge 1$.   
Let $(\psi_1,\psi_2)$ be a Wronskian-normalised basis in which the monodromy matrix takes the form \eqref{18I24.p1}. 
 Let $x \in [0,2\pi)$ be a zero of $\psi_2$, since $\psi_2(x+2\pi) = (-1)^n\psi_2(x)$, we conclude that the number of zeros of $\psi_2$ in a semi-closed interval $I$ of length $2\pi$ is independent of $I$.  
 
\begin{figure}
   \centering
\includegraphics[width=.29\textwidth]{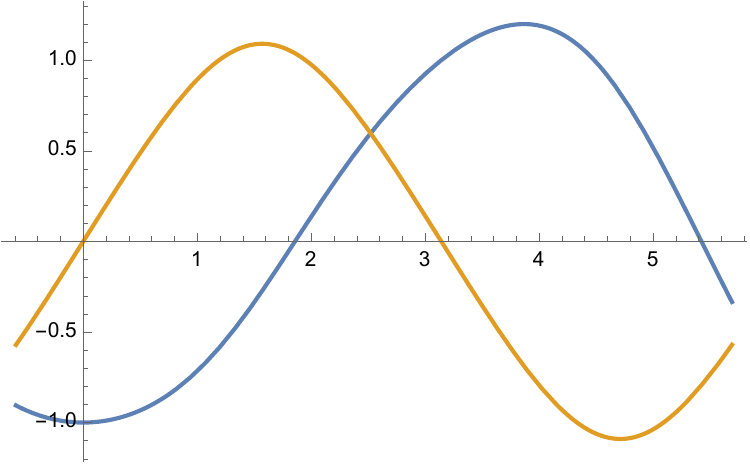}%
\hspace{0.4cm}
\includegraphics[width=.29\textwidth]{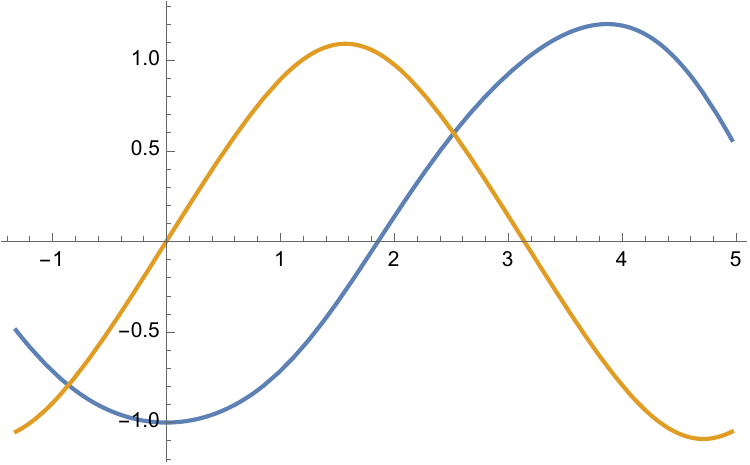}%
\hspace{0.4cm}
\includegraphics[width=.29\textwidth]{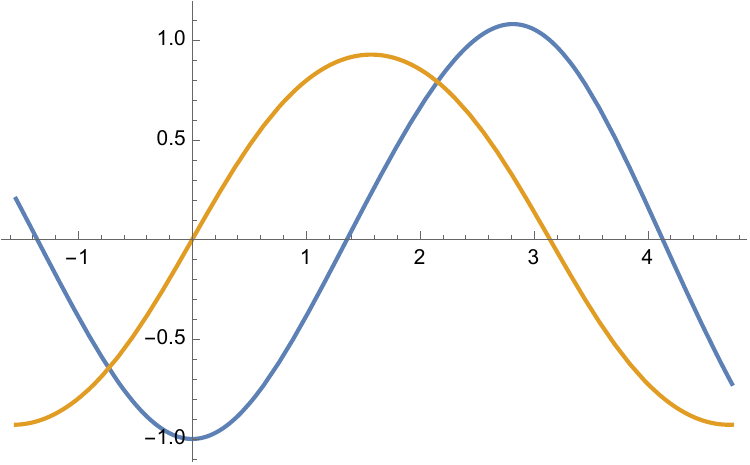}%
   \caption{Canonical parabolic Hill functions of type $\mcP^{q}_2$, with $\psi_1$ in blue and $\psi_2$ in orange.  
   For the first two plots $q = -1$, whereas for the last plot $q=1$.
   When $I= [-0.59, 5.69)$  (left plot) both functions have two zeros, when $I=[-1.32, 4.96)$ (middle plot) the function $\psi_1$ has only one zero, and when $I =[-1.55, 4.74)$ (right plot) the function $\psi_1$ has three zeros. }\label{F10VI24.1}
 \end{figure}

Continuing with the parabolic case, as already explained,  
without loss of generality we can take the Hill functions $(\psi_1,\psi_2)$ to take the canonical form 
\eqref{17I24.2}-\eqref{17I24.3}.
Then for $n\ge 1$ the Hill function $ \psi_2$ has precisely $n+1$ zeros on the closed interval $[0,2\pi]$, with two of them at the ends of the interval. 
Hence $\psi_2$ has precisely $n$ zeros on $I$.
If $\psi$ is proportional to $\psi_2$, it thus has exactly $n$ zeros on $I$. Otherwise it
follows that
 $\psi$
 has at least $n-1$ zeros on $I$ by Lemma~\ref{L14III24.1}. It cannot have more than $n+1$ by the same lemma, otherwise $\psi_2$ would have more than $n$.  
 We conclude that  $\psi$ can only have $n-1$, $n$ or $n+1$ zeros. 
 All three cases can occur, see Figure~\ref{F10VI24.1} for an illustrative plot.   
 
\begin{figure}
  \centering
\includegraphics[width=.28\textwidth]{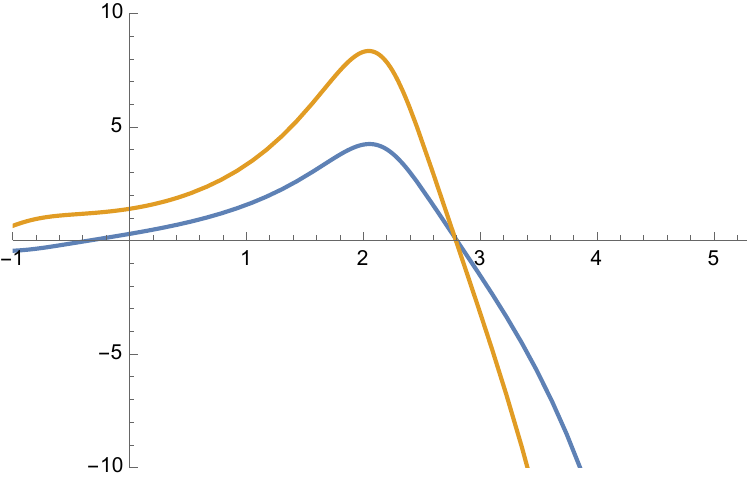}
\includegraphics[width=.28\textwidth]{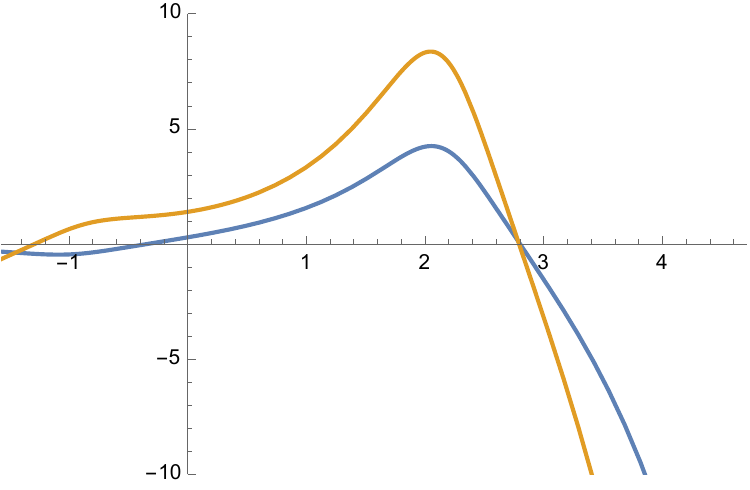}
\includegraphics[width=.28\textwidth]{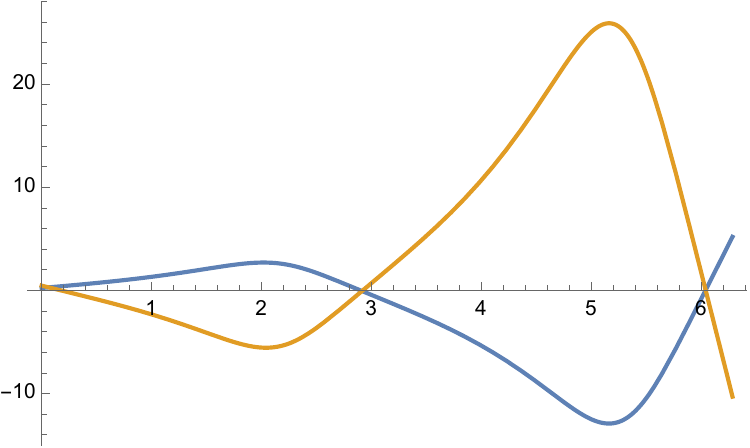}%
  \caption{ 
  Plots of Hill functions of type $\mcH_2$, with $ \psi_1$ in blue and $\psi = \alpha_1 \psi_1 + \psi_2$ in orange, where 
  $(\psi_1, \psi_2)$ are the canonical Hill functions \eqref{16I24.2}-\eqref{16I24.3}.
  In each of the plots one of the functions has two zeros, while the other one has one, two, or three. 
  The first plot is for $\sigma = 2$ and $\alpha_1 =2$ on $\interval = [-1, -1+2\pi)$, the second plot is for $\sigma = 2$
  and $\alpha_1 =2$
  on $\interval = [-\pi/2, 3\pi/2)$ and the last plot is for $\sigma = 1$ and $\alpha_1 =-2$ on $\interval = [0, 2\pi)$. 
  }\label{F10VI24.12}
\end{figure}

Consider, next, a hyperbolic type $\mcH_n$ with $n\ge 1$. 
Due to the form of the monodromy matrix \eqref{27X23.33hq}, the canonical Hill functions have 
the same number of zeros on any $\interval$ by the same argument that was given for the Hill function $\psi_2$ in the parabolic case. 
So, if a function $\psi$
 diagonalising the monodromy matrix has $n$ zeros on a semi-closed interval of length $2\pi$, then it has $n$ zeros on any such interval, and any other Hill function has $n-1$, $n$ or $n+1$ zeros. 
See Figure \ref{F10VI24.12} for a plot demonstrating this for the  Hill functions $  \psi_1$ and the linear combination $\psi = \alpha_1 \psi_1 + \psi_2$, where $(\psi_1, \psi_2)$ 
are the canonical Hill functions given by \eqref{16I24.2} and \eqref{16I24.3}.

Now, in all remaining cases,  $\mu$ can be transformed to a constant.
While this is not necessary for what follows, we note that it suffices to  count zeros on $[0,2\pi)$. Indeed, after the transformation  to a constant has been carried-out, let $I=[a,a+2\pi)$, and let $\psi$ be a Hill function. Then $\psi_a:=\psi(x-a)$ is a Hill function which has the same number of zeros on $[0,2\pi)$ as $\psi$ had on $[a,a+2\pi)$.

In any case, we have:

\begin{enumerate}
  \item
    Suppose that $\mu$
    can be transformed to a negative constant $ \mc \in (-\infty,0)$, i.e.
     $\mu$ is of $\mcE_{\mc}$-type.
     Then, after this transformation,  every Hill function  $\psi$ is a linear combination of $\cos\big(\frac{\sqrt{|\mc|} x}{2}\big)$ and $\sin \big(\frac{\sqrt{|\mc|} x}{2}\big)$,
     i.e. proportional to $\sin(\sqrt{|\mc|} x/2+x_0)$
    for some $x_0$. Hence $\psi$ has $n$ zeros or $n+1$ zeros
  on $\interval $,
    where $n = \lfloor \sqrt{|\mc|} \rfloor$, depending upon the value of   $x_0$.

  \item Suppose that $\mu$ can be transformed to a positive constant
  , i.e. $\mu$ is of $\mcH_0$-type,
   then
  \begin{equation}\label{11I24.2}
    \psi(x) = \alpha e^{\frac{\sqrt{\mu}}{2}x} +  \beta e^{-\frac{\sqrt{\mu}}{2}x}
    \,,
  \end{equation}
 with $\alpha$ and $\beta$ in $\R$,  not simultaneously zero, which has either one or  no zeros on $\interval $.

  \item
  Suppose that $\mu$ can be transformed to $\mu = 0$,
  i.e.\ $\mu$ is of $\mcP_0$-type,
  then
  \begin{equation}
    \psi(x) = \alpha x
    +
    \beta \,,
  \end{equation}
  with $\alpha$ and $\beta$ in $\R$, not simultaneously zero, which has either one or  no zeros on $\interval $.
\qed
%
\end{enumerate}

As an immediate corollary, we obtain:

\begin{theorem}
  \label{T19I24.1}
  A mass-aspect function $\mu$ can be transformed to a constant by
  an asymptotic symmetry if and only if
  \begin{enumerate}
    \item either $\mono$ is elliptic,
    \item  or there exists a solution of the Hill equation which has
    no zeros on a half-closed interval of length $2\pi$.
    \qed
  \end{enumerate}
\end{theorem}

\subsection{Changing orientation}

So far we have only considered orientation-preserving diffeomorphisms of the circle.
It is natural to enquire what happens when orientation is reversed:

\begin{proposition}
  \label{p20I24.1}
The type of the mass-aspect function does not change under orientation-changing diffeomorphisms of $S^1$.
\end{proposition}

\proof
It suffices to establish the result for the diffeomorphism $\sigmar(\red{\varphi})=-\varphi$.

Let us start by showing that if  $\mu$ is of $\mcP^q_{n}$-type, then $\mu\circ \sigmar$ is again of $\mcP^{q}_{n}$-type.
For this, let $(\psi_1,\psi_2)$ be a Wronskian-normalised basis of Hill functions with $\mono$ in the canonical form as above. Then $(\psi_1\circ \sigmar,-\psi_2\circ \sigmar)$ is a Wronskian-normalised basis of Hill functions for the Hill equation with $\mu$ replaced by $\mu\circ \sigmar$. We have
\begin{eqnarray}
  \left(
    \begin{array}{c}
      \psi_{1}\circ \sigmar
   \\
      \psi_{2}\circ \sigmar
   \\
    \end{array}
  \right)\Big|_{ 2\pi}
   &=&
  \left(
    \begin{array}{c}
      \psi_{1}
    \\
      \psi_{2}
    \\
    \end{array}
  \right)\Big|_{-2\pi}
  =
  \eta \left(
  \begin{array}{cc}
    1& q
     \\
    0& 1
  \end{array}
  \right) ^{-1}
  \left(
    \begin{array}{c}
      \psi_{1}
    \\
      \psi_{2}
    \\
    \end{array}
  \right) \Big|_{0}
  \nonumber
\\
   &=& \eta \left(
  \begin{array}{cc}
    1& -q
     \\
    0& 1
  \end{array}
  \right)
  \left(
    \begin{array}{c}
      \psi_{1}
    \\
      \psi_{2}
    \\
    \end{array}
  \right) \Big|_{0}
  =
   \eta \left(
  \begin{array}{cc}
    1&  q
     \\
    0& -1
  \end{array}
  \right)
  \left(
    \begin{array}{c}
      \psi_{1}
    \\
      -\psi_{2}
    \\
    \end{array}
  \right) \Big|_{0}
  \nonumber
\\
   &=& \eta \left(
  \begin{array}{cc}
    1&  q
     \\
    0& -1
  \end{array}
  \right)
  \left(
    \begin{array}{c}
      \psi_{1} \circ \sigmar
    \\
      -\psi_{2} \circ \sigmar
    \\
    \end{array}
  \right) \Big|_{0}
  \,.
  \nonumber
\end{eqnarray}
This is equivalent to
\begin{eqnarray}
  \left(
    \begin{array}{c}
      \psi_{1}\circ \sigmar
   \\
     - \psi_{2}\circ \sigmar
   \\
    \end{array}
  \right)\Big|_{ 2\pi}
   &=&  \eta \left(
  \begin{array}{cc}
    1&  q
     \\
    0&  1
  \end{array}
  \right)
  \left(
    \begin{array}{c}
      \psi_{1} \circ \sigmar
    \\
      -\psi_{2} \circ \sigmar
    \\
    \end{array}
  \right) \Big|_{0}
  \,,
  \nonumber
\end{eqnarray}
as desired.

The remaining cases are either obvious or similar.
\qed
%
%

\subsection{Gluing BK solutions with positive mass}
 \label{ss6XI23.1}

\begin{figure}
  \centering
  \includegraphics[width=.9\textwidth]{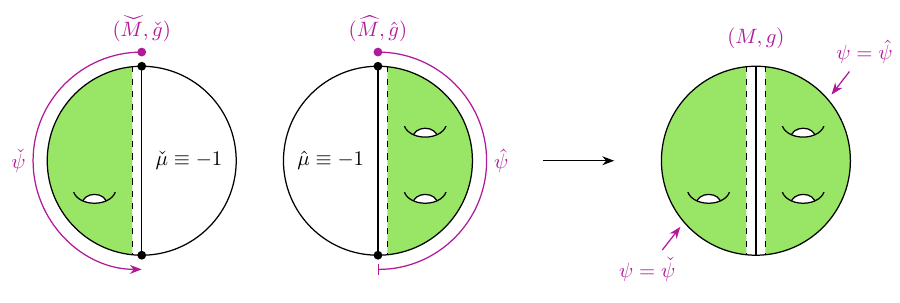}
  \caption{Solutions of the Hill equation on the boundary of the glued manifold can be obtained by concatenating solutions $\hpsi$ and $\chpsi$ from the original manifolds. The metric is that of hyperbolic space outside of the highlighted region.}\label{F15X23.1}
\end{figure}

We wish to determine the monodromy matrix of the Maskit-glued manifolds in terms of the monodromy matrices of the original ones. For this let us denote by $(\chM,\chg)$
and $(\hatM,\hatg)$ the ALH CSC manifolds that are Maskit-glued to $(M,g)$. Then a solution $\psi$ of the Hill equation associated with $(M,g)$ restricts on $[-\frac{\pi}{2},\frac{\pi}{2}]$ to a
solution, say  $\hpsi$, of the
(properly transformed, as dictated by the gluing procedure)
Hill equation  associated with $(\hatM,\hatg)$.
 Similarly, $\psi$  restricts on $[\frac{ \pi}{2},\frac{3\pi}{2}]$ to a solution, say $\chpsi$, of the Hill equation  associated with $(\chM,\chg)$. This shows that solutions  $\psi$ of the Hill equation  associated with $(M,g)$ can  be obtained by concatenating solutions $\chpsi$ defined on $[\frac{\pi}{2}, \frac{3\pi}{2}]$ with solutions $\hpsi$ defined on $[-\frac{\pi}{2},\frac{\pi}{2}]$; this requires the matching of Cauchy data
 $(\psi,\psi')$ at, e.g., $\varphi = \frac{\pi}{2}$.
 The monodromy matrix can then be calculated by comparing $\chpsi(\frac{3\pi}{2})$ with $\hpsi(-\frac{\pi}{2})$, cf.\ Figure~\ref{F15X23.1}.

\begin{remark}
  \label{R19I24.1}
  {\rm
  For future reference, we note that the construction in this section, and the following, applies whenever an ALH metric has a constant  mass-aspect function near a point $\varphi_0$ on the conformal boundary.
  \qed
  }
\end{remark}

We start with a CSC ALH manifold $(\chM,\chg)$ with hyperbolic monodromy and with a Hill function  without zeros.
By Theorem~\ref{T19I24.1}
we can transform $\mu$ to a constant $\chm>0$.

We choose the following Wronskian-normalised basis of solutions of the Hill equation with constant mass parameter $\chm >0$,
 \begin{equation}
  \cpsi_1(x)
    = \frac{1}{|\chm |^{1/4}} e^{\sqrt{\chm}\red{x}/2}\,,
  \qquad
  \cpsi_2 (x)= \frac{1}{|\chm |^{1/4}} e^{-\sqrt{\chm}\red{x}/2}\,.
\end{equation}
The associated monodromy matrix reads
 \begin{equation}\label{2XII23.1a}
   \chmonoz
   =
   \left(
     \begin{array}{cc}
       \exp\big(\sqrt{\chm}\pi\big)& 0
        \\
       0& \exp\big(-\sqrt{\chm}\pi\big)
     \end{array}
   \right)
   \,.
 \end{equation}

We wish to find monotonously increasing solutions of  \eqref{finalmucc}  so that  $\cpsi_1$, transformed using $\chf$ as in \eqref{15X23.5}, becomes a  solution of the Hill equation \eqref{10X23.1} with $ \mu=-1$   on $[-\frac{\pi}2,\frac{\pi} 2]$. Since the sign of $\cpsi_1$ is preserved by this transformation, after performing a rotation of the circle the solution can be made proportional to  $  \cos (\varphi/2)$. 
 \footnote{One could alternatively map $\psi_2$ to $\cos(\varphi/2)$, which would leave to matrices $\mono$ conjugate to the ones we present, without affecting their traces, which are relevant for the determination of the final masses.}
We are thus led to  the ODE
%
\begin{equation}
 \label{2XII23.1b}
 \frac{\chf'(\red{x})}{
  \big(
    \cpsi_1\big( \chf (\red{x})\big)
   \big)^2} =
      \frac{ \cboost}{  2
  \big(
   \cos \left(\frac{\red{x}}{2}\right)
   \big)^2}
 \,,
\end{equation}
where $\cboost>0$ is a constant.
Elementary integration gives
 \begin{equation}
 \chf(\red{x}) = -
 \frac 1 {\sqrt{\chm }}
 {\log \left( -\cboost \tan
   \left(\frac{\red{x}}{2}\right)+ \cshiftH \right)}
   \,,
    \label{7XI23.1a}
 \end{equation}
 where $\cshiftH$ is a constant. 
The function $\chf$ will be defined on $[-\frac{\pi}2,\frac{\pi} 2]$ if and only if
\begin{equation}\label{1XII23.11}
  \cshiftH > \cboost
  \,.
\end{equation}
The choice of normalisation   $\sqrt 2 \cos(x/2)$ leads to
\begin{equation}\label{1XII23.12}
  \cboost = 1
  \,.
\end{equation}
%
In order to be able to extend $\chf$ from $[-\frac{\pi}2,\frac{\pi} 2]$ to a diffeomorphism of $S^1$ the image by   $\chf$  of $[-\frac{\pi}2,\frac{\pi} 2]$ should have length smaller than $2\pi$:
\begin{equation}\label{5XII23.1}
  \chf\big(\frac{\pi}{2}\big)
    -
  \chf\big(-\frac{\pi}{2}\big) < 2\pi
  \qquad
   \Longleftrightarrow
    \qquad
    \cshiftH > \coth(  \sqrt{\chm}\pi)
     \,.
\end{equation}
%
Under the transformation  \eqrefl{15X23.5}, more precisely under
\begin{equation}\label{15X23.5a}
(\cpsi_{1 },\cpsi_{2 })
 \mapsto
 (\chf' )^{-1/2}(\cpsi_1\circ \chf , \cpsi_2\circ \chf )
\end{equation}
with
any $\chf$ which coincides with \eqref{7XI23.1a}
on $[-\frac{\pi}{2},\frac{\pi}{2}]$,
the pair of functions $(\cpsi_1, \cpsi_2)$  is mapped on $[-\frac{\pi}{2},\frac{\pi}{2}] $
to the following pair of solutions of the Hill equation with mass parameter equal to $-1$:
\begin{equation}
   \label{functionshill}
    \left(
   \begin{array}{c}
     \cpsi_1 (x) \\
     \cpsi_2 (x)
   \end{array}
   \right)
   \mapsto
      \sqrt 2 \left(
   \begin{array}{c}
   \cos
   \left(\frac{\red{x}}{2}\right)
    \\
       -  \sin  \left(\frac{\red{x}}{2} \right)
       + \cshiftH \cos
   \left(\frac{\red{x}}{2}\right)
   \end{array}
   \right)
    =
   \cR
   \left(
   \begin{array}{c}
     \sqrt 2 \sin
   \left(\frac{\red{x}}{2}\right) \\
      \sqrt 2  \cos \left(\frac{\red{x}}{2} \right)
   \end{array}
   \right)
   \,,
\end{equation}
where
\begin{equation}\label{7XI23.1bx}
  \cR=  \left(
    \begin{array}{cc}
     0 & 1\\
     -1   &  \cshiftH
    \end{array}
    \right)\,.
\end{equation}
Equivalently,
 \ptcheck{2XII23}
\begin{equation}\label{7XI23.1-x}
    \sqrt 2\left(
   \begin{array}{c}
     \sin
   \left(\frac{\red{x}}{2}\right) \\
       \cos \left(\frac{\red{x}}{2} \right)
   \end{array}
   \right)
    = \cR^{-1}   (\chf' )^{-1/2}
     \left(
   \begin{array}{c}
     \cpsi_1\circ \chf  \\
     \cpsi_2\circ \chf
   \end{array}
   \right)
   \,,
   \qquad
  \cR^{-1}=  \left(
    \begin{array}{cc}
     \cshiftH & -1\\
     1   & 0
    \end{array}
    \right)
     \,.
\end{equation}
Since monodromy is invariant under \eqref{15X23.5a}, we have
\ptcheck{3XII; preserves positivity of the Hill functions}
\begin{eqnarray}
  \left(
   \begin{array}{cc}
     ( (\chf' )^{-1/2} \cpsi_1\circ \chf)'   &  (\chf' )^{-1/2}  \cpsi_1\circ \chf  \\
     ( (\chf' )^{-1/2} \cpsi_2\circ \chf)'   &   (\chf' )^{-1/2} \cpsi_2\circ \chf
   \end{array}
   \right) \Big|_{ \frac{3\pi}{2} }
    & = &
    \chmonoz   \left(
   \begin{array}{cc}
     ( (\chf' )^{-1/2} \cpsi_1\circ \chf)'   &   (\chf' )^{-1/2} \cpsi_1\circ \chf  \\
     ( (\chf' )^{-1/2} \cpsi_2\circ \chf)'   &  (\chf' )^{-1/2}  \cpsi_2\circ \chf
   \end{array}
   \right) \Big|_{-\frac{\pi}{2} }
    \nonumber
\\
   & = &
  \sqrt 2 \chmonoz \cR
   \left(
   \begin{array}{cc}
    \big(  \sin
   \left(\frac{\red{x}}{2}\right)
   \big)'
    &     \sin
   \left(\frac{\red{x}}{2}\right) \\
     \big(  \cos
   \left(\frac{\red{x}}{2}\right)
   \big)'
    &      \cos \left(\frac{\red{x}}{2} \right)
   \end{array}
   \right) \Big|_{-\frac{\pi}{2}}
    \nonumber
\\
   & = &
  \sqrt 2 \chmonoz \cR
   \left(
   \begin{array}{cc}
    \frac{1}{2 \sqrt 2}
    &     -\frac{1}{  \sqrt 2}  \\
     \frac{1}{2 \sqrt 2}
    &     \frac{1}{  \sqrt 2}
   \end{array}
   \right)
   %
     \,.\label{7XI23.98}
\end{eqnarray}

We continue with the analogous problem,  where all checked quantities are replaced by hatted ones, with the interval $[- \frac{\pi}2,\frac{ \pi} 2]$  replaced by $[ \frac{\pi}2,\frac{3\pi} 2]$ (compare Figure~\ref{F15X23.1}), and where the function  $ \cos(x/2)$ is replaced by
$  \sin(x/2)$ in \eqref{2XII23.1b}:
 \begin{equation}
 \hf(\red{\varphi}) = -
 \frac 1 {\sqrt{\hm}}
 {\log \left( \hboost \cot
   \left(\frac{\varphi}{2}\right)+ \hshiftH \right)}
   \,,
    \label{c7XI23.1a}
 \end{equation}
 where $\hshiftH$  and $\hboost>0$ are constants.
The function $\hf$ will be defined on $[ \frac{\pi}2,\frac{3\pi} 2]$  if and only if
\begin{equation}\label{c1XII23.11}
  \hshiftH > \hboost
  \,.
\end{equation}
The choice of normalisation   $\sqrt 2 \sin(x/2)$ leads to
\begin{equation}\label{c1XII23.12}
  \hboost  = 1
  \,.
\end{equation}
Similarly to \eqref{5XII23.1} we have
\begin{equation}\label{5XII23.2}
    \hshiftH > \coth(  \sqrt{\hm}\pi)
     \,.
\end{equation}
%

Under the transformation  \eqrefl{15X23.5a},
with $\chf$ there replaced by any function $\hf$ which coincides with \eqref{c7XI23.1a}
on $[ \frac{\pi}{2},\frac{3\pi}{2}]$,
the pair of functions $(\hpsi_1, \hpsi_2)$  is mapped on $[ \frac{\pi}{2},\frac{3\pi}{2}] $
to the following  solutions of the Hill equation with mass parameter equal to $-1$:
 \ptcheck{3XII}
\begin{equation}
   \label{cfunctionshill}
    \left(
   \begin{array}{c}
     \hpsi_1  \\
     \hpsi_2
   \end{array}
   \right)
   \mapsto
      \sqrt 2 \left(
   \begin{array}{c}
   \sin
   \left(\frac{\red{x}}{2}\right)
    \\
       \hshiftH  \sin  \left(\frac{\red{x}}{2} \right)
       +   \cos
   \left(\frac{\red{x}}{2}\right)
   \end{array}
   \right)
    =
   \hR
   \left(
   \begin{array}{c}
     \sqrt 2 \sin
   \left(\frac{\red{x}}{2}\right) \\
      \sqrt 2  \cos \left(\frac{\red{x}}{2} \right)
   \end{array}
   \right)
   \,,
\end{equation}
where
\begin{equation}\label{c7XI23.1}
  \hR=  \left(
    \begin{array}{cc}
    1  & 0 \\
     \hshiftH   &  1
    \end{array}
    \right)\,.
\end{equation}
Equivalently,
 \ptcheck{3XII and  again 7VII24, mathematica file Checking the R matrices from scratch 7 VII 24 Piotr.nb}
\begin{equation}\label{7XI23.1-xy}
    \sqrt 2\left(
   \begin{array}{c}
     \sin
   \left(\frac{\red{x}}{2}\right) \\
       \cos \left(\frac{\red{x}}{2} \right)
   \end{array}
   \right)
    = \hR^{-1} (\hf' )^{-1/2}
     \left(
   \begin{array}{c}
     \hpsi_1 \circ \hf \\
     \hpsi_2 \circ \hf
   \end{array}
   \right)
   \,,
   \qquad
  \hR^{-1}=    \left(
    \begin{array}{cc}
     1 & 0\\
     -\hshiftH  &1
    \end{array}
    \right)
     \,.
\end{equation}

Similarly to \eqref{7XI23.98} we have
\begin{eqnarray}
  \left(
   \begin{array}{cc}
     ( (\hf' )^{-1/2} \hpsi_1\circ \hf)'   &  (\hf' )^{-1/2}  \hpsi_1\circ \hf  \\
     ( (\hf' )^{-1/2} \hpsi_2\circ \hf)'   &   (\hf' )^{-1/2} \hpsi_2\circ \hf
   \end{array}
   \right) \Big|_{- \frac{\pi}{2} }
    & = &
    \hmonoz^{-1}   \left(
   \begin{array}{cc}
     ( (\hf' )^{-1/2} \hpsi_1\circ \hf)'   &   (\hf' )^{-1/2} \hpsi_1\circ \hf  \\
     ( (\hf' )^{-1/2} \hpsi_2\circ \hf)'   &  (\hf' )^{-1/2}  \hpsi_2\circ \hf
   \end{array}
   \right) \Big|_{ \frac{3\pi}{2} }
    \nonumber
\\
   & = &
  \sqrt 2 \hmonoz^{-1} \hR
   \left(
   \begin{array}{cc}
    \big(  \sin
   \left(\frac{\red{x}}{2}\right)
   \big)'
    &     \sin
   \left(\frac{\red{x}}{2}\right) \\
     \big(  \cos
   \left(\frac{\red{x}}{2}\right)
   \big)'
    &      \cos \left(\frac{\red{x}}{2} \right)
   \end{array}
   \right) \Big|_{ \frac{3\pi}{2}}
    \nonumber
\\
   & = &
  \sqrt 2 \hmonoz^{-1} \hR
   \left(
   \begin{array}{cc}
   - \frac{1}{2 \sqrt 2}
    &     \frac{1}{  \sqrt 2}  \\
    - \frac{1}{2 \sqrt 2}
    &   -  \frac{1}{  \sqrt 2}
   \end{array}
   \right)
     \,.\label{7XI23.96}
\end{eqnarray}
%
Comparing \eqref{7XI23.1-x} with \eqref{7XI23.1-xy}, for $x$ near $\frac{\pi}{2}$ we have
\ptcheck{3XII}
\begin{equation}\label{7XI23.99}
\cR  ^{-1} (\chf' )^{-1/2}
     \left(
   \begin{array}{c}
     \cpsi_1\circ \chf  \\
     \cpsi_2\circ \chf
   \end{array}
   \right)
    =  \hR^{-1} (\hf' )^{-1/2}
     \left(
   \begin{array}{c}
     \hpsi_1 \circ \hf \\
     \hpsi_2 \circ \hf
   \end{array}
   \right)
     \,.
\end{equation}
It follows
 that the functions at the left-hand side of \eqref{7XI23.99} can be smoothly continued from   $[ \frac{\pi}{2},\frac{3\pi}{2}]$ to  $[- \frac{\pi}{2}, \frac{\pi}{2}]$
  by the functions on the right-hand side.
Hence the pair of functions $(\psi_1,\psi_2)$, defined as%
\begin{equation}\label{7XI23.97}
   \left(
   \begin{array}{c}
     \psi_1 \\
     \psi_2
   \end{array}
   \right) \Big|_{[\frac{\pi}{2},\frac{3\pi}{2}]}
   = \cR  ^{-1} (\chf' )^{-1/2}
     \left(
   \begin{array}{c}
     \cpsi_1\circ \chf  \\
     \cpsi_2\circ \chf
   \end{array}
   \right)
   \,,
   \quad
  \left(
   \begin{array}{c}
     \psi_1 \\
     \psi_2
   \end{array}
   \right) \Big|_{[- \frac{\pi}{2}, \frac{\pi}{2}]}
    =     \hR^{-1} (\hf' )^{-1/2}
     \left(
   \begin{array}{c}
     \hpsi_1 \circ \hf \\
     \hpsi_2 \circ \hf
   \end{array}
   \right)
     \,,
\end{equation}
and extended to $\R$ by solving the Hill equation with a $2\pi$-periodic mass-aspect function,
are solutions of the Hill equation on the manifold obtained by Maskit gluing as in Figure~\ref{F15X23.1}.

Let $\mono$ denote the monodromy matrix of the Maskit-glued solution. We have
\ptcheck{3XII}
\begin{align}
  \sqrt 2
     \hR ^{-1} \hmonoz^{-1}
    \hR
   &
   \left(
   \begin{array}{cc}
  -  \frac{1}{2 \sqrt 2}
    &     \frac{1}{  \sqrt 2}  \\
   -  \frac{1}{2 \sqrt 2}
    &    - \frac{1}{  \sqrt 2}
   \end{array}
   \right)
   &
   \nonumber
\\
&
    =
   \hR ^{-1}
  \left(
   \begin{array}{cc}
     ( (\hf' )^{-1/2} \hpsi_1\circ \hf)'   &  (\hf' )^{-1/2}  \hpsi_1\circ \hf  \\
     ( (\hf' )^{-1/2} \hpsi_2\circ \hf)'   &   (\hf' )^{-1/2} \hpsi_2\circ \hf
   \end{array}
   \right) \Big|_{- \frac{\pi}{2} }
    & \mbox{(by \eqref{7XI23.96})}
 \nonumber
\\
& =
    \left(
   \begin{array}{cc}
    \psi_1'   &  \psi_1  \\
   \psi_2'  &   \psi_2
   \end{array}
   \right) \Big|_{- \frac{\pi}{2} }
    & \mbox{(by \eqref{7XI23.97})}
 \nonumber
\\
& =
    \mono^{-1} \left(
   \begin{array}{cc}
    \psi_1'   &  \psi_1  \\
   \psi_2'  &   \psi_2
   \end{array}
   \right) \Big|_{  \frac{3\pi}{2} }
    & \hspace{-2cm}\mbox{(by definition of $\mono$)}
 \nonumber
\\
& =
    \mono^{-1}
    \cR^{-1}
  \left(
   \begin{array}{cc}
     ( (\chf' )^{-1/2} \cpsi_1\circ \chf)'   &  (\chf' )^{-1/2}  \cpsi_1\circ \chf  \\
     ( (\chf' )^{-1/2} \cpsi_2\circ \chf)'   &   (\chf' )^{-1/2} \cpsi_2\circ \chf
   \end{array}
   \right) \Big|_{ \frac{3\pi}{2} }
    & \mbox{(by \eqref{7XI23.97})}
 \nonumber
\\
   & =
  \sqrt 2
    \mono^{-1}
    \cR^{-1} \chmonoz  \cR
   \left(
   \begin{array}{cc}
    \frac{1}{2 \sqrt 2}
    &     -\frac{1}{  \sqrt 2}  \\
     \frac{1}{2 \sqrt 2}
    &     \frac{1}{  \sqrt 2}
   \end{array}
   \right)
    & \mbox{(by \eqref{7XI23.98})}
   \,.
   \nonumber
\end{align}
Setting
\begin{equation}\label{4VII24.1}
  \chmono:= \cR^{-1} \chmonoz \cR
  \,,
  \qquad
   \hmono:= \hR^{-1} \hmonoz \hR
   \,,
\end{equation}
we conclude that
\begin{equation}\label{4VII24.2}
  \fbox{
  $\mono = - \chmono \hmono
  \,.$
  }
\end{equation}

\begin{remark}
  \label{R7VII24.1}
We stress that the key formula \eqref{4VII24.2} holds   regardless of the explicit form of the matrices considered above.
\qedskip
\end{remark}

Introducing $\newcshiftH$ and $\newhshiftH$ through the formulae 
$$
 \cshiftH \equiv   \newcshiftH \coth(  \sqrt{\chm}\pi)
\,,
\
  \hshiftH \equiv \newhshiftH \coth(  \sqrt{\hm}\pi)
   \,,
   \
   \mbox{with $\newcshiftH > 1$ and 
   $\newhshiftH > 1$,}
 $$
we find
\begin{equation}\label{8XII23.1}
\chmono := \cR^{-1} \chmonoz \cR
 = \left(
\begin{array}{cc}
 e^{-\sqrt{\chm} \pi } & 2 \newcshiftH \cosh(  \sqrt{\chm}\pi) \\
 0 & e^{\sqrt{\chm} \pi } \\
\end{array}
\right)
\end{equation}
and
\begin{equation}\label{8XII23.2}
 \hmono = \hR^{-1} \hmonoz \hR
  =\left(
\begin{array}{cc}
 e^{\sqrt{\hm} \pi } & 0 \\
 -2 \newhshiftH 
 \cosh(  \sqrt{\hm}\pi)
  & e^{-\sqrt{\hm} \pi } \\
\end{array}
\right)
   \,.
\end{equation}
Hence
 \begin{eqnarray}
  \mono
   & = &
   - \chmono  \hmono    \nonumber
\\
 & = & 
  \left(
    \begin{array}{cc}
     4 \newcshiftH\newhshiftH \cosh \left(\pi 
       \sqrt{\chm}\right) \cosh \left(\pi 
       \sqrt{\hm}\right)-e^{\pi  \sqrt{\hm}-\pi 
       \sqrt{\chm}} & -2 e^{-\pi 
       \sqrt{\hm}} \newcshiftH\cosh
       \left(\pi  \sqrt{\chm}\right) \\
     2 e^{\pi  \sqrt{\chm}} \newhshiftH \cosh \left(\pi 
       \sqrt{\hm}\right) & -e^{\pi  \sqrt{\chm}-\pi 
       \sqrt{\hm}} \\
    \end{array}
    \right)
   \phantom{xxx}
  \label{2XII23.5}
\end{eqnarray}
\ptcnh{old stuff went to oldPartOfHyperbolic3; consider writing an appendix with the proof that the Hill functions have no zeros; the problem with our previous calculations might be that the choice of eigenvectors we made led to negative Hill functions, or one negative the other positive?}

We have
\begin{equation}\label{3XII23.2-}
   \frac 12 \tr\, \mono  =   2 \,\newcshiftH\,
  \newhshiftH \cosh
  (\sqrt{\chm} \pi) \cosh
  (\sqrt{\hm} \pi ) - \cosh
  (\sqrt{\chm} \pi-\sqrt{\hm} \pi )
  \,,
\end{equation}
which
is larger than one when $\newcshiftH > 1$ and $\newhshiftH > 1$, thus $\mono$ is hyperbolic. Both Hill functions are positive by construction. It follows from Theorem~\ref{T19I24.1} that $\mu$ can be transformed to a positive constant $\mc$ by an asymptotic symmetry.
Using \eqref{3XII23.1}, the value of $\mc$ can be read off from the equation
\begin{equation}\label{3XII23.2}
    \fbox{$
    \cosh(\sqrt{\mc}\pi)
    =   2 \, \newcshiftH\,
  \newhshiftH \cosh
  (\sqrt{\chm} \pi)   \cosh
  (\sqrt{\hm} \pi ) - \cosh
  (\sqrt{\chm} \pi-\sqrt{\hm} \pi )
    $, \quad $\newcshiftH >1$, $\newhshiftH > 1$,}
\end{equation}
where $ \newcshiftH $ and $ \newhshiftH $  are otherwise arbitrary
constants introduced in the gluing.

In the special case $\chm=\hm$ and $\newhshiftH=\newcshiftH $, this becomes
\begin{equation}\label{2XII23.15}
 \fbox{$
    \cosh(\sqrt{\mc}\pi)
    =   2 \newcshiftH^2\cosh^2(\sqrt{\chm} \pi  )
    -1
    \,,
    \qquad
    \newcshiftH > 1
   $. }
\end{equation}

It might be of interest to provide an alternative proof, that the
mass-aspect function of the glued manifold is of $\mcH_0$-type. This  provides a consistency check, and proceeds as follows:
Recall that we are gluing together two ALH CSC metrics which, after applying an asymptotic symmetry, near the conformal boundary at infinity coincide with  funnel metrics, as defined  in Section~\ref{s1XI23.1}. The gluing procedure is insensitive to the nature of the metric away from a small neighborhood of the boundary, we might therefore without loss of generality assume that we are gluing together two hyperbolic bridges. The glued manifold is then complete, with constant negative scalar curvature, non-compact, and geometrically finite. Subsequently,  Theorem~\ref{t26XII23.1} shows that the mass aspect of the glued manifold can be transformed, by an asymptotic symmetry, to a positive constant.

Note that the length of the outermost (as defined by the asymptotic end under consideration) shortest geodesic of the resulting manifold is $2\pi \sqrt\mc$, and that \eqref{2XII23.15} provides an explicit formula for the length of this geodesics.

\begin{Remark}
  \label{R13II23.1}
As already pointed out,   while several  calculations above use the explicit form of the functions $\hpsi$ and $\chpsi$, the key formula \eqref{4VII24.2} does not, and applies regardless of the character of the mass aspect function. 

In some results that follow  we will need to determine the number of zeros of the solutions of the Hill equations obtained as above. Some comments  regardless of the character of the Hill functions in the green zone of Figure~\ref{F15X23.1}, might be useful in this context.

We  note first that
 $\hpsi_1$ has no zeros on $[\pi/2,3 \pi/2]$,
  and near $-\pi/2$,
 therefore the number of zeros of $\hpsi_1$ on $[-\pi/2,3\pi/2]$ is the same as the number of zeros on $\hpsi_1$ on $[-\pi/2,\pi/2]$, and is the same as the number of zeros on $\hpsi_1$ on
 $(-\pi/2,\pi/2)$.
  \ptcheck{3IV; together}

 Next,
 $\chpsi_1$ has one zero on $[-\pi/2,\pi/2]$.
 Therefore the number of zeros of $\chpsi_1$ on $[-\pi/2,3\pi/2]$ equals the number of zeros of $\chpsi_1$ on $(-\pi/2,3\pi/2)$,
 and is smaller by one than the number of zeros of $\chpsi_1$ on $[\pi/2,3\pi/2]$.

 Let $\hN_1$ denote the number of zeros of $\hpsi_1$ on $[-\pi/2,3 \pi/2]$, and let $\chN_1$ be that of $\chpsi_1$ on the same interval.
 Let  $\psi_1$ be obtained by concatenating $\hpsi_1|_{[-\pi/2,\pi/2]}$ with $\chpsi_1|_{[\pi/2,3\pi/2]}$. It follows that the number of zeros of $\psi_1$ on
 $[-\pi/2,3 \pi/2]$
 is equal to $\hN_1+\chN_1-1$.

 A similar analysis shows that the number of zeros of the function $\psi_2$, obtained by concatenating $\hpsi_2|_{[-\pi/2,\pi/2]}$ with $\chpsi_2|_{[\pi/2,3\pi/2]}$, is smaller by one than the sum of the numbers of zeros of $\hpsi_2|_{[-\pi/2,3\pi/2]}$  and  $\chpsi_2|_{[-\pi/2,3\pi/2]}$.

It follows from the analysis in Section~\ref{ss12I24.1} 
that shifting the interval $[-\pi/2,3 \pi/2]$ may remove or introduce one zero. 
We conclude that, after Maskit gluing, the number of zeros of any Hill function
 on a half-closed interval of length $2\pi$,
 varies between $\hN +\chN -2$ and $\hN +\chN$, where $\hN$ and $\chN$ are the original numbers of zeros.
 \qed
 \end{Remark} 

\subsection{Gluing BK solutions with negative mass}
 \label{ess6XI23.1}

Consider a CSC ALH manifold $(\chM,\chg)$ with elliptic monodromy.
(Examples are provided by the space-part of the BK metrics \eqref{twoK} with negative mass $\mc$, which have a conical singularity at $r=0$ except if $\mc =-1$.)
The calculations of Section~\ref{ss6XI23.1} are readily adapted to this case, as follows.

After performing an asymptotic symmetry we can bring the metric to the Birmingham-Kottler form with a constant $\chm<0$ \cite[Section~3.3]{Balog}.
We choose the following Wronskian-normalised basis of solutions of the Hill equation
%
 \begin{equation}
  \label{eII424eqnegneg}
  \cpsi_1 (x)=
    \frac{\sqrt 2 }{|\chm |^{1/4}} \sin\big(\frac{\sqrt{|\chm|}\red{x}}{2}\big)
  \,,
  \qquad
  \cpsi_2 (x) =
    \frac{\sqrt 2 }{|\chm |^{1/4}} \cos\big(\frac{\sqrt{|\chm|}\red{x}}{2}\big)
  \,.
\end{equation}
The associated monodromy matrix reads (cf.~\eqrefl{27X23.31e})
 \begin{equation}\label{e2XII23.1}
   \chmonoz
   =
   \left(
\begin{array}{cc}
     \cos \left(\sqrt{\left| \chm\right| }\pi \right)
      & \sin \left(\sqrt{\left| \chm\right| }\pi\right) \\
 -\sin
   \left(\sqrt{\left| \chm\right| }\pi\right) &
    \cos \left(
    \sqrt{\left| \chm\right| }\pi
    \right) \\
\end{array}
\right)
   \,.
 \end{equation}
In order to map $\chm$ to $-1$, for  $\chm< -1/4$ the simplest choice is 
 \begin{equation}
 \chf|_{[-\frac{\pi}{2}, \frac{\pi}{2}]} (\red{\varphi}) = \frac{\red{\varphi}}{\sqrt{|\chm|}}
   \,;
    \label{e7XI23.1a}
 \end{equation}
 the restriction on $\chm$ guarantees that we can
extend $\chf$ from $[-\frac{\pi}2,\frac{\pi} 2]$ to a diffeomorphism of $S^1$ .
Indeed, under the transformation
\begin{equation}\label{e15X23.5a}
(\cpsi_{1 },\cpsi_{2 })
 \mapsto
 (\chf' )^{-1/2}(\cpsi_1\circ \chf , \cpsi_2\circ \chf )
\end{equation}
with
any $\chf$ which satisfies \eqref{e7XI23.1a},
the pair of functions $(\cpsi_1, \cpsi_2)$  is directly mapped on $[-\frac{\pi}{2},\frac{\pi}{2}] $
to the canonical solutions of the Hill equation with mass parameter equal to $-1$:
\begin{equation}
   \label{efunctionshill}
    \left(
   \begin{array}{c}
     \cpsi_1  \\
     \cpsi_2
   \end{array}
   \right)\Big|_{[-\frac{\pi}{2}, \frac{\pi}{2}]}
   \mapsto
   \left(
   \begin{array}{c}
     \sqrt 2 \sin
   \left(\frac{\red{x}}{2}\right) \\
      \sqrt 2  \cos \left(\frac{\red{x}}{2} \right)
   \end{array}
   \right)
   \,.
\end{equation}
This shows that the matrix corresponding to $\cR$ of \eqref{functionshill} can be taken to be the identity matrix:
\ptcheck{7VII24, mathematica file Checking the R matrices from scratch 7 VII 24 Piotr on the side of negative mass.nb}
\begin{equation}\label{9XII23.21}
  \chm < -1/4
  \qquad
  \Longrightarrow
  \qquad
  \cR = \mathrm{Id}
   \,.
\end{equation}

More generally, for any  $  \chm < 0$ we compose the map \eqref{e7XI23.1a} with a boost
(cf.~\eqref{trafobarvarphi}-\eqref{trafobarvarphicos} below),
\ptcnh{argument of arcsin increasing iff cos phi + beta is positive}
 \begin{eqnarray}
 \chf|_{[-\frac{\pi}{2}, \frac{\pi}{2}]} (\varphi)
  &=&
  -\frac{ i }{\sqrt{|\chm|}}
 \log \left(\frac{i \sqrt{1-\cbeta ^2}
 \sin (\varphi )+\cbeta +\cos (\varphi )}{1 + \cbeta  \cos (\varphi )}\right)
  \nonumber
\\
 & = &
 \frac{1}{\sqrt{|\chm|}}
\arcsin
\Big(
\frac{\sqrt{1-\cbeta ^2}\sin (\red{\varphi})
}
{ 1 + \cbeta  \cos (\red{\varphi})
}
\Big)
 \,,
    \label{see7XI23.1a}
 \end{eqnarray}
where
$\log$ denotes the principal branch of the logarithm, and where
for simplicity we chose $
\cbeta \in [0,1) $ in the last line of \eqref{see7XI23.1a}.
Then
\begin{equation}\label{9XII23.2}
 \max \chf|_{[-\frac{\pi}{2}, \frac{\pi}{2}]}   <
   \pi
   \quad
   \Longleftrightarrow
   \quad
   \left\{
     \begin{array}{ll}
         \cbeta
    >     \cos(\sqrt{|\chm|}\pi), & \hbox{$-1/4\le \chm<0$;} \\
       \cbeta\in [0,1), & \hbox{$\chm< -1/4$,}
     \end{array}
   \right.
\end{equation}
which,
given that $\chf$ is odd,
guarantees that we can
extend $\chf$ from $[-\frac{\pi}2,\frac{\pi} 2]$ to a diffeomorphism of $S^1$.
Note that we can choose $\cbeta =0$ when $  \chm <-1/4$, recovering the simpler formula \eqref{e7XI23.1a}.

Under the transformation \eqref{e15X23.5a},
the pair of functions $(\cpsi_1, \cpsi_2)$  is   mapped on $[-\frac{\pi}{2},\frac{\pi}{2}] $
to
\begin{equation}
   \label{seefunctionshill}
    \left(
   \begin{array}{c}
     \cpsi_1  \\
     \cpsi_2
   \end{array}
   \right)\Big|_{[-\frac{\pi}{2}, \frac{\pi}{2}]}
   \mapsto
     \sqrt 2
   \left(
   \begin{array}{c}
      \ckappa^{-1/2}
     \sin
   \left(\frac{\red{x}}{2}\right) \\
      \ckappa^{1/2}
      \cos \left(\frac{\red{x}}{2} \right)
   \end{array}
   \right)
   \,,
\end{equation}
where
$\ckappa$ is defined in analogy to $\ngkappa$
in \eqref{defgammat} as
\begin{equation}
  \ckappa = \sqrt{\frac{1+\cbeta}{1-\cbeta}}\,.
\end{equation}
This shows that the matrix corresponding to $\cR$ of \eqref{functionshill}  equals
\begin{equation}\label{se9XII23.21}
    \chm < 0
  \qquad
  \Longrightarrow
  \qquad
  \cR =  \left(
    \begin{array}{cc}
      \ckappa^{-1/2}  & 0 \\
     0   &
      \ckappa^{1/2}
    \end{array}
    \right)
    \,,
\end{equation}
where $\ckappa \in [1, \infty)$,
with the additional restriction
\begin{equation}
  \label{19324restk}
  \ckappa >  \sqrt{\frac{1+ \cos(\sqrt{|\chm|}\pi)}{1- \cos(\sqrt{|\chm|}\pi)}}
  = \frac{1+ \cos(\sqrt{|\chm|}\pi)}{|\sin(\sqrt{|\chm|}\pi)|}
\end{equation}
when $-1/4 \leq \chm < 0$.
%
%
\ptcheck{111223}
We obtain
\begin{equation}
  \chmono = \cR^{-1} \cmonoz \cR  = \left(
    \begin{array}{cc}
     \cos (\sqrt{|\chm|}\pi) &
      \ckappa \sin
       (\sqrt{|\chm|}\pi) \\
    - {\ckappa^{-1}} {\sin
       (\sqrt{|\chm|}\pi)}& \cos (\sqrt{|\chm|}\pi) \\
    \end{array}
    \right)\,.
 \label{12XII23.33}
\end{equation}

\subsubsection{Negative with negative}
 \ptcnh{sectioning added; this is EllipticAnyToAny, previous in EllipticLargeToLarge.tex}

Given a second elliptic solution ($\hatM, \hg)$, transformed to a gauge where the mass-aspect function
equals a constant
$\hm  < 0$, with monodromy
 \begin{equation}\label{A2Ae2XII23.1}
   \hmonoz
   =
   \left(
\begin{array}{cc}
     \cos \left(\sqrt{\left| \hm\right| }\pi \right)
      & \sin \left(
      \sqrt{\left| \hm\right| }\pi
      \right) \\
 -\sin
   \left(\sqrt{\left| \hm\right| }\pi
   \right) & \cos \left(
   \sqrt{\left| \hm\right| }\pi
   \right) \\
\end{array}
\right)
   \,,
 \end{equation}
we use any diffeomorphism $\hf$ of $S^1$ such that
 \begin{align}
  \hf|_{[\frac{\pi}{2}, \frac{3 \pi}{2}]} (\red{\varphi})
  &=  \frac{i}{\sqrt{|\hm|}} \log \left(\frac{i \sqrt{1-\hbeta ^2} \sin (\varphi )+\hbeta
  -\cos (\varphi )}{1-\hbeta  \cos (\varphi )}\right) \nonumber \\
 &=-
   \frac{1}{\sqrt{|\hm|}}
  \arcsin
 \Big(
 \frac{\sqrt{1-\hbeta ^2}\sin
   \big(\red{\varphi}\big)
     }
  { 1 - \hbeta  \cos
   \big(\red{\varphi}\big)
   }
   \Big)
   \,,
    \label{A2Ae7XI23.1a+}
 \end{align}
 where
$\log$ denotes again the principal branch of the logarithm, and
  the second line of \eqref{A2Ae7XI23.1a+} holds for $\hbeta \in [0,1) $.
 We have that

 \begin{equation}\label{12XI23.e1}
   \hbeta   > \cos(\sqrt{|\chm|}\pi)
 \end{equation}
 if $-1/4 \leq \chm < 0$; no restriction otherwise.
Under the transformation \eqref{e15X23.5a} with $\chf$ replaced by $\hf$ we have
   \begin{equation}
    \label{A2Aefunctionshill2}
     \left(
    \begin{array}{c}
      \hpsi_1  \\
      \hpsi_2
    \end{array}
    \right)\Big|_{[\frac{\pi}{2}, \frac{3\pi}{2}]}
    \mapsto
      \sqrt 2
    \left(
    \begin{array}{c}
      - \hkappa^{-1/2}
      \cos
    \left(\frac{\red{x}}{2}\right) \\
       \hkappa^{1/2}
       \sin \left(\frac{\red{x}}{2} \right)
    \end{array}
    \right)
    \,,
 \end{equation}
where
\begin{equation}
  \hkappa = \sqrt{\frac{1+\hbeta}{1-\hbeta}}
  \,,
\end{equation}
so that the matrix corresponding to $\hR$ of \eqref{7XI23.1-xy} is given by
\begin{equation}
  \hR = \left(
    \begin{array}{cc}
     0 & - \hkappa^{-1/2} \\
     \hkappa^{1/2} & 0 \\
    \end{array}
    \right)
     \,.
\end{equation}
Here, $\hkappa \in [1, \infty)$,
with the additional restriction
\begin{equation}
  \label{hkappares5324}
  \hkappa >
  \frac{1+ \cos(\sqrt{|\hm|}\pi)}{|\sin(\sqrt{|\hm|}\pi)|}
\end{equation}
when $-1/4 \leq \hm < 0$.
Hence
\begin{equation}
  \label{negative121223}
  \hmono = \hR^{-1} \hmonoz \hR =    \left(
    \begin{array}{cc}
         \cos \left(\sqrt{\left| \hm\right| }\pi \right)
          & \hkappa^{-1} \sin \left(
          \sqrt{\left| \hm\right| }\pi
          \right) \\
     -\hkappa
     \sin\left(\sqrt{\left| \hm\right| }\pi\right) & \cos \left(
       \sqrt{\left| \hm\right| }\pi
       \right)
    \end{array}
    \right)
     \,.
\end{equation}
Equation \eqref{4VII24.2} applies with $\hmono=\hR^{-1} \hmonoz \hR$ and $\cmono=\cR^{-1} \cmonoz \cR $ given by \eqref{12XII23.33},
leading to 
%
\begin{equation}\label{27III24.1}
  \mono = - \cmono \hmono
  =
  \left(
\begin{array}{cc}
  \ckappa \hkappa\chsin
\hsin -\chcos\hcos &
   -\frac{\ckappa \hkappa
   \chsin\hcos+
   \chcos \hsin}{\hkappa} \\
 \frac{\chsin
 \hcos}{\ckappa}+\hkappa
   \chcos\hsin &
   \frac{\chsin\hsin}{ \ckappa \hkappa}-\chcos \hcos \\
\end{array}
\right)
  \,,
\end{equation}
where we wrote, for typographical reasons,
$\hcos$ for $\hcosx$, $\hsin$ for $\hsinx$ and the equivalent expressions for the checked quantities,
with trace
 \ptcheck{12XII}%
 %
 \begin{align}
 \frac{1}{2}\mathrm{tr}(\mono) &=
 \frac{\ckappa^2 \hkappa^2+1}{2 \ckappa\hkappa} \sin
  (\sqrt{|\chm|}\pi) \sin
  (\sqrt{|\hm|}\pi)-\cos (\sqrt{|\chm|}\pi)
  \cos (\sqrt{|\hm|}\pi) \nonumber \\
 &=
    \sqrt{\frac{
    (1+ \cbeta \hbeta )^2}{\left(1-\cbeta^2\right)
    \left(1-\hbeta^2\right)}} \sin
    (\sqrt{|\chm|}\pi) \sin
    (\sqrt{|\hm|}\pi)-\cos (\sqrt{|\chm|}\pi)
    \cos (\sqrt{|\hm|}\pi)
  \,.
  \label{A2A29X23.1a}
 \end{align}

The above formula is not very enlightening. But when both
$\hm<-1/4$ and $\chm<-1/4$
 a simpler formula can be obtained when  \eqref{e7XI23.1a} instead of \eqref{see7XI23.1a} is used.
One finds that $\mono$
is manifestly elliptic, keeping in mind that $\pm \mathbb{I}$ is elliptic in our terminology:
\begin{equation}
  \label{negative20124s}
  \mono = - \cmono  \hmono = - \cmonoz \hmonoz= \left(
    \begin{array}{cc}
     -\cos ((\sqrt{|\chm|}+\sqrt{|\hm|})\pi) & -\sin ((\sqrt{|\chm|}+\sqrt{|\hm|})\pi)
       \\
     \sin ((\sqrt{|\chm|}+\sqrt{|\hm|})\pi) & -\cos ((\sqrt{|\chm|}+\sqrt{|\hm|})\pi)
       \\
    \end{array}
    \right)
     \,.
\end{equation}
Half of its trace
  coincides with
  \eqref{A2A29X23.1a} after setting $\hbeta=\cbeta=0$ there.
The mass parameter, say $\mc$,  can
  be read from $\mono$ using  \eqref{31X23.1},
  \begin{equation}\label{A2A29X23.3}
   \fbox{$
  \cos \left(\sqrt{\left| \mc\right| }\pi  \right) =
   \frac{1}{2 } \tr\, \mono =
  \cos \left((  \sqrt{\left| \chm \right|}+ \sqrt{\left| \hm \right|} + 1) \pi\right)
    $}
    \,,
  \end{equation}
  after taking into account the number of zeros of the solution of the Hill equation. Indeed, letting $\hat N$ be the number of zeros of $\hpsi_1$ and $\check N$ that of $\cpsi_1$, namely
 $$
   \hat N = \lfloor \sqrt{\left| \hm \right|}  \rfloor
    \ \mbox{and} \
   \check N = \lfloor \sqrt{\left| \chm \right|}  \rfloor
   \,,
 $$
 the final mass  $\mc<0$ is obtained by finding the solution $\sqrt{\left| \mc\right| }$ of \eqref{A2A29X23.3} lying in
  $[\hat N+ \check N,\hat N + \check N+1) $. 

It follows from Theorem~\ref{T19I24.1} that
 the mass aspect of the glued solution can be transformed to the negative constant $\mc$ so determined.

Note that in the special case,
\begin{equation}
  \label{special16324}
    \sqrt{\left| \chm \right|}+ \sqrt{\left| \hm \right|}   = p
\end{equation}
where $p$ is an integer, the monodromy matrix \eqref{negative20124s} becomes
\begin{equation}
  \label{negative20243}
  \mono = (-1)^{p +1}\left(
    \begin{array}{cc}
     1 & 0
       \\
     0 & 1
       \\
    \end{array}
    \right)
     \,.
\end{equation}

Parabolic solutions exist with
$\hm, \chm < 0$ when 
\begin{equation}
  \label{193negnegtr1}
  \frac{\left(\ckappa^2 \hkappa^2+1\right) \sin (\pi
   \sqrt{\left| \chm \right|}) \sin (\pi  \sqrt{\left| \hm \right|})}{2 \ckappa
   \hkappa}-\cos (\pi \sqrt{\left| \chm \right|}) \cos (\pi  \sqrt{\left| \hm \right|}) = \pm 1\,.
\end{equation}
In what follows we ignore the (simpler) cases where $\cos $ or $\sin$ vanishes.
The case $\tr(\mono)=  \red{2}$ can then be achieved by setting
\begin{equation}
   \hkappa \ckappa = j\left(\frac{\pi  \sqrt{\left| \chm \right|}}{2}\right) j\left(\frac{\pi
   \sqrt{\left| \hm \right|}}{2}\right) \geq 1\,,
\end{equation}
whereas the case $\tr(\mono)=  -2$ is obtained when
\begin{equation}
  \hkappa \ckappa = - j\left(\frac{\pi
  \sqrt{\left| \hm \right|}}{2}\right) \Big/  j\left(\frac{\pi  \sqrt{\left| \chm \right|}}{2}\right)  \geq 1
  \,,
\end{equation}
with $j(x)$ being either   both  $\cot(x)$ or both $\tan(x)$; if  $-1/4 \leq \hm, \chm  < 0$,
we have the additional restrictions \eqref{hkappares5324}, \eqref{19324restk}.

As an example, let $\hm, \chm  < -1/4$ and $\hkappa =1$,
then \eqref{27III24.1} becomes
%
\ptcheck{19VII24} 
\begin{align}
  \label{27III24.21}
  \mono &=
  \left(
    \begin{array}{cc}
      1 + \varepsilon_1 (\chcos +  \hcos)
      & - j\big(
         \frac{\pi  \sqrt{\left| \hm \right|}}{2}\big) (\chcos
          +
       \hcos) \\
        \displaystyle
        (
      \chcos  + \hcos)/
   j\big(\frac{\pi  \sqrt{\left| \hm \right|}}{2}\big)  &
     1 -  \varepsilon_1 ( \chcos
    +
      \hcos )\\
    \end{array}
    \right)
    & \mathrm{for} \quad \mathrm{tr}(\mono) = 1\,,
\end{align}
with $\varepsilon_1 = \pm 1$, 
and \ptcheck{19724}
\begin{align}
  \label{27III24.22}
  \mono &=
  \left(
    \begin{array}{cc}
     - 1 + \varepsilon_1 (\chcos - \hcos)
      & - j\big(
         \frac{\pi  \sqrt{\left| \hm \right|}}{2}\big) (\chcos
          -
       \hcos) \\
        \displaystyle
        (
      \chcos  - \hcos)/
   j\big(\frac{\pi  \sqrt{\left| \hm \right|}}{2}\big)  &
    - 1 - \varepsilon_1 ( \chcos
    -
      \hcos )\\
    \end{array}
    \right)
    & \mathrm{for} \quad \mathrm{tr}(\mono) = -1\,,
\end{align}
with again $\varepsilon_1 = \pm 1$, and $j(x)$ being $\cot(x)$ for $\varepsilon_1 = 1$ and 
$\tan(x)$ otherwise in both cases. 
%

We will need the following simple observation:

\begin{lemma}
  \label{L9IV24.1}
Let $\mono$ be of $\mcP_n^q$ type. If $\mono_{21}\ne 0$, then  $q=(-1)^{n+1}\mathrm{sgn}(\mono_{21})$.
\end{lemma}

\proof
Let $e_2 = (1,0)^T$, let $\lambda\in \{\pm1\}$ be the (double) eigenvalue of $\mono$, set
$$
e_1 : = \epsilon (\mono -\lambda\,\Id)  e_2 = \epsilon
\left(
  \begin{array}{c}
    \mono_{11} -\lambda\\
    \mono_{21} \\
  \end{array}
\right)
 \,,
 \qquad
 \epsilon \in \{\pm 1\}
 \,,
$$
where $\epsilon$ has to be chosen so that $\{e_1,e_2\}$ is positively oriented, as needed to preserve the Wronskian normalisation condition.
By the Cayley-Hamilton theorem $(\mono -  \lambda \, \Id)^2 =0$, hence $(\mono - \lambda \, \Id) e_1=0$. By definition of $e_1$ we have $\mono e_2 = \lambda e_2 + \epsilon e_1$, so that in the basis $(e_1,e_2)$ the matrix $\mono$ takes the form
\begin{equation}\label{9IV23.11}
   \left(
     \begin{array}{cc}
       \lambda & \epsilon \\
       0 & \lambda \\
     \end{array}
   \right)
   =
   \lambda   \left(
     \begin{array}{cc}
       1 & {\epsilon \lambda}\\
       0 & 1 \\
     \end{array}
   \right)
   \,.
\end{equation}
The orientation of this basis can be determined by calculating the determinant
\begin{equation}\label{9IV23.12}
  |e_1 e_2| =  \left|
     \begin{array}{cc}
       \left(\mono_{11} -\lambda\right)\epsilon & 1 \\
       \mono_{21} \epsilon & 0 \\
     \end{array}
   \right| = - \mono_{21} \epsilon
   \,,
\end{equation}
and the result follows.
\qedskip
%

\begin{figure}
  \centering
  \includegraphics[width=.5\textwidth]{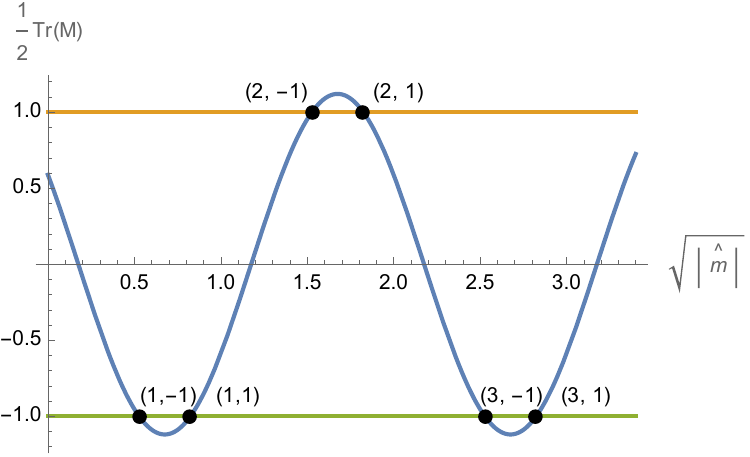}
  \caption{
    The left-hand side of \eqref{193negnegtr1}
    with
  $\sqrt{|\chm|} = 1.3$, 
  $\hkappa = 1$, and  $\ckappa = 1.8$, with
  $\sqrt{\left|\hm\right|} $
  varying between $0$ and $3.4$. The labels $(n,q)$ at the intersection points indicate the parabolic type $\mcP^q_n$ of the associated mass-aspect function. The first intersection point is at $\sqrt{|\hm|}\approx 0.53$, 
  and note that the  plot  is in principle only valid for $\sqrt{|\hm|}>1/2$ because of the constraint \eqref{hkappares5324}.
}\label{F22324.1}
\end{figure}

\begin{remark}
  \label{R11IV24.1}
If $\mono_{21}$ vanishes then $\mono$ is directly upper-triangular, in which case  $q= (-1)^n\mathrm{sgn}(\mono_{12})$.
\qedskip
\end{remark}
Applying the lemma, assuming that $\cos (\pi
\sqrt{\left| \chm \right|})+\cos (\pi  \sqrt{\left| \hm \right|})$ does not vanish (otherwise  \eqref{27III24.21}  is elliptic), we find that the monodromy matrix $\mono$ given by \eqref{27III24.21} is similar to
\begin{equation}
  \left(
   \begin{array}{cc}
     1 & q \\
     0 & 1
   \end{array}
   \right) \quad
 \mathrm{with} \quad q =
 -
  \mathrm{sgn}\Big(
    \frac{\cos (\pi
    \sqrt{\left| \chm \right|})+\cos (\pi  \sqrt{\left| \hm \right|})}
    {\tan \big(\frac{\pi  \sqrt{\left| \hm \right|}}{2}\big)}
    \Big)\,.
 \end{equation}
Likewise, assuming that $\cos (\pi
\sqrt{\left| \chm \right|})-\cos (\pi  \sqrt{\left| \hm \right|})$ does not vanish, the monodromy matrix \eqref{27III24.22} is similar to
\begin{equation}
    \label{negneg17}
 - \left(
     \begin{array}{cc}
       1 & q \\
       0 & 1
     \end{array}
     \right) \quad
   \mathrm{with} \quad q =
    \mathrm{sgn}
    \Big(
      \frac
      {\cos (\pi
      \sqrt{\left| \chm \right|}) - \cos (\pi  \sqrt{\left| \hm \right|})}
      {\tan \big(\frac{\pi  \sqrt{\left| \hm \right|}}{2}\big)}
      \Big)\,,
   \end{equation}
   see
Figure~\ref{F22324.1}.
\ptcheck{19VII; the table}

 \subsubsection{Positive with negative}
 \label{ss8XII23.1}

 We pass now to the case where, near their conformal boundaries at infinity, possibly after applying an asymptotic isometry, $(\chM,\chg)$   is a BK solution with positive mass parameter and $(\hatM,\hatg)$ is a BK solution with negative mass parameter.  After bringing the mass aspect to minus $1$ in the overlap region near $\varphi = \pi/2$, in both cases we use the  Wronskian-normalised basis of solutions of the Hill equation
 \begin{equation}
  \psi_1 =
     \sqrt 2  \sin(  \varphi/2)
  \,,
  \qquad
  \psi_2 =
    \sqrt 2  \cos( \varphi/2)
  \,.
\end{equation}
From Section~\ref{ss6XI23.1} we find (cf. \eqref{8XII23.1}):
\begin{equation}\label{8XII23.3}
\chmono:= \cR^{-1} \chmonoz \cR
 = \left(
\begin{array}{cc}
 e^{-\sqrt{\chm} \pi } & 2 \newcshiftH  \cosh(  \sqrt{\chm} \pi ) \\
 0 & e^{\sqrt{\chm} \pi } \\
\end{array}
\right)
\,.
\end{equation}
If
$\hm <-1/4$
 we can use \eqref{e2XII23.1},
\begin{equation}\label{em2XII23.1}
   \hmono = \hmonoz
   =
   \left(
\begin{array}{cc}
     \cos \left(\sqrt{\left| \hm\right| }\pi \right)
      & \sin \left(
      \sqrt{\left| \hm\right| }\pi
      \right) \\
 -\sin
   \left(\sqrt{\left| \hm\right| }\pi
   \right) & \cos \left(
   \sqrt{\left| \hm\right| }\pi
   \right) \\
\end{array}
\right)
   \,.
\end{equation}
Using  $\mono = -   \hmono \cmono$ and writing, for typographical reasons,
$\chcosh$ for $\chcoshx$,  $\hcos$ for $\hcosx$, etc.,
we find
%
%
\ptcheck{19VII24; with mathematica in monodromy-new.nb } 
\begin{equation}
  \mono =
  \left(
\begin{array}{cc}
 2\newcshiftH \chcosh \hsin-e^{- \pi \sqrt{\chm}} 
  \hcos & -2\newcshiftH \chcosh
   \hcos-e^{- \pi \sqrt{\chm}}
   \hsin \\
 e^{\pi  \sqrt{\chm}} \hsin & -e^{\pi 
   \sqrt{\chm}} \hcos \\
\end{array}
\right) 
\,,
 \label{16III24.1}
\end{equation}
so that
\begin{equation}\label{8XII23.4}
 \frac 12  \tr\, \mono =      \chcoshx
    \big(
     \newcshiftH  \hsinx-
     \hcosx
     \big)
  \,.
\end{equation}
Since the determinant of $\mono$ equals $1$,  the  eigenvalues
$$
 \lambda_\pm =  \frac 12  \tr\, \mono \pm \sqrt{\left( \frac 12  \tr\, \mono\right)^2 -1}
$$
 of $\mono$
will be distinct and real, i.e.\ $\mono$ will be hyperbolic, if and only if
\begin{equation}\label{8XII23.5}
 \Big|\frac 12  \tr\, \mono\Big|  > 1
   \qquad
    \Longleftrightarrow
    \qquad
      \chcoshx
    \big|
    \newcshiftH  \hsinx-
    \hcosx
     \big| > 1
  \,.
\end{equation}
We see that:
\begin{figure}
  \centering
  \includegraphics[width=.5\textwidth]{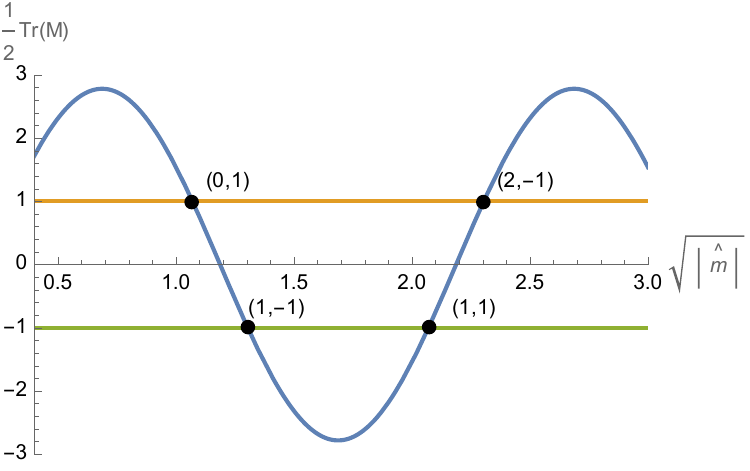}
  \caption{ The right-hand side of \eqref{17324genmonodromy} with
  $\sqrt{\chm} \pi = 1$ and
  $ \newcshiftH \hkappa =1.5$, with
  $\sqrt{\left|\hm\right|} $
  varying between $0.4$ and $3 $. The intersection points are labeled by pairs $(n,q)$ indicating the type $\mcP_n^q$.
  The plot starts at 0.4 because of the constraint \eqref{hkappares5324}.
}\label{F28I24.1}
\end{figure}
\begin{enumerate}
  \item[(1)] For any $\chm>0$, and for any  $\hm$ such that $\sin (\left| \hm\right| \pi)\ne 0$, for all $\newcshiftH$ large enough
  the inequalities \eqref{8XII23.5} are satisfied. For such $\chm$'s and $\newcshiftH$'s the monodromy matrix of the glued manifold will have positive  or negative eigenvalues depending upon the sign of $  \sin (\left| \hm\right| \pi)$.
  \item[(2)] For any $\chm>0$  and $\newcshiftH>1$, we can find a range of $\hm$'s   so that we have instead
\begin{equation}\label{8XII23.5b}
      \chcoshx
    \big|
    \newcshiftH  \hsinx-
    \hcosx
     \big| < 1
  \,,
\end{equation}
leading to elliptic monodromy of the glued solution.
\end{enumerate}

Suppose that  $(\chM,\chg)$   is a conformally compact ALH CSC solution, and that  $(\hatM,\hatg)$ is a BK solution with negative mass, hence containing a conical singularity.
 Then the glued manifold will carry an ALH CSC metric which will be conformally compact and smooth except for one conical singularity.
  In case (1) our construction certainly provides large families of such solutions with hyperbolic monodromy and with a mass-aspect function which \emph{cannot} be transformed to a constant.

Under (2) our construction provides a wealth of such solutions with elliptic
monodromy, in which case
the mass-aspect function can always be transformed to a constant~\cite[Section~3.3]{Balog}.
More on  this and related issues can be found in Section~\ref{s27X23.1} below, see in particular Theorem~\ref{t19I24.2}.

If
$-1/4 \leq\hm <0$, and in fact for any
 $\hm < 0$
 we can use instead
\eqref{negative121223},
\begin{equation}
  \hmono = \hR^{-1} \hmonoz \hR =    \left(
    \begin{array}{cc}
         \hcosx
          & {\hkappa}^{-1}
         \hsinx \\
     -\hkappa \hsinx & \hcosx
    \end{array}
    \right)
     \,.
\end{equation}
One finds (the explicit form of $\mono$ is not very enlightening and is too long to be usefully displayed here)
%
\begin{equation}
 \frac 12 \tr\, \mono =
- \frac{1}{2}\mathrm{tr}\left(\cmono \hmono \right) =\cosh (\sqrt{\chm}\pi)
   \left(\hkappa\, 
   \newcshiftH
   \hsinx-\hcosx\right)
   \,,
    \quad
    \hkappa \in [1, \infty)\,,
    \
    \newcshiftH >1
     \,,
    \label{19I24.11}
 \end{equation}
  and with $\hkappa$ satisfying the additional restriction \eqref{hkappares5324}
  when $-1/4 \leq \hm <0$.
An identical discussion concerning the properties of trace of $\mono$ applies.

In Section~\ref{t19I24.2} we will need to understand the parabolic solutions so obtained.
Then,
$\tr \,\mono/2 = \pm 1 $ will be satisfied by setting
\begin{align}
  \label{17324genmonodromy}
  \text{sech}\left(\pi
  \sqrt{\chm}\right) &= \pm \left(\hkappa \newcshiftH \sin (\pi  \sqrt{|\hm|})
  - \cos (\pi
  \sqrt{|\hm|}) \right)
   \nonumber
   \\
  &= \pm \sqrt{1+\newcshiftH^2 \hkappa^2}  \sin \big(\pi  \sqrt{|\hm|}
  - \mathrm{arccot}(\newcshiftH \hkappa)
  \big) \in (0, 1]\,,
\end{align}
leading to the monodromy matrix 
%
\ptcheck{190724, with mathematica}
\begin{equation}
   \mono 
    = \left(
      \begin{array}{cc}
       \left(\sqrt{y^2-1}+y\right) \cos (\pi  \sqrt{|\hm|})\pm2 &
         \pm \frac{2 \newcshiftH}{1-\hkappa
         \newcshiftH \tan (\pi 
         \sqrt{|\hm|})}+\frac{\left(\sqrt{y^2-1}-y\right) \sin
         (\pi  \sqrt{|\hm|})}{\hkappa} \\
       \hkappa \left(\sqrt{y^2-1}+y\right) \sin (\pi 
         \sqrt{|\hm|}) & -\left(\sqrt{y^2-1} + y\right) \cos (\pi 
         \sqrt{|\hm|}) \\
      \end{array}
      \right)
    ,
\end{equation}
where
\begin{equation}
  \label{22324eqnot}
1  <   y = \cosh(\pi \sqrt{\chm}) =  \pm \frac{1}{\newcshiftH\, \hkappa\, \sin (\pi  \sqrt{|\hm|})-\cos (\pi
  \sqrt{|\hm|})} < \infty\,.
\end{equation}
By Lemma~\ref{L9IV24.1} this matrix is of parabolic type,
\begin{equation}
 \bar \mono = A^{-1} \,\mono\, A = \pm\left(
  \begin{array}{cc}
    1 & q \\
    0 & 1
  \end{array}
  \right) \quad
\mathrm{with} \quad q = \mp \,
 \mathrm{sgn}(
  \sin(\sqrt{|\hm|}\pi) )\,,
\end{equation}
 see Figure~\ref{F28I24.1}. 

 \begin{remark}
   \label{R17III24.1}
   Some remarks on the labeling of the intersection points in graphs such as  Figure~\ref{F28I24.1} are in order.

   We start by noting that for any parabolic solution we have  $\tr\,\mono=2(-1)^n$, so $\tr\,\mono=2$ requires an even number of zeros on any interval of length $2\pi$, with zeros occurring simultaneously at the beginning and at the end of an interval of length $2\pi$ counted only once.  Similarly,  a parabolic solution with $\tr\,\mono=-2$ requires an odd number of zeros.

   Now, for a parabolic solution with  $\psi_2(x_0)=0$ we have
   \begin{eqnarray}
     \psi_1(x_0+2\pi) &=&
      (-1)^n\big( \psi_1(x_0) + q \psi_2(x_0)\big)
      =
      (-1)^n \psi_1(x_0)
      \,, \\
     \psi_2(x_0+2\pi) &=& (-1)^n \psi_2(x_0) = 0
     \,.
   \end{eqnarray}
   So if  $n$ is even, then the sign of $\psi_1$ remains the same at each pair of $2\pi$-spaced zeros of $\psi_2$, and changes otherwise.

   If $\psi_1(\hat x_0)=0$ we have instead
   \begin{eqnarray}
     \psi_1(\hat x_0+2\pi) &=&
      (-1)^n\big( \psi_1(\hat x_0) + q \psi_2(\hat x_0)\big)
      =
     q  (-1)^n \psi_2(\hat x_0) \ne 0
      \,, \\
     \psi_2(\hat x_0+2\pi) &=& (-1)^n \psi_2(\hat x_0)  \ne 0
     \,.
   \end{eqnarray}
   So the sign of   $ \psi_2(\hat x_0+2\pi)$ can be determined according to the value of $n$ as above,
    while that of $  \psi_1(\hat x_0+2\pi)$ determines $q$.

   We emphasise that the above properties are independent of whether or not the Hill functions are in the canonical form \eqref{16I24.2}-\eqref{16I24.3}.

   Note also that  the locations  of the zeros of a Hill function vary smoothly under smooth variations of $f$ and of $\mu$;
   this follows from the implicit function theorem,
   as the derivative of a Hill function does not vanish at its zeros. When varying $\hm$
    at fixed remaining parameters it is convenient to rotate the circle so that the $\mcP_n^q$ function $\psi_2$ has a zero at $x=0$.
     Then $\psi_2$ has $n+1$ zeros in $[0,2\pi] $ when counting the zero at $x=2\pi$ and $\psi_1$ has $n$ zeros in $[0,2\pi] $, with $\psi_1(2\pi)\ne 0$.

   Once $n$ has been determined, the value of $q$ can  be found from our explicit expressions for $\mono$, see Lemma~\ref{L9IV24.1}.
   \qed
 \end{remark}

\subsection{Gluing BK solutions with zero mass}
 \label{pess6XI23.1}

Consider a CSC ALH manifold $(\chM,\chg)$ with parabolic monodromy. For definiteness, we will only consider these manifolds which can be transformed to a BK solution with zero mass, i.e.\ hyperbolic cusps. The calculations so far are readily adapted to this case, as we show in what follows.

We choose the following Wronskian-normalised basis of solutions of the Hill equation:
 \begin{equation}
  \cpsi_1 (\red{x}) =x
  \,,
  \qquad
  \cpsi_2 (\red{x})=
    1
  \,.
\end{equation}
The associated monodromy matrix is upper-triangular
 \begin{equation}\label{pe2XII23.1}
   \chmonoz
   =
   \left(
\begin{array}{cc}
    1
      & 2\pi \\
 0  &
    1
\end{array}
\right)
   \,.
 \end{equation}
Any  function satisfying
 \begin{equation}
 \chf|_{[-\frac{\pi}{2}, \frac{\pi}{2}]} (\red{\varphi}) =
 \tan
  \left( \frac{1}{2}  \arcsin
  \Big(
  \frac{\sqrt{1-\cbeta ^2}\sin
    \big(\red{\varphi}\big)
      }
   { 1 + \cbeta  \cos
    \big(\red{\varphi}\big)
    }
    \Big)  \right)
    \label{pe7XI23.1a}
 \end{equation}
with $\cbeta  \in [0, 1)$
transforms
the pair of functions $(\cpsi_1, \cpsi_2)$ to
\begin{equation}
   \label{pefunctionshill}
    \left(
   \begin{array}{c}
     \cpsi_1  \\
     \cpsi_2
   \end{array}
   \right)
    \Big|_{[-\frac{\pi}{2}, \frac{\pi}{2}]}
   \mapsto
   \sqrt 2 \left(
   \begin{array}{c}
    \big(\frac{1-\cbeta}{1+\cbeta} \big)^{1/4}\sin
  \left( \frac{\red{x}}{2}  \right)  \\
       \bigl(\frac{1+\cbeta}{1-\cbeta} \bigr)^{1/4} \cos
  \left( \frac{\red{x}}{2}  \right)
   \end{array}
   \right) =
   \cR
   \left(
   \begin{array}{c}
     \sqrt 2 \sin
   \left(\frac{\red{x}}{2}\right) \\
      \sqrt 2  \cos \left(\frac{\red{x}}{2} \right)
   \end{array}
   \right)
   \,,
\end{equation}
where
\begin{equation}\label{7XI23.1}
  \cR=
   \left(
    \begin{array}{cc}
      \big(\frac{1-\cbeta}{1+\cbeta} \big)^{1/4} & 0\\
     0   &  \bigl(\frac{1+\cbeta}{1-\cbeta} \bigr)^{1/4}
    \end{array}
    \right)\,.
\end{equation}
This leads to the following monodromy matrix
\begin{equation}
 \cmono= \cR^{-1} \chmonoz \cR = \left(
\begin{array}{cc}
 1 & \red{2 \pi \ckappa}\\
 0 & 1 \\
\end{array}
\right)
\,,
 \label{12XII23.31}
\end{equation}
where
\begin{equation}
 \red{\ckappa}=  \sqrt{ \frac{1+\cbeta}{1-\cbeta}}
  \,
  \in [1,\infty)
  \,.
\end{equation}
and $\hkappa$ will be
defined analogously, with all checks replaced by hats.

\subsubsection{Cusp with positive}

We glue a boosted BK solution $(\hatM,\hatg)$ with mass $\hm$, with monodromy given by \eqref{8XII23.2}, to  a parabolic solution with monodromy \eqref{12XII23.31}. Proceeding as in the previous sections, we find
\begin{equation}
  \frac 12 \tr\, \mono
  = - \frac 12 \tr \big( \cmono \hmono)
  =
  \cosh (\sqrt{\hm}\pi)
   \Big( \red{2 \pi \ckappa}\, \newhshiftH  -1
  \Big)
  \,.
\end{equation}
We conclude that the final manifold is, near the conformal boundary at infinity, a BK metric with positive mass $\mc$ satisfying
 \ptcheck{12XII}
\begin{equation}
 \fbox{$
\cosh (\sqrt{\mc}\pi)
  =
  \cosh (\sqrt{\hm}\pi)
   \Big( \red{2 \pi \ckappa}\,\newhshiftH  -1
  \Big)
  $, where $\newhshiftH > 1$, $\check{\ngkappa}\in [1,\infty)$.}
\end{equation}

\subsubsection{Cusp with cusp}
Given a second BK solution  ($\hatM, \hg)$ with zero mass
we use any diffeomorphism $\hf$ of $S^1$ satisfying
 \begin{equation}
 \hf|_{[\frac{\pi}{2}, \frac{3 \pi}{2}]} (\red{x})=
 - \tan
  \left( \frac{1}{2}  \arcsin
  \Big(
  \frac{\sqrt{1-\hbeta ^2}\sin
    \big(x\big)
      }
   { 1 - \hbeta  \cos
    \big(x\big)
    }
    \Big)  \right)
   \,,
    \label{pe7XI23.1a+}
 \end{equation}
 with a boost parameter
 $\hbeta \in [0,1)$.

Under the transformation \eqref{e15X23.5a} with $\chf$ replaced by $\hf$ we find now
\begin{equation}
   \label{ppefunctionshill2}
    \left(
   \begin{array}{c}
     \hpsi_1  \\
     \hpsi_2
   \end{array}
   \right)
   \mapsto
   \hR
   \left(
   \begin{array}{c}
     \sqrt 2 \sin
   \left(\frac{\red{x}}{2}\right) \\
      \sqrt 2  \cos \left(\frac{\red{x}}{2} \right)
   \end{array}
   \right)
   \,,
\end{equation}
with
\begin{equation}\label{12XII23.2}
  \hR = \left(
    \begin{array}{cc}
     0 &
       -\sqrt[4]{\frac{1-\hbeta}{1+\hbeta }} \\
     \sqrt[4]{\frac{1+\hbeta }{1-\hbeta}}
       & 0 \\
    \end{array}
    \right)
    \equiv \left(
    \begin{array}{cc}
     0 &
       -\hkappa^{-1/2}\\
   \hkappa^{1/2}
       & 0 \\
    \end{array}
    \right)
  \,.
\end{equation}
%
This leads to
\begin{equation}
  \hmono= \hR^{-1} \hmonoz \hR = \left(
    \begin{array}{cc}
     1 & 0 \\
     -\red{2 \pi \hkappa}& 1 \\
    \end{array}
    \right)\,.
\end{equation}

Consider thus a Maskit gluing of two hyperbolic cusps, each of them a complete hyperbolic manifold with vanishing mass. Letting $\cbeta>0$ and $\hbeta>0$ denote the boost parameters arising during the gluing procedure, using \eqref{12XII23.31} we obtain a manifold which, asymptotically, can be transformed to the BK form, with mass parameter $\mc>0$ given by
 \ptcheck{12XII}
 \begin{equation}
 \fbox{
  $\cosh(\sqrt{\mc} )= -\frac 12 \tr (\cmono\hmono)=
  (2 \pi)^2\red{\ckappa}\, \red{\hkappa}-1$, where $\check{\ngkappa},\, \red{\hkappa}\in [1,\infty)$.
   }
    \label{12XII23.1}
 \end{equation}
It follows in vacuum (see Theorem \ref{t26XII23.1})
that the glued manifold will contain an outermost closed geodesic, of length $2\pi \sqrt{ \mc}$, which separates both cusps from the ALH region.

\subsubsection{Cusp with negative}
We glue a boosted BK solution $(\hatM,\hatg)$ with mass $\hm <0$, with monodromy given by \eqref{negative121223}, to  a parabolic solution with monodromy \eqref{12XII23.31}.
Proceeding as before we find the monodromy matrix of the glued manifold
\ptcheck{19724, with mathematica}
\begin{equation}\label{21I24.1}
\mono = \left(
\begin{array}{cc}
 2 \pi  \ckappa \hkappa \sin \left(\pi 
   \sqrt{| \hm| }\right)-\cos \left(\pi  \sqrt{|
   \hm| }\right) & -\frac{2 \pi  \ckappa
   \hkappa \cos \left(\pi  \sqrt{| \hm|
   }\right)+\sin \left(\pi  \sqrt{| \hm|
   }\right)}{\hkappa} \\
 \hkappa \sin \left(\pi  \sqrt{| \hm|
   }\right) & -\cos \left(\pi  \sqrt{| \hm| }\right)
   \\
\end{array}
\right)
\end{equation}
with half-trace equal to
\begin{equation}
  \frac 12 \tr\, \mono
  =
  \pi
  \check{\ngkappa}\red{\hkappa}
      \hsinx-\hcosx
  \,.
   \label{31I24.1}
\end{equation}
We will get a parabolic monodromy if
\begin{equation}
  \hcosx
  =
  \pm 1 +
  \pi
  \red{\ckappa}\red{\hkappa}
      \hsinx
  \,.
\end{equation}

\begin{figure}
   \centering
   \includegraphics[width=.5\textwidth]{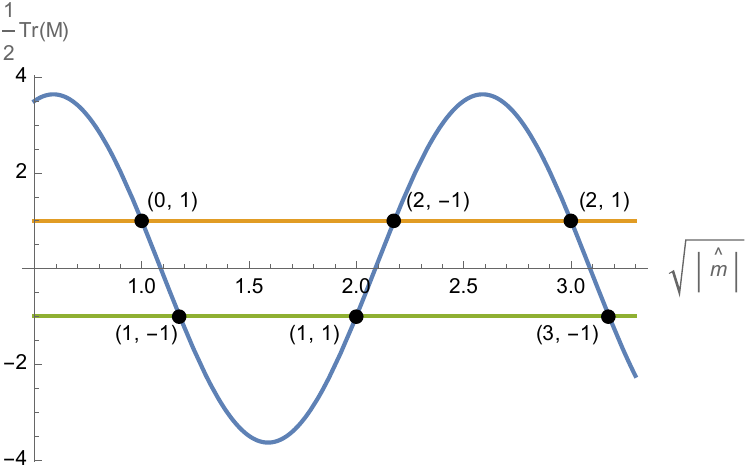}
   \caption{The right-hand side of \eqref{31I24.1} with
   ${\pi \red{\ckappa}} {\red{ \hkappa}}=3.5$
   and with
   $\sqrt{\left|\hm\right|}\in [0.5,3.5] $. The labels $(n,q)$ indicate the parabolic type $\mcP_n^q$ of the intersection points.
   The plot starts at 0.5 because of the constraint \eqref{hkappares5324}.
   }\label{F28I24.1a}
 \end{figure}

The matrix \eqref{21I24.1} is obviously parabolic if $\sqrt{|\hm|} \in \mathbb{N}$
with
\begin{equation}
q= 1
 \,,
\end{equation}
and one  checks (cf.\ Remark~\ref{R13II23.1}) that the \emph{remaining parabolic} solutions have monodromy  conjugate to
\begin{equation}
\pm\left(
  \begin{array}{cc}
    1 & q \\
    0 & 1
  \end{array}
  \right) \quad
\mathrm{with} \quad q = \mp
 \mathrm{sgn}(
  \sin(\sqrt{|\hm|}\pi) )=-1\,,
\end{equation}
see
Figure~\ref{F28I24.1a}. 

If $\left|\tr\, \mono\right| < 2$, we conclude that the final
 manifold is, near the conformal boundary at infinity,
 a BK metric with negative mass $\mc <0$ satisfying
\begin{equation}
 \fbox{$
   \displaystyle
 \cos \left(\sqrt{\left| \mc\right| }\pi  \right) =
 \pi
 \red{\ckappa}\red{\hkappa}
    \hsinx-\hcosx
  $, where $\red{\ckappa} \in [1, \infty)$ and
  $\red{\hkappa}\in [1,\infty)$.}
\end{equation}
If $-1/4 \leq \hm < 0$, $\hkappa$ is subject to the additional restriction
\eqref{hkappares5324}.

\section{All time-symmetric vacuum solutions near the conformal boundary}
 \label{s27X23.1}
 Let us apply the coordinate transformation \eqref{change2} to the metric
 $b$ given by \eqref{27VII23.91s}:
one finds
  \begin{align}
   g - \hat b
     &=
    \frac{f''(\hat{\varphi})^2
    }{\hat{r}^2
    f'(\hat{\varphi})^2}
   (\hat{\theta}^2)^2
     -   \frac{f^{(3)}(\hat{\varphi}) f''(
      \hat{\varphi})}{ {\hat r}^3
     f'(\hat{\varphi})^2}\hat{\theta}^1\hat{\theta}^2 \nonumber \\
     &~~
     + \left(\frac{f''(\hat{\varphi})^2-f^{(3)}(\hat{\varphi}) f'(\hat{\varphi})}{\hat{r}^2  f'(\hat{\varphi})^2}  +\frac{f^{(3)}(\hat{\varphi})^2}{4 \hat{r}^4
     f'(\hat{\varphi})^2}\right) ( \hat{\theta}^1)^2\,.
      \label{5IX23.3ex}
  \end{align}
This is a metric with constant scalar curvature $-2$, as $g$ in \eqref{5IX23.3ex} differs from \eqref{27VII23.91s} by a coordinate transformation. This is true \emph{locally} regardless of whether or not the map $\varphi \mapsto f(\varphi)$ is a diffeomorphism. The associated mass-aspect function is obtained from \eqref{finalmucc},
  \begin{align}
     \hat \mu &=  - 2\red{S(f)(\hat \varphi)}
     \,.
     \label{Afinalmucc}
  \end{align}
Locally, any function $\hat \mu$ can be obtained from this formula.

However, it is not clear whether or not there are any global obstructions here.
In order to settle this we consider an asymptotically locally hyperbolic metric of the form
\begin{equation}
  g = x^{-2} \left( dx^2 + g_{\varphi\varphi}(x, \varphi) d\varphi^2
  \right)\,,
 \end{equation}
 where $x = 1/r$.
 By expanding the function $g_{\varphi\varphi}$ as
 \begin{equation}
  \label{expansion}
 g_{\varphi\varphi} (x, \varphi) = 1+ f_1(\red{\varphi}) x + f_2(\red{\varphi}) x^2 + f_3(\red{\varphi}) x^3 + f_4(\red{\varphi}) x^4 + O(x^5)
 \end{equation}
 and solving the CSC equation (equivalently, the time-symmetric vacuum constraint equation)
 \begin{equation}
  R = -2
 \end{equation}
 order by order, one finds that the expansion  \eqref{expansion} stops
 \begin{equation}
  g_{\varphi\varphi}(x, \varphi) = 1+  f_2(\red{\varphi}) x^2 + f_4(\red{\varphi}) x^4\,,
 \end{equation}
 where $f_2$ is arbitrary and
 \begin{equation}
f_{4}(\red{\varphi}) = \frac{f_2(\red{\varphi})^2}{4}\,.
 \end{equation}
 The fact that the Fefferman-Graham expansion stops in two spatial dimensions has been of course already pointed out in
 \cite{Skenderis:1999nb, Banados:1998gg}.

Returning to the coordinate $r=1/x$ and using the coframe \eqref{11XI23.2} leads to
%
\begin{equation}
  g = r^{-2}  dr^2 + \big(
   r^2 + f_2(\varphi)   +  \frac{f_2(\varphi)^2}{4 r^2}
   \big)
    d\varphi^2
    = (\theta^2)^2 + \big(
   1 + \frac{f_2(\varphi)}{2r^2}
   \big)^2
    (\theta^1)^2
  \,.
   \label{9I24.1}
 \end{equation}
From this we can read off the mass-aspect function
\begin{equation}\label{11XI23.1}
  \mu  = \mu_{22} + 2 \mu_{11} = 2 f_2
  \,.
\end{equation}
 We conclude that:

\begin{Proposition}
  \label{P11XI23.1}
For any function $  \mu$
there exists an ALH CSC
 metric defined near the conformal boundary at infinity for which $\mu$ is the mass-aspect function.
\end{Proposition}
The question then arises, which mass-aspect functions can be obtained at boundaries of well behaved ALH  general relativistic initial data sets with matter satisfying the dominant energy condition. The results of Section~\ref{s16I22.3} suggest strongly that the exotic $\mcH_n$-type and $\mcP_n^q$-type mass aspects with $n\ge 1$ will not occur in such data sets. So some sort of singularity in the metric is expected in such cases. Consistently with this, we have:
\begin{theorem}
  \label{t19I24.2}
  \begin{enumerate}
    \item
  Mass-aspect functions of $\mcE_{-1}$-, $\mcH_0$- and $\mcP_0$-type can be realised by smooth,
   geometrically finite, complete ALH CSC manifolds.
\item   Mass-aspect functions of $\mcP_n^{q}$-type and  of $\mcH_n$-type with $n \geq 1$,
   and of $\mcE_{m}$-type with $-1\ne m<0$,
   can be realised by ALH CSC metrics on  manifolds which have the same properties  except for admitting one conical point.
  \item Mass-aspect functions of  $\mcP_n^{q}$-type, $n\ge 1$, can be realised by  conformally compact ALH CSC metrics which are smooth except for two conical points.
  \end{enumerate}
\end{theorem}

\proof
In view of Corollary~\ref{C26XII23.1}, and taking into account that  negative-mass BK metrics provide an exhaustive list of examples for the $\mcE_{\mc}$  family, it remains to construct examples with  mass-aspect functions of $\mcH_n$-type  and $\mcP_n^q$-type  with $n\ge 1$.

For this consider the gluing, as in Section~\ref{ss8XII23.1}, of a BK metric with negative mass $\hm$ with a BK metric with positive mass $\chm$.
Recall \eqref{19I24.11},
\begin{equation}
 \frac 12 \tr\, \mono
   =\cosh (\sqrt{|\chm|}\pi)
   \left(
   \hkappa
   \,
   \newcshiftH
   \hsinx-\hcosx
   \right)
   \,.
    \label{19I24.11a}
 \end{equation}
where $  \newcshiftH>1$ and $\hm <0$. Both $\hm$ and $  \newcshiftH $ can be freely chosen within the given ranges. The construction of  Section~\ref{ss8XII23.1} provides Hill functions which have between $n:=\lfloor \sqrt{|\hm|}\rfloor$
and  $n+4$ zeros;  we will return to this shortly.

In order to obtain a candidate for a parabolic monodromy matrix we need to solve  the equation
\begin{equation}\label{19I24.p2a}
  \tr\, \mono = \pm 2
  \qquad
  \Longleftrightarrow
  \qquad
   \cosh (\sqrt{\chm} \pi)
    \big(
     \hkappa
     \newcshiftH
     \sin (\sqrt{\left| \hm\right| }\pi)-
     \cos ( \sqrt{\left|\hm\right|}  \pi)
     \big) = \pm 1
     \,.
\end{equation}
For any given $\chm >0 $ and  $  \newcshiftH >1$ this equation with the plus sign has two solutions as
$\lfloor \sqrt{|\hm|}\rfloor $ varies between $n$ and $n+2$, and two solutions with the minus sign, providing the desired mass-aspect functions of $\mcP_n^q$-type; see Figure~\ref{F28I24.1} for a typical plot.

Finally, varying $\chm >0 $,  $ \hkappa \newcshiftH >1$ and $\hm$ within the ranges
\begin{equation}\label{19I24.p2ah}
  \pm  \cosh (\sqrt{\chm} \pi)
    \big(
      \hkappa \newcshiftH  \sin (\sqrt{\left| \hm\right| }\pi)-
     \cos ( \sqrt{\left|\hm\right|}  \pi)
     \big) >1
     \,
\end{equation}
covers the family of mass-aspect functions of $\mcH_n$-type.

See Figure~\ref{F22324.1} 
for the justification of point 3.
\qedskip

\appendix

\section{Boosting asymptotically locally hyperbolic metrics}
\label{boostappendix}

 The aim of this appendix
is to analyse the behaviour of the mass-aspect function under boosts of a hyperbolic background.
 For simplicity, we also assume here that
 the metric has an asymptotic expansion at infinity of the form \eqref{29VII23.81}
 with a well-defined mass-aspect function.

Recall that
the upper sheet of the hyperboloid
\begin{subequations}
   \label{hyp}
\begin{equation}
\label{hyp1}
 \mcHtwo:= \{t^2 - x^2 -y^2=1
  \}
\end{equation}
in $\mathbb{R}^{1, 2}$
\begin{equation}
   \label{hyp2}
ds^2 = - dt^2 +dx^2 +dy^2
\end{equation}
\end{subequations}
can be parameterised as
\begin{equation}\label{15I22.1}
  t =  \sqrt{1+ r^2}
  \,,
  \quad
   x =  r \cos(\red{x})
  \,,
  \quad
   y =  r \sin(\red{x})
   \,.
\end{equation}
In the  coordinates $(r,\varphi)$, the metric $\hypmet$ induced on $\mcHtwo$ reads
\begin{equation}\label{15I22.2}
  \hypmet =
    \frac{d r^2}{1+ r^2} + r^2 d\varphi^2
  \,.
\end{equation}

Consider a boost $x^\mu \mapsto \bar x^\mu = \Lambda(\beta)^\mu{}_\nu  x^\nu$ with velocity $\beta \in (-1,1)$:
\begin{subequations}
   \label{boosts1}
\begin{eqnarray}
 \label{15II21.5a}
  \sqrt{1+\bar r^2} &=&  \gamma (\sqrt{1+ r^2} - \beta r \,\cos \varphi)
  \,,
\\
  \bar r \cos \bar \varphi &=& \gamma
   ( r \cos \varphi
  - \beta \sqrt{1+ r^2} ) \label{boostscosinus}
  \,,
\\
  \bar r \sin \bar \varphi &=& r \sin \varphi
  \,,
 \label{15II21.5c}
\end{eqnarray}
\end{subequations}
where $\gamma = 1/\sqrt{1-\beta^2}$, and where the inverse map can be obtained from the above by replacing $\beta$ with $-\beta$.
This implies
 \ptcheck{5IX23; rechecked that the sum of squares is one on 10XII23}
\begin{eqnarray}
  r  &=&  \gamma \bar{r} (1 + \beta \cos \bar \varphi) + O(\bar{r}^{-1})
  \,,
\\
   \sin \varphi &=&  \frac{ \sin \bar \varphi}{\gamma (1+\beta \cos \bar \varphi)}
    + O (\bar{r}^{-2}) \label{trafobarvarphi}
    \,,
\\
    \cos \varphi &=& \frac{\cos \bar \varphi + \beta}{1+ \beta \cos \bar \varphi} + O (\bar{r}^{-2}) \label{trafobarvarphicos}
\end{eqnarray}
and
\ptcrnh{equivalently, in the limit, $\tan (\bar\varphi/2) = \sqrt{\frac{1+\beta}{1-\beta}}
\tan (\varphi/2)$}
\begin{eqnarray}
   d r  &=&   \gamma (1+\beta \cos\bar  \varphi) d\bar r
   - \gamma \beta \sin \bar \varphi \, \bar r d\bar \varphi
    + O (\bar r^{-2}) d \bar r
     + O (\bar r^{-1}) d \bar \varphi\,, \\
     d \varphi &=& \frac{d \bar \varphi}{\gamma(1+ \beta \cos(\bar \varphi))} + O (\bar r^{-2}) d \bar \varphi
     +  O (\bar r^{-3}) d \bar r
  \,.
\end{eqnarray}

We would like to understand how a metric, which asymptotes to a hyperbolic metric for large $r$ and which has a well-defined
mass aspect tensor, transforms under boosts.
This is a direct calculation  if the mass aspect tensor is defined with respect to a hyperbolic background $\mathring b$, since then the background is invariant under boosts. However, we used   \eqref{29VII23.81} to define $\mu_{ij}$:
\begin{align}
    g &= b +  r^{-2}  \mu_{i j}(\red{x})  \theta^i \theta^j
     + O(r^{-3}) \label{29VII23.81appendix}
      \,,
 \end{align}
with
\begin{equation}
b = \frac{dr^2}{r^2} +r^2 d \varphi^2
\end{equation}
and
 where the size of the error terms is understood as the size of components in the $b$-orthonormal coframe \eqref{27VII23.93}:
 \begin{equation}
    \{{\theta}^2 = \frac{d {r}}{{r}}\,,  {\theta}^1 = {r} d {\varphi}\}\,.
 \end{equation}
After performing the boost, we need to take into account the fact that the new background metric is
\begin{equation}
 \bar{b} = \frac{d \bar r^2}{\bar r^2} + \bar r^2 d \bar \varphi^2
\end{equation}
where $(\bar r,\bar \varphi)$ have been defined in \eqref{boosts1}.
The corresponding $\bar b$-ON coframe reads
\begin{equation}
  \label{barcoframe}
\{\bar \theta^{\twor}= \frac{d\bar r}{ \bar r} \,,\ \bar \theta^{\onephi} = \bar r d\bar \varphi
     \}\,.
\end{equation}
 Using the change of coordinates \eqref{15II21.5a}-\eqref{15II21.5c}, the fact that
 \begin{equation}
   \theta^i \theta^j = {\bar \theta}^i {\bar \theta}^j + O({\bar r}^{-1})
   \,,
 \end{equation}
and the rewriting of \eqref{29VII23.81appendix} as
\begin{equation}
 g= \frac{dr^2}{r^2+1} +r^2 d\varphi^2 + r^{-2}\theta^{\twor} \theta^{\twor}+  r^{-2}  \mu_{i j}(\red{x})  \theta^i \theta^j
     + O(r^{-3})
      \,,
 \end{equation}
 where the sum of the first two terms is invariant under $\Lambda(\beta)$,
 we readily obtain
     \begin{multline}\label{15I22.3}
      \bar g =   \bar{b} + \bar{r}^{-2}\left(
                     -1 + \frac{(1+ \mu_{\twor \twor}(\omofvarphi ))}{\gamma^2(1+ \beta \cos \bar{\varphi})^2}\right) (\bar \theta^{\twor})^2
                        \\
                     +
                     \bar{r}^{-2} \frac{\mu_{\onephi \onephi}( \omofvarphi )}{\gamma^2(1+\beta \cos\bar\varphi)^2}
                       (\bar \theta^{\onephi})^2
                     + \bar{r}^{-2}\frac{2 \mu_{12}(\omofvarphi ) }{\gamma^2(1+\beta \cos\bar \varphi)^2} \bar \theta^{\onephi}    \bar \theta^{\twor}
                       +  O(\bar r^{-3})\,,
     \end{multline}
     where $\varphi= \varphi(\bar \varphi)$  is understood now by
     taking the limit $r\to\infty$ in \eqref{trafobarvarphi},
     \begin{equation}
       \sin  (\omofvarphi):= \frac{\sin \bar{\varphi}}{\gamma (1+ \beta \cos(\bar\varphi))}
       \,.
     \end{equation}
     Here, the error terms are again to be understood as error terms for the metric coefficients  with respect to the  $\bar b$-ON coframe \eqref{barcoframe}.
The resulting mass-aspect function, say $\bar \mu$, reads
\begin{equation}\label{15I22.4}
  \bar \mu(\bar \varphi)
  =  \frac{(2
  \mu_{\onephi \onephi}(\omofvarphi )+
  \mu_{\twor \twor}(\omofvarphi )
  +1)}{\gamma^2(\beta  \cos (\bar{\varphi})+1)^2}-1
  \equiv
   \frac{( \mu(\omofvarphi )
  +1)}{\gamma^2(\beta  \cos (\bar{\varphi})+1)^2}-1
                \,.
\end{equation}

In the special case of  a Birmingham-Kottler metric,
\begin{equation}
g = \frac{dr^2}{r^2 -\mc } + r^2 d \varphi^2
= b +  \left( \frac{\mc}{r^2}+O(r^{-4}) \right) \theta^{\twor} \theta^{\twor}\,,
\end{equation}
the original mass-aspect function $\mu =\mu_{\twor \twor}=\mc$
becomes
\begin{equation}
    \bar{\mu}(\bar \varphi) =
    \frac{(1+   \mc)}{\gamma^2(\beta  \cos (\bar{\varphi})+1)^2}-1
             \,.
\end{equation}
The resulting Hamiltonian mass, say $\bar \Hzero$, is
\begin{equation}
   \label{massafterboost}
\bar \Hzero = \frac{1}{2\pi}\int_{\bar{\varphi}= 0}^{\bar{\varphi}= 2 \pi} \bar{\mu} d \bar{\varphi}
 = \frac{(1 + \mc)}{\gamma^2 (1- \beta^2)^{3/2}} - 1
 =
 \gamma (1 + \mc) - 1
  \,.
\end{equation}
\emph{Not} unexpectedly, $\bar \Hzero$ is $\gamma$-indendent when $\mc=-1$.

 \ptcrnh{should be finished by calculating the energy-momentum vector that results; needs both KIDs and the formula for the momentum vector; one could even apply an isometry of AdS to obtain the center of mass? .... perhaps later}
\ptcrnh{HillGeo.tex commented out, an inconclusive suggestion of Arick Shao; replaced by the beginning of an attempt using CMC, not clear if it goes anywhere and calculations not finished, so file HillCMC commented out too}

\bibliographystyle{amsplain}
\bibliography{ChruscielWutte-minimal}
\end{document}